\documentclass[11pt]{article}
\usepackage{xcolor}

\usepackage{jheppub}
\usepackage{amssymb,amsfonts,amsmath,amsthm,lineno}
\usepackage{comment}
\usepackage{enumerate}
\usepackage{mathrsfs}
\usepackage{bbold}
\usepackage{tikz}
\usetikzlibrary{cd, shapes.geometric, arrows}
\usepackage{hyperref}
\usepackage{cleveref}
\usepackage{slashed}
\usepackage{color}
\usepackage{empheq}
\usepackage{soul}
\usepackage{booktabs}
\usepackage{multirow}
\usepackage{todonotes}

\usepackage{tabularx}
\usepackage{placeins}

\usepackage{rotating}

\usepackage{amsmath}
\makeatletter
\newcommand{\Vast}{\bBigg@{4.75}}
\makeatother

\newcommand{\dd}{\mathrm{d}}
\newcommand{\cO}{\mathcal{O}}

\newcommand{\Ptwo}{P_2(\cos\theta)}

\def\g{\gamma}

\newcommand{\be}{\begin{equation}}
\newcommand{\ee}{\end{equation}}
\newcommand{\bea}{\begin{eqnarray}}
\newcommand{\eea}{\end{eqnarray}}

\newcommand{\cM}{\mathcal{M}}
\newcommand{\cE}{\mathcal{E}}

\newcommand\qt\tau

\makeatletter
\renewcommand{\@seccntformat}[1]{\csname the#1\endcsname.\,\,}
\makeatother
\allowdisplaybreaks

\let \savenumberline \numberline
\def \numberline#1{\savenumberline{#1.}}

\makeatletter
\def\@fpheader{\relax}
\makeatother

\def\bea{\begin{eqnarray}}
\def\eea{\end{eqnarray}}

\def\nn{\nonumber}

\usetikzlibrary{shapes,arrows,chains}
\usetikzlibrary{decorations.markings}
\usetikzlibrary{decorations.pathmorphing}
\usetikzlibrary{shapes.multipart}
\tikzset{snake it/.style={decorate, decoration=snake}}

\usepackage{tikz-cd}
\usetikzlibrary{cd,decorations.markings, arrows.meta}

\usepackage{xcolor}
\definecolor{ogreen} {RGB}{71,191,145}
\definecolor{edit} {RGB}{123,150,145}
\definecolor{purple} {RGB}{148,0,211}

\title{Spin and Quadrupole Sectors in Nonrelativistic Gravity}

\author[]{Utku Zorba}

\emailAdd{utku.zorba@iuc.edu.tr}
\affiliation[]{Department of Engineering Sciences, Istanbul University-Cerrahpaşa\\
Avcilar 34320 Istanbul, T\"{u}rkiye
}
\abstract{We study the large-\(c\) expansion of general relativity in ADM variables.
Using a unified even \(\omega\)-expansion, the ADM formulation gives a common
starting point for Galilean and Carrollian limits. We focus on the
Galilean branch and derive the ADM action and field equations up to NNLO.
We then construct stationary vacuum solutions of the Galilean ADM equations up to NNLO in weak and strong branches. In the weak branch, we find NLO Kerr-type, Hartle--Thorne-type and mixed-type solutions. The NLO weak equations also allow a simple extension to higher
mass multipoles. At NNLO, the weak Kerr-type and extended
Hartle--Thorne-type sectors solve the equations separately, but their naive sum is not a solution. The nonlinear NNLO equations generate mixed \(J^2Q\) source terms, which
require additional corrections to the NNLO lapse and NNLO spatial tensor
field. This gives a mixed weak-branch Galilean solution of the truncated NNLO equations in the ADM
gauge, whose full relativistic completion is not fixed by the truncated data alone. In the strong branch, Kerr-type data solve the equations through NNLO while the strong Hartle--Thorne-type data solve the  NLO equations. We also explain how the ADM data can be reconstructed order by order into approximate spacetime metrics. These reconstructed metrics should not be interpreted by themselves as unique full relativistic completions. Since these metrics include spin, quadrupole and mixed spin--quadrupole effects, they may be useful for studying the spacetime around rotating compact objects such as black holes and neutron stars.}

\begin{document}

\maketitle
\vfill\eject

\section{Introduction}
\label{sec:introduction}

The large-\(c\) expansion of general relativity provides a systematic way
to derive nonrelativistic gravity. In the covariant approach, the
Einstein--Hilbert action is expanded order by order in \(1/c\). The NLO
theory obtained in this way is a nonrelativistic theory based on type II
Newton--Cartan geometry
\cite{VandenBleeken:2017rij,Hansen:2019pkl,VandenBleeken:2019gqa,
Hansen:2020pqs,Hartong:2022lsy}. This construction has also pointed to
several open directions, including the Hamiltonian formulation of the
resulting theories, the role of asymptotic charges, and the expansion of
rotating solutions such as Kerr. The large-\(c\) expansion of the Kerr metric has already been studied in \cite{Hartong:2023yxo}, where the resulting nonrelativistic geometry
was related to Lense--Thirring and Hartle--Thorne-type data after an
appropriate coordinate choice. The relation between covariant large-\(c\)
expansions and post-Newtonian gravity has also been developed
systematically in Ref.~\cite{Hartong:2023ckn}, including the role of
post-Newtonian gauge choices and the matching between near-zone and
exterior-zone expansions. A related \(3+1\) formulation was developed in
\cite{Elbistan:2022plu}, where similar nonrelativistic actions were
obtained using a Kol--Smolkin-type parametrization
\cite{Kol:2007bc,Kol:2010ze,Kol:2010si}, and their relation to
post-Newtonian gravity was discussed up to NNLO. Related first-order and
algebraic constructions of Galilean and Carrollian gravity were also
studied in \cite{Guerrieri:2020vhp,Ekiz:2022wbi}.

In this work we continue the ADM strategy that was first implemented for
the Carrollian expansion in \cite{Bal:2026xup}. The odd large-\(c\)
expansion of the ADM action has also been studied in \cite{Elbistan:2025vyh}.  The ADM decomposition is particularly useful for stationary backgrounds,
because it separates the lapse, shift and spatial metric fields and gives
a direct order-by-order read-off of the nonrelativistic variables. We
first organize the ADM fields by means of a unified \(\omega\)-expansion.
With one scaling of \(\omega\), this expansion gives the Galilean
large-\(c\) theory, while with the opposite scaling it gives the
Carrollian small-\(c\) theory. Thus the same ADM starting point provides
a common framework for both non-Lorentzian limits. The Carrollian
branch, including its stationary solutions, has been studied in
\cite{Hansen:2021fxi,Bal:2026xup}. Here we focus on the Galilean branch and analyze its ADM action, field equations and stationary vacuum solutions.

The Galilean and Carrollian limits are complementary in their physical scope. The Carrollian limit is naturally associated with null hypersurfaces and the conformal null boundary of asymptotically flat spacetimes, and therefore plays a direct role in flat-space holography \cite{Donnay:2022aba,Ruzziconi:2026bix}. Galilean and Newton--Cartan structures, on the other hand, are relevant both for nonrelativistic gravity and for non-AdS holographic settings such as Lifshitz holography \cite{Christensen:2013lma}; in the present context, the Galilean expansion also connects with post-Newtonian physics \cite{Hartong:2023ckn} and organizes mass, spin and higher multipole data order by order. This makes the ADM expansion useful both for exploring holographic and asymptotic aspects of non-Lorentzian gravity and for applications to rotating compact objects.

First, we construct the Galilean expansion of the ADM action and field
equations up to NNLO. Second, we use this formulation to find explicit
stationary solutions and to understand how relativistic rotating and
multipolar geometries are encoded in the Galilean variables. Stationary
axisymmetric spacetimes provide a natural setting for studying multipole
moments and deviations from Kerr, with the Ernst formulation
\cite{Ernst:1968,Ernst:1968II} and the Manko--Novikov class
\cite{Manko:1992} serving as useful reference points. In this way, we
organize stationary spin and quadrupole sectors of the Galilean ADM
theory and distinguish the data fixed by the truncated Galilean equations
from the data fixed by matching to relativistic expansions, together
with their weak- and strong-branch realizations. In this sense, the purpose of the construction is not merely to reproduce
large-\(c\) expansions of known relativistic metrics, but to use the ADM
Galilean equations as an intrinsic order-by-order solution-building
framework.

A large-\(c\) expansion of a known relativistic solution automatically
satisfies the consistently expanded Einstein equations at each retained
order, once the variables, coordinates and gauge are matched. The
converse, however, is not guaranteed: a solution of the truncated
Galilean ADM equations need not admit a full relativistic completion,
and such a completion, if it exists, need not be unique.

In the weak branch, the leading spatial data are flat and the solutions
can be built order by order. At NLO, the equations are closely related to
Galilean-invariant nonrelativistic gravity and are linear in the NLO
fields. We find weak Kerr-type and Hartle--Thorne-type sectors at this
order, and their mixed weak superposition is also allowed because the NLO
equations are linear in the relevant fields. The same linear structure
also allows a simple extension to higher mass multipoles at NLO. These
backgrounds may be viewed as stationary solutions of Galilean-invariant
nonrelativistic gravity in the sense of
\cite{Hansen:2019pkl,Hansen:2020pqs}.

At NNLO the structure becomes more restrictive. At the same order, the weak
Kerr-type sector and the extended Hartle--Thorne-type sector solve the
Galilean equations separately. The extended Hartle--Thorne-type sector
requires the post-linear \(Q^2\) completion \cite{Frutos-Alfaro:2015lua}. However, the naive sum of these two  NNLO
sectors is not a solution by itself. The NNLO equations contain nonlinear
source terms built from the lower-order fields, and these generate mixed
spin--quadrupole terms proportional to \(J^2Q\). We show that these
residual terms can be cancelled by adding specific corrections to
the NNLO lapse and  spatial tensor. This leads to a mixed weak-branch NNLO solution of the truncated Galilean ADM equations in the ADM gauge. This solution is not obtained
by a naive superposition of the separate Kerr-type and quadrupolar
sectors. Its full relativistic completion is not fixed by the truncated
data alone and, if it exists, may require additional higher-order
fields, higher-shift data, and coordinate or gauge choices.

The strong branch has a different character. The leading ADM data are
already nontrivial, and we find strong Kerr-type stationary solutions at
both NLO and NNLO. By contrast, the strong Hartle--Thorne-type sector is
available only at NLO within the ordinary Hartle--Thorne metric. Under
the strong scaling, the Hartle--Thorne metric provides the required NLO
data, including an independent quadrupole parameter, but it does not
supply independent \(\mathcal{O}(c^{-4})\) data needed  for a generic NNLO 
strong-branch completion. Constructing such a solution would therefore
require solving the NNLO equations directly, or starting from a
higher-order Hartle--Thorne-type relativistic metric. 

The ADM data can also be used to reconstruct the corresponding spacetime metric components. This allows us to compare the
Galilean solutions with expanded relativistic metrics and to extract
standard quantities associated with stationary axisymmetric geometries.
For example, frame dragging follows from
\(\Omega_{\rm drag}=-g_{t\phi}/g_{\phi\phi}\). In the weak
mixed solution, the mixed \(J^2Q\) terms enter the reconstructed metric
through the higher-order spatial components at $\mathcal{O}(c^{-4})$. As a result, their direct
effect on \(\Omega_{\rm drag}\) does not appear at the NNLO order
considered here, but only at the next order, where additional higher-shift
data are also required. We therefore
interpret the reconstructed metric only as a finite-order ADM
reconstruction. It does not by itself define a unique full relativistic
solution; establishing such a completion requires the missing
higher-order data and gauge choices, together with an additional check
of the Einstein equations. Geodesic motion may provide a more sensitive
probe. However, a detailed analysis of the
resulting geodesic motion, orbital frequencies and possible precession
effects is left for future work.

The paper is organized as follows. In Sec.~\ref{sec:adm-expansion} we
develop the ADM expansion of general relativity. We first introduce the
unified \(\omega\)-expansion of the ADM variables and show how the
Galilean and Carrollian branches arise from different scalings. We then
expand the ADM action and the corresponding ADM constraint and evolution
equations order by order. After setting up this general framework, in
Sec.~\ref{sec:stationary-solutions} we focus on the Galilean branch and
study stationary vacuum solutions. We begin with the stationary reduction
of the Galilean field equations and the possible static leading--order backgrounds. We then discuss the weak branch, including the NLO weak
Kerr-type, weak Hartle--Thorne-type and mixed weak solutions, together
with their NLO multipolar extension. We next move to NNLO, where we
construct the weak Kerr-type solution, the weak extended
Hartle--Thorne-type solution, and  the truncated mixed weak-branch solution fixed by the \(J^2Q\) source terms. We also discuss the strong
branch, including the strong Kerr-type solutions and the limitations of
the strong Hartle--Thorne-type data at the NNLO. In Sec.~\ref{sec:conclusions} we
summarize the results and discuss possible future directions. The appendices collect the expansion details, the field equations, the ADM read-off maps, the relation to the NLO nonrelativistic gravity action, and the relativistic Kerr, Hartle--Thorne and post-linear Hartle--Thorne expansions used in the main text.


\section{Expansion of GR in ADM formalism} \label{sec:adm-expansion}

The expansion of the metric tensor, including higher-order corrections to the fields and the associated curvature terms, is highly nontrivial \cite{VandenBleeken:2017rij,Hansen:2019pkl,Hansen:2021fxi}. Since even the NLO expansion is technically involved, extending the analysis beyond this order becomes considerably more demanding. In the present work, we therefore employ the ADM decomposition, following the strategy used in our Carrollian small-$c$ expansion analysis \cite{Bal:2026xup}. This provides a unified framework in which the expansion can be organized systematically. We start from the ADM action \cite{Arnowitt:1962hi} \footnote{We suppressed the gravitational coupling in the definition of the action.}
\begin{equation}
S_{\text{ADM}} = 
    \int\,dt\,d^{3}x\, \hat N\,\sqrt{\hat \g}\, \left(\hat R\, + \hat K_{ij} \hat K^{ij}-\hat K^2\, \right)\,,
  \label{ADMAction}
\end{equation}
where $\hat{R}$  is Ricci scalar corresponding to the spatial metric $\hat \gamma_{ij}$ and the extrinsic curvature is defined as  
\begin{align}
	\hat K_{ij}&=\hat N^{-1}\left(\frac{1}{2\,c} \partial_t \hat \gamma_{ij}\,- \,\hat\nabla_{(i}\hat N_{j)}\right) \,.\label{Krel}
\end{align}
The parentheses denote symmetrization with unit weight. The corresponding relativistic metric in ADM gauge is then
\begin{equation} 
	\label{metrel}
	ds^2 = -c^2\, \hat N^2dt^2 + \hat \gamma_{ij} (dx^i + c\,\hat N^i dt) (dx^j + c\,\hat N^j dt)\,.
\end{equation}
Before moving on, let us discuss the structure of the ADM action. The explicit form of the action is 
\begin{equation}
S_{\mathrm{ADM}}
=
\int dt\,d^{3}x\,
\hat N\sqrt{\hat \gamma}\,
\left(
\hat R
+
\frac{\hat \gamma^{ikjl}}{\hat N^2}\,\left(
\frac{1}{4c^2}\,\dot{\hat \gamma}_{ij}\,\dot{\hat \gamma}_{kl}
-\frac{1}{c}\,\dot{\hat \gamma}_{ij}\,\hat \nabla_{(k}\hat N_{l)}
+ \hat \nabla_{(i}\hat N_{j)}\,\hat \nabla_{(k}\hat N_{l)}
\right)
\right)\,, \nn
\end{equation}
where $\hat \g^{ikjl} = \hat \g^{ik} \hat \g^{jl}- \hat \g^{ij}\, \hat \g^{kl}$. The structure of the action by itself exhibits the organization of the terms in inverse powers of $c$\,. Since we will stick to the even expansion (see for the odd expansion \cite{Ergen:2020yop,Elbistan:2025vyh}), we will consider the rescaling of the shift vector in a convenient way. This only change the dimension of the physical shift vector. We begin with the rescaling of the shift vector as 
\bea 
\hat N ^ i \to c^{-1}\, \hat N^i\,,
\eea 
so that the nontrivial vector part can talk to the kinetic terms. After rescaling of the $\hat N^i$, we rewrite the metric as 
\begin{equation} 
	\label{metscale}
	ds^2 = -c^2\, \hat N^2dt^2 + \hat \gamma_{ij} (dx^i + \hat N^i dt) (dx^j + \hat N^j dt)\,.
\end{equation}
Then action takes the form 
\begin{equation}
S_{\text{ADM}} =
    \int\,dt d^{3}x\, \hat N\,\sqrt{\hat \g}\, \left(\hat R\, + \,c^{-2}\,(\hat K_{ij} \hat K^{ij}-\hat K^2)\, \right)\,,
  \label{ADMActionscaled}
\end{equation}
where the new extrinsic curvature is 
\begin{align}
	\hat K_{ij}&=\hat N^{-1}\left(\frac{1}{2} \partial_t \hat \gamma_{ij}\,- \,\hat\nabla_{(i}\hat N_{j)}\right) \,.\label{Krelscaled}
\end{align}
The structure of the action reflects the familiar organization of the Einstein--Hilbert action in ADM variables. The leading-order term is the universal spatial-curvature part and controls the static sector, whereas the kinetic part becomes relevant for time-dependent configurations and appears at subleading order. At the same time, this decomposition makes the relation to Carrollian expansions transparent. From the small-$c$ point of view, the leading-order theory corresponds to electric Carrollian gravity, while the spatial Ricci-scalar term appears at subleading order and is captured by magnetic, or NLO, Carrollian gravity \cite{Hansen:2021fxi,Bal:2026xup}. The ADM decomposition is therefore useful for exhibiting the Galilean and Carrollian organizations within a common framework. In the sequel, we make this structure explicit by introducing a unified expansion framework for Galilean and Carrollian gravity and their higher-order generalization. Since the Carrollian version of the ADM expansion was investigated in detail in \cite{Bal:2026xup}, here we use it mainly as a structural comparison and focus on the construction and analysis of solutions in the Galilean (large-$c$) sector.

\subsection{The \texorpdfstring{$\omega$}{omega}-expansion}

\begin{figure}[htbp]
    \centering
    \includegraphics[width=0.7\textwidth]{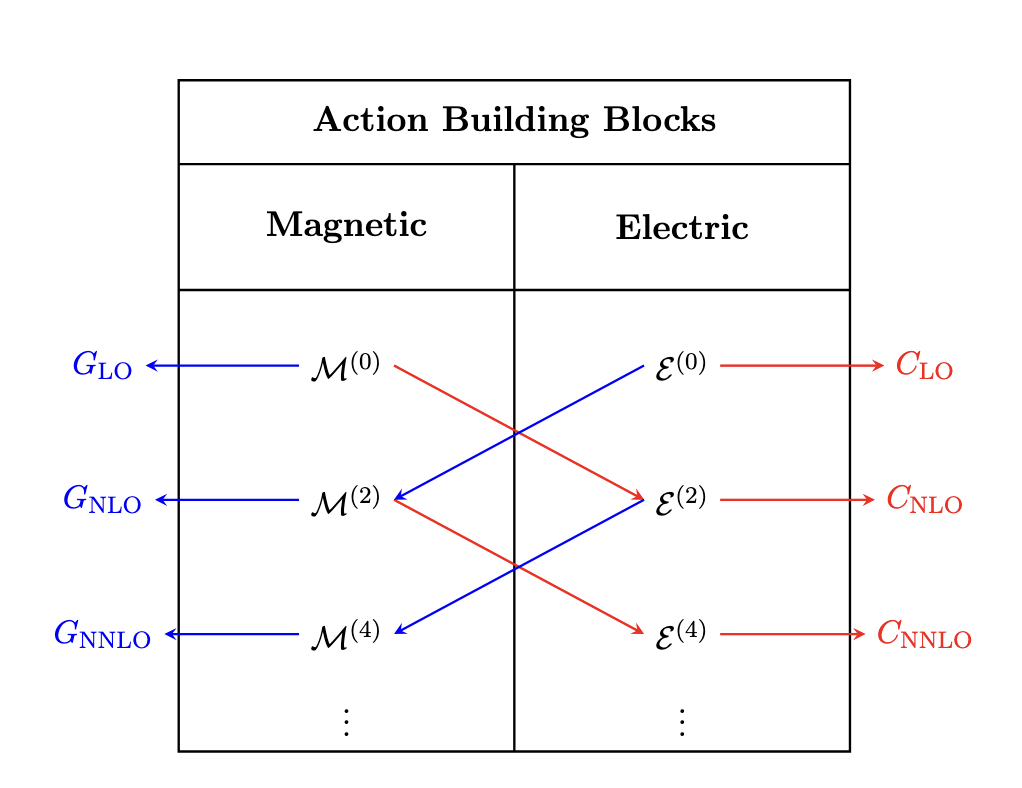}
    \caption{Building blocks of the $\omega$-expansion. The Galilean hierarchy is obtained by following the blue lines, while the Carrollian hierarchy is obtained by following the red lines.}
    \label{figure1}
\end{figure}

In this section we introduce the parameter $\omega$ instead of $c$. This parameter enables us to discuss the Galilean and Carrollian expansions in a unified way. Although we will not consider the details of the Carrollian expansion, it is useful to incorporate the two expansions into a common framework. This parametrization also makes the physical interpretation clearer. Therefore, we start with the decomposition
\begin{equation}
  L_{\rm ADM}=\mathcal M+\omega^\alpha \mathcal E ,
\end{equation}
where the magnetic and electric parts are defined by
\begin{equation}
  \mathcal M=\hat N\sqrt{\hat\gamma}\,\hat R,
  \qquad
  \mathcal E=\hat N\sqrt{\hat\gamma}
  \left(\hat K_{ij}\hat K^{ij}-\hat K^2\right).
\end{equation}
These terms will be our building blocks for constructing the Galilean and Carrollian actions. Here we introduce the scaling exponent $\alpha$ in order to keep track of the Galilean or Carrollian expansion. For the Galilean theory we set $\alpha=2$ and $\omega=c^{-1}$, while for the Carrollian theory we set $\alpha=-2$ and $\omega=c$. The choice of $\alpha$ enables us to identify which part of the action is leading in the corresponding expansion. For instance, if $\alpha=2$, then the Lagrangian takes the form
\begin{equation}
\mathcal L_{\rm ADM} = \left( \cM  + c^{-2} \cE \right)\,,
\end{equation}
and in the large-$c$ expansion the kinetic part becomes subleading. By contrast, if $\alpha=-2$ and $\omega=c$, the same ADM action is organized as a small-$c$ expansion. In this case the electric term $\cE$ is leading, while the curvature term $\cM$ becomes subleading; see the Carrollian expansion \cite{Bal:2026xup}. Since we have a unified framework, we can now expand the fields with respect to small-$\omega$ parameter 
\begin{align}
  \hat{\gamma}_{ij}
  &=
  \gamma_{ij}
  +\omega^2\beta_{ij}
  +\omega^4\epsilon_{ij},
  \label{eq:omega-ansatz-gamma}
  \\
  \hat N
  &=
  N
  +\omega^2 M
  +\omega^4 P,
  \label{eq:omega-ansatz-lapse}
  \\
  \hat N^i
  &=
  N^i
  +\omega^2 A^i
  +\omega^4 Z^i.
  \label{eq:omega-ansatz-shift}
\end{align}
The inverse spatial metric expands as
\begin{equation}
\hat\gamma^{ij}
=\gamma^{ij}-\omega^{2}\beta^{ij}+\omega^{4}\big(\beta^{ik}\beta_{k}{}^{j}-\epsilon^{ij}\big)+\mathcal O(\omega^{6}).
\end{equation}
From this point forward, all spatial indices on the expansion coefficients, such as
$N^i$, \(\beta_{ij}\), \(\epsilon_{ij}\), \(A^i\), and \(Z^i\), are raised (lowered) using the leading-order spatial metric \(\gamma^{ij}\) (\(\gamma_{ij}\)) and all the trace expressions are given by $X= \gamma^{ij} X_{ij}$.  The covariant derivative compatible with
\(\gamma_{ij}\) is denoted by \(\nabla_i\).  With this ansatz, the extrinsic curvature expands as
\begin{equation}
  \hat K_{ij}
  =
  K_{ij}
  +\omega^2 L_{ij}
  +\omega^4 F_{ij}.
  \label{eq:omega-ansatz-k-expansion}
\end{equation}
The coefficients are
\begin{align}
  K_{ij}
  &=
  N^{-1}
  \left(
    \frac{1}{2}\partial_t\gamma_{ij}
    -\nabla_{(i}N_{j)}
  \right),
  \label{eq:omega-ansatz-k}
  \\
  L_{ij}
  &=
  N^{-1}
  \left(
    \frac{1}{2}\partial_t\beta_{ij}
    -\nabla_{(i}A_{j)}
    -\beta_{k(i}\nabla_{j)}N^k
    -\frac{1}{2}N^k\nabla_k\beta_{ij}
    -M K_{ij}
  \right),
  \label{eq:omega-ansatz-l}
  \\
  F_{ij}
  &=
  N^{-1}
  \left[
    \frac{1}{2}\partial_t\epsilon_{ij}
    -\nabla_{(i}Z_{j)}
    -\beta_{k(j}\nabla_{i)}A^k
    -\epsilon_{k(j}\nabla_{i)}N^k
  \right.
  \nonumber\\
  &\hspace{3.6em}\left.
    -\frac{1}{2}N^k\nabla_k\epsilon_{ij}
    -\frac{1}{2}A^k\nabla_k\beta_{ij}
    -\left(L_{ij}M+K_{ij}P\right)
  \right].
  \label{eq:omega-ansatz-f}
\end{align}
Using the expansion ansatz, the magnetic $\cM$ term can be written as
\begin{equation}
  \hat{\mathcal{M}}
  =
  \sqrt{\gamma}
  \left[
    \mathcal{M}^{(0)}
    +\omega^2\mathcal{M}^{(2)}
    +\omega^4\mathcal{M}^{(4)}
  \right]
  +O(\omega^6).
  \label{magneticexp}
\end{equation}
The three coefficients are
\begin{align}
  \mathcal{M}^{(0)}
  &=
  NR,
  \label{magneticzeroth}
  \\
  \mathcal{M}^{(2)}
  &=
  \left(
    M+\frac{1}{2}N\beta
  \right)R
  +N R^{(2)},
  \label{magneticsecond}
  \\
  \mathcal{M}^{(4)}
  &=
  \left[
    P+\frac{1}{2}M\beta
    +N
    \left(
      \frac{1}{2}\epsilon
      +\frac{1}{8}\beta^2
      -\frac{1}{4}\beta_{ij}\beta^{ij}
    \right)
  \right]R
  \nonumber\\
  &\hspace{2em}
  +\left(
    M+\frac{1}{2}N\beta
  \right)R^{(2)}
  +N R^{(4)}\,,
  \label{magneticfourth}
\end{align}
where the explicit forms of $R^{(2)}$ and $R^{(4)}$ are given in Eqs.~\eqref{eq:ricci-scalar-r2-expanded} and \eqref{eq:ricci-scalar-r4-expanded}, respectively, in Appendix~\ref{expansionapp}. For the electric part, we find
\begin{equation}
  \hat{\mathcal{E}}
  =
  \sqrt{\gamma}
  \left[
    \mathcal{E}^{(0)}
    +\omega^2\mathcal{E}^{(2)}
    +\omega^4\mathcal{E}^{(4)}
  \right]
  +O(\omega^6).
  \label{eq:electric-block-expansionbulk}
\end{equation}
Then the coefficients are
\begin{align}
  \mathcal{E}^{(0)}
  &=
  N\mathcal{Q}^{(0)},
  \label{eq:electric-e0bulk}
  \\
  \mathcal{E}^{(2)}
  &=
  \left(
    M+\frac{1}{2}N\beta
  \right)\mathcal{Q}^{(0)}
  +N\mathcal{Q}^{(2)},
  \label{eq:electric-e2bulk}
  \\
  \mathcal{E}^{(4)}
  &=
  \left[
    P+\frac{1}{2}M\beta
    +N
    \left(
      \frac{1}{2}\epsilon
      +\frac{1}{8}\beta^2
      -\frac{1}{4}\beta_{ij}\beta^{ij}
    \right)
  \right]\mathcal{Q}^{(0)}
  \nonumber\\
  &\quad
  +\left(
    M+\frac{1}{2}N\beta
  \right)\mathcal{Q}^{(2)}
  +N\mathcal{Q}^{(4)}\,,
  \label{eq:electric-e4bulk}
\end{align}
where the $\mathcal Q^{(n)}$ are given in Eq.~\eqref{eq:q-coefficients} in Appendix~\ref{expansionapp}. The structure of the action is then
\bea 
\mathcal L_{\rm ADM} &=& \left( \cM  + \omega^{\alpha}\, \cE \right)\,, \nn \\
&=& \sqrt{\gamma}
  \left[
    \mathcal{M}^{(0)}
    +\omega^2\mathcal{M}^{(2)}
    +\omega^4\mathcal{M}^{(4)} + \omega^\alpha \left(\mathcal{E}^{(0)}
    +\omega^2\mathcal{E}^{(2)}
    +\omega^4\mathcal{E}^{(4)}\right)
  \right] \,. 
\eea 
The Galilean and Carrollian theories are obtained by choosing $\alpha=2$  and  $\alpha=-2$, respectively. For $\alpha=2$, the expansion gives the Galilean hierarchy. For $\alpha=-2$, it gives the Carrollian hierarchy, after factoring out the overall power of $\omega$ that fixes the leading order. This organization is summarized in Fig.~\ref{figure1}.
Following the blue arrows in Fig.~\ref{figure1}, one obtains the LO, NLO and NNLO Galilean theories listed in Table~\ref{tab:theories_expansion}. Similarly, following the red arrows gives the corresponding Carrollian hierarchy. The remaining details of the expansion are collected in Appendix~\ref{expansionapp}. The NLO Galilean theory is the ADM form of Newtonian gravity \cite{Hansen:2019pkl,Hansen:2020pqs}. For completeness, we review this relation in Appendix~\ref{newtonianaction}.
\begin{table}[htbp]
    \centering
    \renewcommand{\arraystretch}{1.8}
    \begin{tabular}{@{} c l l @{}}
        \toprule
        \textbf{Order} & \textbf{Galilean theories} & \textbf{Carrollian theories} \\
        \midrule

        \textbf{LO}
        & $\mathcal L^{\rm Gal}_{\rm LO} = \sqrt{\gamma}\,\mathcal{M}^{(0)}$
        & $\mathcal L^{\rm Car}_{\rm LO} = \sqrt{\gamma}\,\mathcal{E}^{(0)}$ \\

        \textbf{NLO}
        & $\mathcal L^{\rm Gal}_{\rm NLO} = \sqrt{\gamma}\left[ \mathcal{M}^{(2)} + \mathcal{E}^{(0)} \right]$
        & $\mathcal L^{\rm Car}_{\rm NLO} = \sqrt{\gamma}\left[ \mathcal{E}^{(2)} + \mathcal{M}^{(0)} \right]$ \\

        \textbf{NNLO}
        & $\mathcal L^{\rm Gal}_{\rm NNLO} = \sqrt{\gamma}\left[ \mathcal{M}^{(4)} + \mathcal{E}^{(2)} \right]$
        & $\mathcal L^{\rm Car}_{\rm NNLO} = \sqrt{\gamma}\left[ \mathcal{E}^{(4)} + \mathcal{M}^{(2)} \right]$ \\

        \bottomrule
    \end{tabular}
    \caption{Expansion of the Galilean and Carrollian theories up to NNLO.}
    \label{tab:theories_expansion}
\end{table}

\subsection{Hamilton's field equations}

We introduce the Hamiltonian formalism in order to write the field equations in a simple form. The Hamiltonian of the ADM action is 
\begin{equation}
\begin{aligned}
H=&\int d^3x\sqrt{\hat \gamma }\left(\hat N\,\mathcal{H}  +  \hat N_i \mathcal{H}^i\right)\,,
\end{aligned}
\end{equation} \label{hamiltonanrel}
where
\begin{equation}
\begin{aligned}
\mathcal{H}
&= \hat R + \,\hat{K}^2 - \hat{K}_{ij}\hat{K}^{ij},\\[4pt]
\mathcal{H}_i
&= -2\hat\nabla_j
  \left(
    \hat K^j{}_{i}
    -\delta^j_i\hat K
  \right)
\end{aligned}
\end{equation}
There is no momentum conjugate to $\hat N$, hence the constraint equation
for $\hat N$ is
\begin{equation}
\frac{\delta H}{\delta \hat N} = 0
\quad\Rightarrow\quad
\boxed{
\mathcal{H} = 0.
}
\end{equation}
Similarly, varying with respect to $\hat N^i$ gives momentum constraint equation
\begin{equation}
\frac{\delta H}{\delta \hat N^i} = 0
\quad\Rightarrow\quad
\boxed{
\mathcal{H}_i = 0.
}
\end{equation}
Hamilton's equations are
\begin{align}
\partial_t \hat \gamma_{ij} &=  2 \hat N  \hat K_{ij} +2 \hat \nabla_{(i} \hat N_{j)}\, \\
\partial_t \hat K_{ij} &=  \hat \nabla_i \hat \nabla_j \hat N
-  \hat N\Big(\hat R_{ij} + \hat K \hat K_{ij} - 2\, \hat K_{ik}\, \hat K^k{}_j\Big)
+  \mathcal{L}_{\vec{\hat N}}\,\hat K_{ij}\,,
\end{align}
where $\mathcal{L}_{\vec{N}}$ is the Lie derivative along the shift:
\bea
\mathcal{L}_{\vec{\hat N}} \hat K_{ij}
= \hat N^k \hat \nabla_k \hat K_{ij} + \hat K_{ik}\, \hat \nabla_j \hat N^k + \hat K_{kj} \hat \nabla_i N^k. 
\eea 
Now that we have the constraints and Hamilton's equations, we rescale the shift vector as
$\hat N^i \to c^{-1}\hat N^i$ and assign one extra inverse power of $c$ to time derivatives,
$\partial_t \to c^{-1}\partial_t$. We then introduce the expansion parameter $\omega$ and scaling exponent $\alpha$ to organize the expansion of Hamilton's field equations.

\paragraph{Hamiltonian constraint}
\begin{equation}
\mathcal{H}
:= \hat R +\omega^\alpha
\left(
\hat K^2
-
\hat K_{ij}\hat K^{ij}
\right)
=0.
\end{equation}

\paragraph{Momentum constraint}
\begin{equation}
\mathcal{H}_i
:=
-2\hat\nabla_j
\left(
\hat K^j{}_{i}
-
\delta^j_i\hat K
\right)
=0.
\end{equation}

\paragraph{Evolution of \(\hat\gamma_{ij}\)}
For later convenience, we define
\begin{equation}
\hat{\mathcal{B}}_{ij}
:=
\partial_t \hat \gamma_{ij}
-
2 \hat N \hat K_{ij}
-
2 \hat \nabla_{(i} \hat N_{j)}
=0\,.
\end{equation}

\paragraph{Evolution of \(\hat K_{ij}\)}
Similarly, we define
\begin{equation}
 \hat{\mathcal{A}}_{ij}
 :=
 \omega^\alpha
 \left[
 \partial_t \hat K_{ij}
 +
 \hat N
 \left(
 \hat K \hat K_{ij}
 -
 2\,\hat K_{ik}\hat K^k{}_j
 \right)
 -
 \mathcal{L}_{\vec{\hat N}}\hat K_{ij}
 \right]
 -
 \hat \nabla_i \hat \nabla_j \hat N
 +
 \hat N\,\hat R_{ij}
 =0 \,.
\end{equation}
\subsection{Expansion of Hamilton's field equations}
In this subsection we expand the Hamilton's equations given above. We will introduce the electric and magnetic part as we did for the action previously. 

\paragraph{Hamiltonian constraint:} The Hamiltonian constraint can be decomposed into magnetic and electric blocks as
\[
  \hat{\mathcal{H}}
  =
  \hat{\mathcal{H}}^{\rm M}
  +\omega^\alpha \hat{\mathcal{H}}^{\rm E}
  =
  0,
\]
where
\[
  \hat{\mathcal{H}}^{\rm M}
  :=
  \hat R,
  \qquad
  \hat{\mathcal{H}}^{\rm E}
  :=
  \left(\hat K^2 - 
    \hat K_{ij}\hat K^{ij}
  \right)
  = - \hat{\mathcal{Q}}.
\]
The magnetic and electric sectors are expanded independently:
\begin{align}
  \hat{\mathcal{H}}^{\rm M}
  &=
  \overset{(0)}{\mathcal{H}}{}^{\rm M}
  +\omega^2\overset{(2)}{\mathcal{H}}{}^{\rm M}
  +\omega^4\overset{(4)}{\mathcal{H}}{}^{\rm M}
  +O(\omega^6),
  \\
  \hat{\mathcal{H}}^{\rm E}
  &=
  \overset{(0)}{\mathcal{H}}{}^{\rm E}
  +\omega^2\overset{(2)}{\mathcal{H}}{}^{\rm E}
  +\omega^4\overset{(4)}{\mathcal{H}}{}^{\rm E}
  +O(\omega^6).
\end{align}
The coefficients of the expansion are
\begin{align}
  \overset{(0)}{\mathcal{H}}{}^{\rm M}
  &:=
  R,
  &
  \overset{(2)}{\mathcal{H}}{}^{\rm M}
  &:=
  R^{(2)},
  &
  \overset{(4)}{\mathcal{H}}{}^{\rm M}
  &:=
  R^{(4)},
  \\
  \overset{(0)}{\mathcal{H}}{}^{\rm E}
  &:=
  -\mathcal{Q}^{(0)},
  &
  \overset{(2)}{\mathcal{H}}{}^{\rm E}
  &:=
  -\mathcal{Q}^{(2)},
  &
  \overset{(4)}{\mathcal{H}}{}^{\rm E}
  &:=
  -\mathcal{Q}^{(4)}.
\end{align}
Here the curvature and extrinsic-curvature expansions are given in
Eqs.~\eqref{eq:ricci-scalar-r2-expanded}--\eqref{eq:ricci-scalar-r4-expanded}
and \eqref{eq:q-coefficients}. Since we now have the building blocks for the field equations,
we can again use the same mechanism as we did for the action. For instance, the Galilean
and Carrollian Hamiltonian constraints are summarized in Table~\ref{tab:field_equations}.

\paragraph{Momentum Constraint:}

Dropping the irrelevant overall factor $-2$, the momentum constraint can be written as
\[
\mathcal{H}_i
  :=
  \hat\nabla_j
  \left(
    \hat K^j{}_{i}
    -\delta^j_i\hat K
  \right)
  =
  0.
\]
Therefore, using the expansion ansatz, we have 
\begin{equation}
  \mathcal{H}_i
  =
  \mathcal{H}^{(0)}_i
  +\omega^2\mathcal{H}^{(2)}_i
  +\omega^4\mathcal{H}^{(4)}_i
  +O(\omega^6),
\end{equation}
The momentum equations through order \(\omega^4\) are
\[
  \boxed{
  \mathcal{H}^{(0)}_i=0,
  \qquad
  \mathcal{H}^{(2)}_i=0,
  \qquad
  \mathcal{H}^{(4)}_i=0.
  }
\]
Through NNLO, the Galilean hierarchy contains the momentum-constraint
equations up to order \(\omega^2\), whereas the Carrollian hierarchy
contains them up to order \(\omega^4\).\footnote{This difference comes from the overall factor \(\omega^\alpha\) multiplying the kinetic ADM block in the Hamiltonian
\eqref{hamiltonanrel} after the shift rescaling. Thus the same expansion
of the momentum constraint appears at different orders in the Galilean and
Carrollian hierarchies.} The corresponding Galilean and Carrollian momentum-constraint equations are summarized in Table~\ref{tab:field_equations}.

\paragraph{Metric Evolution Equation:}

The defining ADM relation is
\[
  \hat{\mathcal{B}}_{ij}
  :=
  \partial_t\hat\gamma_{ij}
  -2\hat N\hat K_{ij}
  -2\hat\nabla_{(i}\hat N_{j)}
  =
  0.
\]
From the expansion, we get
\[
  \hat{\mathcal{B}}_{ij}
  =
  \mathcal{B}^{(0)}_{ij}
  +\omega^2\mathcal{B}^{(2)}_{ij}
  +\omega^4\mathcal{B}^{(4)}_{ij}
  +O(\omega^6).
\]
The first three equations are
\begin{align}
  \mathcal{B}^{(0)}_{ij}
  &:=
  \partial_t\gamma_{ij}
  -2NK_{ij}
  -2\nabla_{(i}N_{j)}
  =
  0,
  \\
  \mathcal{B}^{(2)}_{ij}
  &:=
  \partial_t\beta_{ij}
  -2NL_{ij}
  -2MK_{ij}
  -2\nabla_{(i}A_{j)}
  \nonumber\\
  &\quad
  -2\beta_{k(i}\nabla_{j)}N^k
  -N^k\nabla_k\beta_{ij}
  =
  0,
  \\
  \mathcal{B}^{(4)}_{ij}
  &:=
  \partial_t\epsilon_{ij}
  -2NF_{ij}
  -2ML_{ij}
  -2PK_{ij}
  -2\nabla_{(i}Z_{j)}
  \nonumber\\
  &\quad
  -2\beta_{k(j}\nabla_{i)}A^k
  -2\epsilon_{k(j}\nabla_{i)}N^k
  -N^k\nabla_k\epsilon_{ij}
  -A^k\nabla_k\beta_{ij}
  =
  0.
\end{align}
These equations are equivalent to the definitions of \(K_{ij}\), \(L_{ij}\), and
\(F_{ij}\) given earlier in \cref{eq:omega-ansatz-k,eq:omega-ansatz-l,eq:omega-ansatz-f}. The corresponding Galilean and Carrollian equations are summarized in Table~\ref{tab:field_equations}. Notice that the relations \(\mathcal B^{(n)}_{ij}=0\) are ADM defining relations for the extrinsic-curvature coefficients rather than independent variational field equations. In particular, \(\mathcal B^{(4)}_{ij}=0\) contains \(F_{ij}\), and hence the vector \(Z_i\). This does not imply that \(Z_i\) enters the Galilean NNLO dynamics. Indeed, the Galilean NNLO action contains only the combination \(\mathcal M^{(4)}+\mathcal E^{(2)}\), and the electric block \(\mathcal E^{(2)}\) depends on \(K_{ij}\) and \(L_{ij}\), but not on \(F_{ij}\). Therefore, although we denote the order-\(\omega^4\) relation by \(\overset{(4)}{\mathcal B}{}^{\rm Gal}_{ij}\) in Table~\ref{tab:field_equations}, it should be understood only as the order-\(\omega^4\) kinematical definition of \(F_{ij}\). It is not imposed as an independent variational field equation in the Galilean NNLO solution analysis below. When \(Z_i\) appears in the appendices, it is obtained directly from the
large-\(c\) expansion of the corresponding relativistic metric, but it does
not contribute to the Galilean NNLO variational equations.

\paragraph{\texorpdfstring{\(K_{ij}\) evolution equation}{Kij Evolution Equation}:}

For the \(K_{ij}\) evolution equation we use the split
\[
  \hat{\mathcal{A}}_{ij}
  =
  \hat{\mathcal{A}}^{\rm M}_{ij}
  +\omega^{\alpha}\hat{\mathcal{A}}^{\rm E}_{ij}
  =
  0,
\]
with
\begin{align}
  \hat{\mathcal{A}}^{\rm M}_{ij}
  &:= \hat N\hat R_{ij}
   - \hat\nabla_i\hat\nabla_j\hat N\,,
  \\
  \hat{\mathcal{A}}^{\rm E}_{ij}
  &:=\mathcal{D}_t\hat K_{ij}
  + \hat N
  \left(
    \hat K\hat K_{ij}
    -2\hat K_i{}^k\hat K_{jk}
  \right)\,,
\end{align}
where we define 
\bea
\mathcal{D}_t\hat K_{ij} = \left( \partial_t \hat K_{ij}  
-  \mathcal{L}_{\vec{\hat N}}\,\hat K_{ij} \right)\,. 
\eea 
The magnetic block has the expansion
\begin{align}
  \hat{\mathcal{A}}^{\rm M}_{ij}
  &=
  \overset{(0)}{\mathcal{A}}{}^{\rm M}_{ij}
  +\omega^2\overset{(2)}{\mathcal{A}}{}^{\rm M}_{ij}
  +\omega^4\overset{(4)}{\mathcal{A}}{}^{\rm M}_{ij}
  +O(\omega^6),
\end{align}
where $\overset{(n)}{\mathcal{A}}{}^{\rm M}_{ij}$ are given in Eq.~\eqref{Kmagevolexpeom}. Similarly, the electric block expands as
\begin{align}
  \hat{\mathcal{A}}^{\rm E}_{ij}
  &=
  \overset{(0)}{\mathcal{A}}{}^{\rm E}_{ij}
  +\omega^2\overset{(2)}{\mathcal{A}}{}^{\rm E}_{ij}
  +\omega^4\overset{(4)}{\mathcal{A}}{}^{\rm E}_{ij}
  +O(\omega^6),
\end{align}
where $\overset{(n)}{\mathcal{A}}{}^{\rm E}_{ij}$ are given in Eq.~\eqref{Kelevalexeom}. Therefore, we can collect the constraint and evolution equations for the Galilean and Carrollian theories order by order in Table~\ref{tab:field_equations}. The explicit form of the field equations is given in Appendix~\ref{eomexpapp}. Now that we have the field equations order by order, we can now discuss the possible background solutions.  
\begin{table}[htbp]
    \centering
    \small
    \renewcommand{\arraystretch}{2.0}
    \setlength{\tabcolsep}{4pt}
    \begin{tabular}{@{} c c p{0.36\textwidth} p{0.36\textwidth} @{}}
        \toprule
        \textbf{Order}
        & \textbf{Component}
        & \textbf{Galilean field equations}
        & \textbf{Carrollian field equations} \\
        \midrule

        \multirow{4}{*}{\textbf{LO}}
        & $\mathcal{H}$
        & $\begin{aligned}
            \overset{(0)}{\mathcal{H}}{}^{\rm Gal}
            &:= \overset{(0)}{\mathcal{H}}{}^{\rm M}=0
          \end{aligned}$
        & $\begin{aligned}
            \overset{(0)}{\mathcal{H}}{}^{\rm Car}
            &:= \overset{(0)}{\mathcal{H}}{}^{\rm E}=0
          \end{aligned}$ \\

        & $\mathcal{H}_i$
        & \textit{absent}
        & $\begin{aligned}
            \overset{(0)}{\mathcal{H}}{}^{\rm Car}_{i}
            &:= \mathcal{H}^{(0)}_{i}=0
          \end{aligned}$ \\

        & $\mathcal{B}_{ij}$
        & $\begin{aligned}
            \overset{(0)}{\mathcal{B}}{}^{\rm Gal}_{ij}
            &:= \mathcal{B}^{(0)}_{ij}=0
          \end{aligned}$
        & $\begin{aligned}
            \overset{(0)}{\mathcal{B}}{}^{\rm Car}_{ij}
            &:= \mathcal{B}^{(0)}_{ij}=0
          \end{aligned}$ \\

        & $\mathcal{A}_{ij}$
        & $\begin{aligned}
            \overset{(0)}{\mathcal{A}}{}^{\rm Gal}_{ij}
            &:= \overset{(0)}{\mathcal{A}}{}^{\rm M}_{ij}=0
          \end{aligned}$
        & $\begin{aligned}
            \overset{(0)}{\mathcal{A}}{}^{\rm Car}_{ij}
            &:= \overset{(0)}{\mathcal{A}}{}^{\rm E}_{ij}=0
          \end{aligned}$ \\
        \midrule

        \multirow{4}{*}{\textbf{NLO}}
        & $\mathcal{H}$
        & $\begin{aligned}
            \overset{(2)}{\mathcal{H}}{}^{\rm Gal}
            &:=
            \overset{(2)}{\mathcal{H}}{}^{\rm M}
            +
            \overset{(0)}{\mathcal{H}}{}^{\rm E}=0
          \end{aligned}$
        & $\begin{aligned}
            \overset{(2)}{\mathcal{H}}{}^{\rm Car}
            &:=
            \overset{(2)}{\mathcal{H}}{}^{\rm E}
            +
            \overset{(0)}{\mathcal{H}}{}^{\rm M}=0
          \end{aligned}$ \\

        & $\mathcal{H}_i$
        & $\begin{aligned}
            \overset{(2)}{\mathcal{H}}{}^{\rm Gal}_{i}
            &:= \mathcal{H}^{(0)}_{i}=0
          \end{aligned}$
        & $\begin{aligned}
            \overset{(2)}{\mathcal{H}}{}^{\rm Car}_{i}
            &:= \mathcal{H}^{(2)}_{i}=0
          \end{aligned}$ \\

        & $\mathcal{B}_{ij}$
        & $\begin{aligned}
            \overset{(2)}{\mathcal{B}}{}^{\rm Gal}_{ij}
            &:= \mathcal{B}^{(2)}_{ij}=0
          \end{aligned}$
        & $\begin{aligned}
            \overset{(2)}{\mathcal{B}}{}^{\rm Car}_{ij}
            &:= \mathcal{B}^{(2)}_{ij}=0
          \end{aligned}$ \\

        & $\mathcal{A}_{ij}$
        & $\begin{aligned}
            \overset{(2)}{\mathcal{A}}{}^{\rm Gal}_{ij}
            &:=
            \overset{(2)}{\mathcal{A}}{}^{\rm M}_{ij}
            +
            \overset{(0)}{\mathcal{A}}{}^{\rm E}_{ij}=0
          \end{aligned}$
        & $\begin{aligned}
            \overset{(2)}{\mathcal{A}}{}^{\rm Car}_{ij}
            &:=
            \overset{(2)}{\mathcal{A}}{}^{\rm E}_{ij}
            +
            \overset{(0)}{\mathcal{A}}{}^{\rm M}_{ij}=0
          \end{aligned}$ \\
        \midrule

        \multirow{4}{*}{\textbf{NNLO}}
        & $\mathcal{H}$
        & $\begin{aligned}
            \overset{(4)}{\mathcal{H}}{}^{\rm Gal}
            &:=
            \overset{(4)}{\mathcal{H}}{}^{\rm M}
            +
            \overset{(2)}{\mathcal{H}}{}^{\rm E}=0
          \end{aligned}$
        & $\begin{aligned}
            \overset{(4)}{\mathcal{H}}{}^{\rm Car}
            &:=
            \overset{(4)}{\mathcal{H}}{}^{\rm E}
            +
            \overset{(2)}{\mathcal{H}}{}^{\rm M}=0
          \end{aligned}$ \\

        & $\mathcal{H}_i$
        & $\begin{aligned}
            \overset{(4)}{\mathcal{H}}{}^{\rm Gal}_{i}
            &:= \mathcal{H}^{(2)}_{i}=0
          \end{aligned}$
        & $\begin{aligned}
            \overset{(4)}{\mathcal{H}}{}^{\rm Car}_{i}
            &:= \mathcal{H}^{(4)}_{i}=0
          \end{aligned}$ \\

        & $\mathcal{B}_{ij}$
        & $\begin{aligned}
            \overset{(4)}{\mathcal{B}}{}^{\rm Gal}_{ij}
            &:= \mathcal{B}^{(4)}_{ij}=0
          \end{aligned}$
        & $\begin{aligned}
            \overset{(4)}{\mathcal{B}}{}^{\rm Car}_{ij}
            &:= \mathcal{B}^{(4)}_{ij}=0
          \end{aligned}$ \\

        & $\mathcal{A}_{ij}$
        & $\begin{aligned}
            \overset{(4)}{\mathcal{A}}{}^{\rm Gal}_{ij}
            &:=
            \overset{(4)}{\mathcal{A}}{}^{\rm M}_{ij}
            +
            \overset{(2)}{\mathcal{A}}{}^{\rm E}_{ij}=0
          \end{aligned}$
        & $\begin{aligned}
            \overset{(4)}{\mathcal{A}}{}^{\rm Car}_{ij}
            &:=
            \overset{(4)}{\mathcal{A}}{}^{\rm E}_{ij}
            +
            \overset{(2)}{\mathcal{A}}{}^{\rm M}_{ij}=0
          \end{aligned}$ \\

        \bottomrule
    \end{tabular}
    \caption{Schematic organization of the constraints, evolution equations and kinematical ADM relations at LO, NLO and NNLO.}
    \label{tab:field_equations}
\end{table}

\FloatBarrier
\section{Stationary vacuum solutions in the Galilean expansion} \label{sec:stationary-solutions}

In this section we explore nontrivial stationary vacuum solutions of the Galilean (large-$c$)  field equations obtained in the previous section. In the previous section the Carrollian hierarchy was included only as a structural comparison. Hence, we now restrict our attention to the Galilean branch of Table~\ref{tab:field_equations}. For the Galilean expansion we set
\begin{equation}
  \omega=c^{-1},
  \qquad
  \alpha=2,
\end{equation}
so that the ADM variables are expanded as
\begin{align}
  \hat{\gamma}_{ij}
  &=
  \gamma_{ij}
  +c^{-2}\beta_{ij}
  +c^{-4}\epsilon_{ij}
  +\mathcal O(c^{-6}),
  \\
  \hat N
  &=
  N
  +c^{-2}M
  +c^{-4}P
  +\mathcal O(c^{-6}),
  \\
  \hat N^i
  &=
  N^i
  +c^{-2}A^i
  +c^{-4}Z^i
  +\mathcal O(c^{-6}).
\end{align}
We focus on stationary configurations where  all fields are time independent. This sector is simple enough to allow for an explicit order-by-order analysis. It still retains the effects of rotation and higher multipole moments through the shift vector and the subleading spatial fields.

For the stationary solutions considered here, we distinguish two branches. In the weak branch, the leading-order geometry is flat, with
\[
N=1,\qquad
\gamma_{ij}=\mathrm{diag}\left(1,r^2,r^2\sin^2\theta\right),
\qquad
N^i=0,
\]
and the Newtonian potential enters at subleading order through $M$. In the strong branch, the leading-order fields already contain a nontrivial gravitational background,
\[
N=\sqrt{f},\qquad
\gamma_{ij}=\mathrm{diag}\left(f^{-1},r^2,r^2\sin^2\theta\right),
\qquad
f=1-\frac{2Gm}{r},
\]
and the leading shift need not vanish. For the relativistic Kerr and Hartle--Thorne seeds used in Appendices~\ref{kerrapp} and~\ref{htapp}, the weak branch corresponds to taking $c\to\infty$ with $G,m,J$ (and $Q$ for Hartle--Thorne) fixed, whereas the strong branch is obtained with $m\to cm$, $G\to cG$, $J\to cJ$, and, for the Hartle--Thorne sector, $Q\to c^{-1}Q$. The weak branch is therefore adapted to rotating and multipolar solutions around flat leading data, whereas the strong branch describes expansions around nontrivial Schwarzschild-type leading geometries.

We explore different types of solutions by adapting this branched structure. First, we solve the Galilean field equations directly within the stationary ansatz. Second, whenever possible, we compare the resulting fields with the large-$c$ expansion of known exact or approximate relativistic vacuum metrics. This comparison fixes integration constants and provides a useful interpretation of the Galilean data. However, these two steps should be kept distinct. Solving the truncated Galilean equations fixes the data intrinsic to the finite-order Galilean hierarchy, whereas matching to a relativistic expansion fixes a choice of relativistic representative, coordinates and gauge. Thus a Galilean solution constructed below should not automatically be interpreted as a unique full relativistic completion. In this way Kerr, Hartle--Thorne and mixed rotating-multipolar sectors can be treated within the same ADM expansion scheme.
A summary of the stationary solutions constructed below is given in Table~\ref{tab:stationary-solutions}.

\begin{table}[t]
\centering
\small
\setlength{\tabcolsep}{4pt}
\begin{tabularx}{\linewidth}{c c l X}
\hline
Branch & Order & Sector & Result \\
\hline
Weak & NLO
& Kerr, Hartle--Thorne, mixed
& Solutions; higher mass multipoles allowed \\

Weak & NNLO
& Kerr
& Solution \\

Weak & NNLO
& Extended Hartle--Thorne
& Solution after $Q^2$ completion \\

Weak & NNLO
& Mixed
& Corrected solution; naive sum fails due to $J^2Q$ terms \\

Strong & NLO--NNLO
& Kerr
& Solution through NNLO \\

Strong & NLO
& Hartle--Thorne
& Solution; generic NNLO extension not constructed \\
\hline
\end{tabularx}
\caption{Summary of the stationary Galilean solutions constructed in
Sec.~\ref{sec:stationary-solutions}.}
\label{tab:stationary-solutions}

\end{table}

\subsection{Stationary reduction of the Galilean field equations}

We now consider  the Galilean field equations under the stationary assumption. At this stage we do not impose any condition on the leading shift \(N^i\). This is important because the weak and strong branches have different leading ADM data. In particular, we will set the LO shift vector $N^i =0$ in the weak branch in order to further simplify the field equations. In fact, this condition is motivated by the observation that, if one expands the relativistic Kerr or Hartle--Thorne metric in large-\(c\), the first nontrivial rotational effects arise at the order of \(\mathcal{O}(c^{-2})\). Therefore, to capture the slow rotational effects in the nonrelativistic gravitational sectors, we can consistently set LO shift vector $N^i =0$ in this weak branch sector. However, the full analysis of the weak branch with nonzero $N^i$ would be interesting and can be explored elsewhere. As a result, in what follows the leading shift vanishes in the weak branch, while we  keep it nonzero in the strong branch.

In the stationary sector the ADM kinematical relations reduce to
\begin{align}
K_{ij}
&=
-N^{-1}\nabla_{(i}N_{j)},
\\
L_{ij}
&=
N^{-1}
\left(
-\nabla_{(i}A_{j)}
-\beta_{k(i}\nabla_{j)}N^k
-\frac12 N^k\nabla_k\beta_{ij}
-MK_{ij}
\right).
\end{align}
The expression for \(F_{ij}\) is analogous and is not needed explicitly in the Galilean NNLO variational equations. These relations should be understood as the order-by-order definitions of \(K_{ij}\), and  \(L_{ij}\), rather than as independent variational field equations.

We keep the stationary equations in block form in the main text.  Their explicit stationary forms are collected in Appendix~\ref{stationaryredeom}, and the corresponding simplifications in the weak branch are also collected in Appendix~\ref{weakredeom}. In the solution analysis below, we impose these Galilean variational equations at the relevant order. The relations \(\mathcal B^{(n)}_{ij}=0\) are used only to define the extrinsic-curvature coefficients order by order.

\subsection{Static leading - order backgrounds}

Let us first discuss the leading-order static backgrounds, since these backgrounds will be used as seeds for the higher-order stationary equations. Using the results of Appendix~\ref{stationaryredeom}, then the LO equations reduce to
\begin{align}
R &= 0\,, \qquad  N R_{ij} -\nabla_i\nabla_j N = 0.
\label{LOeomred}
\end{align}
These equations determine the leading spatial metric \(\gamma_{ij}\) and the leading lapse \(N\). The basic static solutions can be written compactly as
\begin{subequations}
\label{static_seeds}
\begingroup
\small
\setlength{\jot}{6pt}
\begin{align}
\text{\bf Flat solution:}\qquad
&
N=1,
&
\gamma_{ij}
&=
\mathrm{diag}
\left(
1,r^2,r^2\sin^2\theta
\right),
\label{flat_sol}
\\
\text{\bf Strong-gravity solution:}\qquad
&
N=\sqrt{f},
&
\gamma_{ij}
&=
\mathrm{diag}
\left(
f^{-1},r^2,r^2\sin^2\theta
\right),
\label{strong_grav_sol}
\\
\text{\bf C-metric solution:}\qquad
&
N=\frac{\sqrt{Q_{\rm C}}}{\Omega_{\rm C}},
&
\gamma_{ij}
&=
\Omega_{\rm C}^{-2}
\mathrm{diag}
\left(
Q_{\rm C}^{-1},
\frac{r^2}{P_{\rm C}},
P_{\rm C}r^2\sin^2\theta
\right).
\label{cmetric_sol}
\end{align}
\endgroup
\end{subequations}
Here $f = 1- \frac{2Gm}{r}$ and 
\bea
 P_{\rm C}=1+2aGm\cos\theta\,, \quad Q_{\rm C}&=
  \left(1-a^2r^2\right)f,
  \quad
  &\Omega_{\rm C}=1+ar\cos\theta .
\eea 
The parameters \(a\) , \(m\) and $G$ denote the acceleration, mass parameters and gravitational constant, respectively. 

We will use these static backgrounds as seeds to explore the higher-order stationary solutions. The flat solution will be used as the seed for the weak-gravity branch, while the strong-gravity solution will be used as the seed for the strong-gravity branch. The C-metric solution is included to show that accelerated static backgrounds also fit into the leading Galilean equations; however, we will not investigate accelerated solutions in the present work. Note that these solutions are also background solutions of magnetic Carroll gravity and NLO Carrollian gravity \cite{Hansen:2021fxi,Kolar:2025ebv,Bal:2026xup}.

\subsection{Weak branch}

In the weak branch we use the flat leading data \eqref{flat_sol}. Thus
\begin{equation}
N=1,
\qquad
\gamma_{ij}
=
\mathrm{diag}\left(1,r^2,r^2\sin^2\theta\right).
\end{equation}
The first nontrivial scalar, vector and spatial-metric fields appear at NLO through \(M\), $N^{i}$ and \(\beta_{ij}\). Although the shift coefficient \(A_i\) appears at order \(c^{-2}\) in the ADM expansion, it does not enter the weak-branch Galilean equations at NLO. It contributes first at NNLO through the electric block, or equivalently through \(L_{ij}=-\nabla_{(i}A_{j)}\). Before imposing the weak-branch simplification \(N^i=0\), the order-by-order field content is naturally organized as
\begin{equation}
\label{eq:weak-branch-field-content}
\begin{aligned}
\text{LO:}\qquad
&
N=1,
\qquad
\gamma_{ij}
=
\mathrm{diag}\left(1,r^2,r^2\sin^2\theta\right),
\\
\text{NLO:}\qquad
&
M,
\qquad N^i,
\qquad
\beta_{ij},
\\
\text{NNLO:}\qquad
&
P,
\qquad A_i,
\qquad
\epsilon_{ij}.
\end{aligned}
\end{equation}
This branch is naturally adapted to the large-\(c\) expansion of asymptotically flat stationary metrics with fixed gravitational constant $G$, mass $m$, spin $J$ and quadrupole moment $Q$\,. 

In addition, as we discussed earlier,  we can set $N^i = 0$ in the NLO and NNLO field equations. Therefore \(K_{ij}=0\), and the first nontrivial stationary vector field is the NLO shift \(A_i\),
\begin{equation}
L_{ij}
=
-\nabla_{(i}A_{j)} .
\end{equation}
However, it does not contribute to the field equations at NLO, but only  at  NNLO because of the expansion hierarchy in the Galilean sector (see Table~\ref{tab:theories_expansion}) . For the weak branch, the NLO equations simplify to
\begin{subequations}
\label{eq:weak-nlo-eombulk}
\begin{align}
\overset{(2)}{\mathcal H}{}^{\rm Gal}
&=
\nabla_i\nabla_j\beta^{ij}
-
\nabla^2\beta
=
0,
\label{eq:weak-nlo-ham}
\\
\overset{(2)}{\mathcal A}{}^{\rm Gal}_{ij}
&=
-\nabla_i\nabla_jM
+\frac12
\left(
\nabla_k\nabla_i\beta_j{}^k
+\nabla_k\nabla_j\beta_i{}^k
-\nabla^2\beta_{ij}
-\nabla_i\nabla_j\beta
\right)
=
0 .
\label{eq:weak-nlo-evol}
\end{align}
\end{subequations}
The NNLO equations reduce to
\begin{subequations}
\label{eq:weak-nnlo-eombulk}
\begin{align}
\overset{(4)}{\mathcal H}{}^{\rm Gal}
&=
\nabla_i\nabla_j \epsilon^{ij}
-
\nabla^2 \epsilon
\nonumber\\
&\quad
+\nabla_k\!\left(
\beta^{ij}\nabla^k\beta_{ij}
-\beta^{ij}\nabla_i\beta_j{}^{k}
+\beta^{ki}\nabla_i\beta
-\beta^{k\ell}\nabla_i\beta^{i}{}_{\ell}
\right)
\nonumber\\
&\quad
+\frac12\,(\nabla_i\beta_{jk})(\nabla^j\beta^{ik})
-\frac14\,(\nabla_i\beta_{jk})(\nabla^i\beta^{jk})
-\frac14\,(\nabla_i\beta)(\nabla^i\beta)
=
0,
\label{eq:weak-nnlo-eombulk-H}
\\[0.5em]
\overset{(4)}{\mathcal H}{}^{\rm Gal}_{i}
&=
\nabla_jL_i{}^j - \nabla_iL = 0,
\label{eq:weak-nnlo-eombulk-Hi}
\\[0.5em]
\overset{(4)}{\mathcal A}{}^{\rm Gal}_{ij}
&=
-\nabla_i\nabla_jP
+
R^{(4)}_{ij}
+\frac12
\left(
\nabla_i\beta_j{}^k
+\nabla_j\beta_i{}^k
-\nabla^k\beta_{ij}
\right)\nabla_kM
\nonumber\\
&\quad
+\frac{M}{2}
\left(
\nabla_k\nabla_i\beta_j{}^k
+\nabla_k\nabla_j\beta_i{}^k
-\nabla^2\beta_{ij}
-\nabla_i\nabla_j\beta
\right)
=
0.
\label{eq:weak-nnlo-eombulk-A}
\end{align}
\end{subequations}

\subsubsection{NLO weak solutions: Kerr, Hartle--Thorne, and Mixed-type solutions}

At NLO, the weak-branch Galilean equations involve only the scalar field \(M\)
and the spatial metric correction \(\beta_{ij}\). Since the equations
\eqref{eq:weak-nlo-eombulk} are linear in these fields, the Kerr-type and
quadrupolar Hartle--Thorne-type sectors can be treated independently. The comparison with the corresponding relativistic metrics is used not to
define the truncated Galilean equations, but to fix the integration
constants and to give a relativistic interpretation of the Galilean data;
see Appendices~\ref{kerrapp} and \ref{htapp}.

Let us also clarify the role of the Newtonian potential in the weak branch.
Although one of the LO equations reads
\begin{equation}
  N R_{ij}-\nabla_i\nabla_j N=0 ,
\end{equation}
and its trace resembles a Laplace equation, the leading lapse \(N\) is not the
Newtonian potential in the weak large-\(c\) expansion. Indeed, the ADM lapse is
expanded as
\begin{equation}
  \hat N = N+c^{-2}M+\cdots ,
\end{equation}
so that, for the weak leading data \(N=1\),
\begin{equation}
  g_{tt}=-c^2\hat N^2+\cdots
        =-c^2-2M+\mathcal O(c^{-2}) .
\end{equation}
Therefore the Newtonian potential is encoded in the subleading lapse \(M\),
rather than in the leading lapse \(N\). A non-trivial leading lapse would
modify \(g_{tt}\) already at order \(c^2\), and hence belongs to the strong
time-dilation sector rather than to the weak Newtonian branch.

This is also reflected in the field equations. In the weak branch, the trace
of the NLO spatial equation
\(\overset{(2)}{\mathcal A}{}^{\rm Gal}_{ij}\) in
\eqref{eq:weak-nlo-eombulk} gives
\begin{equation}
  -\nabla^2 M
  +\left(\nabla_i\nabla_j\beta^{ij}-\nabla^2\beta\right)=0 .
\end{equation}
Using the NLO Hamiltonian constraint
\(\overset{(2)}{\mathcal H}{}^{\rm Gal}\) in
\eqref{eq:weak-nlo-eombulk},
\begin{equation}
  \nabla_i\nabla_j\beta^{ij}-\nabla^2\beta=0 ,
\end{equation}
one obtains
\begin{equation}
  \nabla^2 M=0
\end{equation}
in vacuum. Upon coupling to nonrelativistic matter, this equation becomes the
usual Poisson equation. Thus, in the weak branch, the Newtonian potential is
captured by the subleading lapse \(M\) at NLO, rather than by the leading
lapse \(N\) at LO, in agreement with the covariant large-\(c\) expansion
\cite{VandenBleeken:2017rij,Hansen:2019pkl,Hansen:2020pqs}.

Since the weak-branch NLO equations \eqref{eq:weak-nlo-eombulk} are linear on
the flat background, the vacuum solution can be extended to
higher mass multipoles. In what follows we use the shorthand notation
\begin{equation}
  P_\ell \equiv P_\ell(\cos\theta).
\end{equation}
Together with the corresponding multipolar expansion of \(M(r,\theta)\),
the asymptotically decaying branch gives the family
\begin{align}
  M(r,\theta)
  &=
  -\frac{C_0}{2r}
  -
  \frac12
  \sum_{\ell\geq 1}
  \frac{C_\ell}{r^{\ell+1}}P_\ell,
  \label{eq:NLO-multipole-M}
  \\
  \beta_{rr}(r,\theta)
  &=
  \frac{C_0}{r}
  +
  \sum_{\ell\geq 1}
  \frac{C_\ell}{r^{\ell+1}}P_\ell,
  \label{eq:NLO-multipole-beta-rr}
  \\
  \beta_{\theta\theta}(r,\theta)
  &=
  \sum_{\ell\geq 1}
  \frac{C_\ell}{r^{\ell-1}}P_\ell,
  \label{eq:NLO-multipole-beta-thth}
  \\
  \beta_{\phi\phi}(r,\theta)
  &=
  \sin^2\theta
  \sum_{\ell\geq 1}
  \frac{C_\ell}{r^{\ell-1}}P_\ell .
  \label{eq:NLO-multipole-beta-phph}
\end{align}
Here \(C_0\) is the monopole parameter, while \(C_\ell\) with
\(\ell\geq1\) describe higher mass multipoles. If one
imposes equatorial reflection symmetry, only the even modes
\(\ell=0,2,4,\ldots\) are kept. For example, keeping the first few even
multipoles, one obtains
\begin{align}
  M(r,\theta)
  &=
  -\frac{C}{2r}
  -\frac{D}{2r^3}P_2
  -\frac{E}{2r^5}P_4
  +\cdots ,
  \\
  \beta_{rr}(r,\theta)
  &=
  \frac{C}{r}
  +\frac{D}{r^3}P_2
  +\frac{E}{r^5}P_4
  +\cdots ,
  \\
  \beta_{\theta\theta}(r,\theta)
  &=
  \frac{D}{r}P_2
  +\frac{E}{r^3}P_4
  +\cdots ,
  \\
  \beta_{\phi\phi}(r,\theta)
  &=
  \left[
  \frac{D}{r}P_2
  +\frac{E}{r^3}P_4
  +\cdots
  \right]\sin^2\theta .
\end{align}
Here \(C\equiv C_0\) fixes the monopole, \(D\equiv C_2\) the quadrupole
and \(E\equiv C_4\) the hexadecapole-type mass multipole. 
Higher multipoles can be included at NLO without changing the linear
equations, but their NNLO completion would require solving the
corresponding nonlinear backreaction equations.

\paragraph{Weak and static Hartle--Thorne-type sector.} The weak and static Hartle--Thorne-type solution corresponds to keeping only the monopole and quadrupole pieces, with
\begin{equation}
  C=2Gm,
  \qquad
  D=-2GQ.
\end{equation}
Therefore the quadrupolar Hartle--Thorne-type NLO weak data are
\begin{align}
M
&=
-\frac{Gm}{r}
+
\frac{GQ}{r^3}P_2,
\\
\beta_{ij}
&=
\mathrm{diag}
\left(
\frac{2Gm}{r}
-
\frac{2GQ}{r^3}P_2,
\;
-\frac{2GQ}{r}P_2,
\;
-\frac{2GQ}{r}P_2\sin^2\theta
\right).
\label{eq:weak-ht-nlo-data}
\end{align}
These fields solve the NLO weak Galilean equations but to fix the coefficient we use the Hartle--Thorne metric \cite{Hartle:1967he, Hartle:1968si}. We expand this metric with respect to large-$c$ and obtain the order by order ADM data. The details can be found in the Appendix~\ref{htapp}.

\paragraph{Kerr-type sector.} Apart from the multipole expansion given above, there is another nontrivial solution of the NLO weak field equations \eqref{eq:weak-nlo-eombulk}. For the reason explained below, it will be called Kerr-type solution.  The Kerr-type NLO weak data are
\begin{align}
M
&=
-\frac{Gm}{r},
\\
\beta_{ij}
&=
\mathrm{diag}
\left(
\frac{2Gm}{r}
-\frac{J^2}{m^2r^2}\sin^2\theta,
\;
\frac{J^2}{m^2}\cos^2\theta,
\;
\frac{J^2}{m^2}\sin^2\theta
\right).
\label{eq:weak-kerr-nlo-data}
\end{align}
These fields solve the NLO weak Galilean equations \footnote{See \cite{Ergen:2020yop} for the discussion on weak and strong Kerr solutions in odd expansion of GR.}. At this order, the Kerr-type data contain the spin-squared spatial
deformation through $\beta_{ij}$. Because of the linearity of the field equations, the coefficients of these $J^2$ terms are not fixed intrinsically by the Galilean equations alone. We fix their normalization by matching to the weak large-$c$ expansion of the Kerr metric, see Appendix~\ref{kerrapp}. A related large-$c$ expansion of Kerr was studied in \cite{Hartong:2023yxo}; the comparison involves a different coordinate/read-off convention, as discussed in Appendix~\ref{kerrapp}. The linear in spin shift $A_\phi$, which gives the frame-dragging sector, enters the weak-branch equations only at NNLO.

\paragraph{Mixed NLO weak sector.}
Since the NLO weak equations are linear in \(M\) and \(\beta_{ij}\), the Kerr-type and quadrupolar Hartle--Thorne-type sectors can be superposed at this order. The mixed NLO weak data are
\begin{align}
M &= -\frac{Gm}{r} + \frac{GQ}{r^3}P_2, \\
\beta_{rr} &= \frac{2Gm}{r} - \frac{2GQ}{r^3}P_2 - \frac{J^2}{m^2r^2}\sin^2\theta, \notag \\
\beta_{\theta\theta} &= \frac{J^2}{m^2}\cos^2\theta - \frac{2GQ}{r}P_2, \notag \\
\beta_{\phi\phi} &= \left[ \frac{J^2}{m^2} - \frac{2GQ}{r}P_2\right]\sin^2\theta. \label{eq:weak-mixed-nlo-data}
\end{align}
We checked that these fields solve the NLO weak Galilean equations as well. At this order there is no obstruction to the superposition within the linear truncated NLO system, and the construction can be generalized to the mass multipoles. This statement does not by itself select a unique full relativistic completion. The obstruction appears only at NNLO, where nonlinear source terms generate mixed $J^2Q$ contributions. This will be discussed separately in the mixed weak-branch solution.

\subsubsection{NNLO weak solutions: Kerr-type and Extended Hartle--Thorne-type}

We now turn to the NNLO weak sector. At this order the equations are no longer linear in the NLO fields. In particular, the NNLO Hamiltonian and evolution equations contain nonlinear source terms built from \(M\), \(\beta_{ij}\), and the vector contribution through \(L_{ij}\). The compact form of the weak-branch NNLO equations was given in \eqref{eq:weak-nnlo-eombulk}. This is the first order at which the rotational vector sector contributes to the weak-branch dynamics.

For the weak branch, the NNLO momentum equation reduces to
\begin{equation}
  \nabla^jL_{ij}-\nabla_iL=0,
  \qquad
  L_{ij}=-\nabla_{(i}A_{j)}.
\end{equation}
Equivalently, on the flat leading spatial background, this gives
\begin{equation}
  \nabla^2A_i-\nabla_i\nabla_jA^j=0.
  \label{eq:weak-vector-equation}
\end{equation}
For a purely azimuthal toroidal ansatz,
\begin{equation}
  A_i dx^i=A_\phi(r,\theta)d\phi,
\end{equation}
with no \(\phi\)-dependence, the divergence vanishes. Therefore \eqref{eq:weak-vector-equation} reduces to the vector Laplace equation
\begin{equation}
  \nabla^2A_i=0.
\end{equation}
Regular angular dependence and asymptotically vanishing boundary conditions then lead to the toroidal harmonic expansion
\begin{equation}
  A_\phi(r,\theta)
  =
  \sum_{\ell=1}^{\infty}
  \frac{a_\ell}{r^\ell}\,
  \sin\theta\,\partial_\theta P_\ell,
  \label{eq:toroidal-vector-solution}
\end{equation}
where the growing radial branch has been discarded. The stationary rotating solutions considered below correspond to the \(\ell=1\) sector \footnote{Note that the spatial indices are raised and lowered by the spatial metric $\gamma$\,.},
\begin{equation}
  A_\phi
  =
  -\frac{2GJ}{r}\sin^2\theta .
  \label{eq:weak-Aphi}
\end{equation}

The \(\ell=1\) mode is the minimal rotational sector. It has the same
linear-in-spin structure as the usual Lense--Thirring frame-dragging
term. More generally, the homogeneous vector equation also allows higher
toroidal harmonics. These modes correspond to independent current
multipole data of the source. For example, in an equatorially symmetric
configuration the odd modes \(\ell=3,5,\ldots\) may be interpreted as
higher current multipoles. In the present work we keep only the
\(\ell=1\) mode, since our aim is to construct the minimal
spin--quadrupole sector involving the angular momentum \(J\) and the mass
quadrupole moment \(Q\). Including higher toroidal modes would enlarge
the solution space, but their physical interpretation and possible
higher-order backreaction should be analyzed separately.

\paragraph{Kerr-type sector.}
Together with the NLO Kerr-type data, the NNLO weak Kerr-type fields are
\begin{align}
  P
  &=
  \frac{GJ^2}{m r^3}\cos^2\theta
  -\frac{G^2m^2}{2r^2},
  \\
   A_\phi
  &=
  -\frac{2GJ}{r}\sin^2\theta\,, \nn \\
  \epsilon_{rr}
  &=
  \frac{
  \left(\frac{J^2}{m^2}-2Gmr\right)
  \left(\frac{J^2}{m^2}\sin^2\theta-2Gmr\right)}
  {r^4},
  \nonumber\\
  \epsilon_{\theta\theta}
  &=
  0,
  \qquad
  \epsilon_{\phi\phi}
  =
  \frac{2GJ^2}{m r}\sin^4\theta .
  \label{eq:weak-kerr-nnlo-data}
\end{align}
These fields solve the NNLO weak Galilean equations. The Kerr-type NNLO data are obtained by matching to the weak large-$c$ expansion of the Kerr metric in the ADM read-off convention used in Appendix~\ref{kerrapp}. A related large-$c$ expansion of Kerr was studied in \cite{Hartong:2023yxo}. The comparison with that result involves a different coordinate/read-off convention, and the corresponding coordinate transformation is discussed in Appendix~\ref{kerrapp}.
 The vector contribution enters the weak-branch Galilean data at NNLO
through
\[
  A_\phi
  =
  -\frac{2GJ}{r}\sin^2\theta .
\]
This has the same linear-in-spin structure as the standard
Lense--Thirring frame-dragging term. In the present expansion, however,
we regard it as part of the weak Kerr-type NNLO sector rather than as a
separate NLO Lense--Thirring solution \cite{Lense:1918}.

\paragraph{Hartle--Thorne-type sector.}
If one reads the NNLO fields from the ordinary Hartle--Thorne metric (see Eq. \eqref{eq:weak-ht-epsilon} in Appendix~\ref{htapp}), one obtains
\begin{align}
  P
  &=-\frac{1}{2}
\left(
-\frac{Gm}{r}
+
\frac{GQ}{r^3}P_2
\right)^2
+
\frac{G^2mQ}{r^4}P_2,
  \nonumber\\
   A_\phi
  &=
  -\frac{2GJ}{r}\sin^2\theta\,, \nn \\
  \epsilon_{rr}
  &=
  \frac{4G^2m^2}{r^2}
  -\frac{10G^2mQ}{r^4}P_2,
  \nonumber\\
  \epsilon_{\theta\theta}
  &=
  -\frac{5G^2mQ}{r^2}P_2,
  \nonumber\\
  \epsilon_{\phi\phi}
  &=
  -\frac{5G^2mQ}{r^2}P_2\sin^2\theta .
  \label{eq:weak-ht-nnlo-naive}
\end{align}
However, these fields do not solve the full NNLO weak Galilean equations. 
The reason is a mismatch between the truncation of the ordinary 
Hartle--Thorne metric and the nonlinear structure of the NNLO equations. 
The ordinary Hartle--Thorne geometry is linear in the independent quadrupole 
parameter \(Q\), whereas the NNLO weak Galilean equations contain nonlinear 
source terms built from the NLO fields, schematically
\begin{equation}
  \nabla\beta\,\nabla\beta,
  \qquad
  \beta\,\nabla\nabla\beta .
\end{equation}
Since the NLO quadrupolar field is proportional to \(Q\), these terms source 
\(Q^2\) contributions at NNLO.  Thus, unless one consistently truncates the 
NNLO equations to linear order in \(Q\), the corresponding post-linear 
quadrupolar corrections must be included.  The naive Hartle--Thorne read-off 
in \eqref{eq:weak-ht-nnlo-naive} does not contain these \(Q^2\) terms and 
therefore does not satisfy the full NNLO weak Galilean equations. This motivates considering a post-linear extension of the Hartle--Thorne
sector, where the missing \(Q^2\) corrections are included \cite{Frutos-Alfaro:2015lua}. In the next
subsection we therefore turn to the extended Hartle--Thorne metric and use
it as the relativistic input for the weak NNLO Galilean read-off.

\paragraph{Extended Hartle--Thorne-type sector.}
Here the word extended refers to the post-linear quadrupole--quadrupole completion of the ordinary Hartle--Thorne metric  \cite{Frutos-Alfaro:2015lua}. This extension is necessary because the ordinary Hartle--Thorne metric is linear in the quadrupole parameter \(Q\), whereas the NNLO weak equations generate \(Q^2\) source terms from the NLO quadrupolar fields. The required correction modifies the NNLO lapse and spatial metric by adding the \(Q^2\) terms. The consistent weak post-linear Hartle--Thorne-type data are
\begin{align}
  P
  &=
  -\frac{G^2m^2}{2r^2}
  +\frac{2G^2mQ}{r^4}P_2
  +\frac{G^2Q^2}{2r^6}(P_2)^2,\nn 
  \\
  A_\phi
  &=
  -\frac{2GJ}{r}\sin^2\theta\,, \nn \\
  \epsilon_{rr}
  &=
  \frac{4G^2m^2}{r^2}
  -\frac{10G^2mQ}{r^4}P_2
  +\frac{G^2Q^2}{12r^6}
  \left(
  8(P_2)^2
  -16P_2
  +77
  \right),
  \nonumber\\
  \epsilon_{\theta\theta}
  &=
  -\frac{5G^2mQ}{r^2}P_2
  +\frac{G^2Q^2}{36r^4}
  \left(
  44(P_2)^2
  +8P_2
  -43
  \right),
  \nonumber\\
  \epsilon_{\phi\phi}
  &=
  \left[
  -\frac{5G^2mQ}{r^2}P_2
  +\frac{G^2Q^2}{36r^4}
  \left(
  44(P_2)^2
  +8P_2
  -43
  \right)
  \right]\sin^2\theta .
  \label{eq:weak-ht-postlinear-epsilon}
\end{align}
Together with the NLO Hartle--Thorne-type fields, these post-linear NNLO data solve the weak NNLO Galilean equations. This shows explicitly that the NNLO Galilean equations detect the incompleteness of the ordinary Hartle--Thorne seed and require the quadrupole--quadrupole sector.

\subsubsection{Post-linear mixed weak-branch solution}
\label{sec:mixed-weak-branch}

We now combine the weak Kerr-type spin sector with the extended Hartle--Thorne-type quadrupolar sector. The previous subsections show that these two sectors solve the NNLO weak Galilean equations separately. However, because the NNLO equations contain nonlinear source terms built from the NLO fields, their naive superposition is not automatically a solution. In particular, mixed spin--quadrupole source terms of order \(J^2Q\) remain. The purpose of this subsection is to show that the NNLO Galilean equations require a corrected mixed weak branch solution. To show this, we start from the naive superposition of the NNLO Kerr-type and post-linear Hartle-Thorne type data and then determine the additional terms required by the mixed $J^2Q$ sources.

At NLO, the mixed weak data are obtained by combining the Kerr-type and quadrupolar Hartle--Thorne-type sectors as,
\begin{align}
M
&=
-\frac{Gm}{r}
+
\frac{GQ}{r^3}P_2,
\\
\beta_{ij}
&=
\mathrm{diag}
\left(
\beta_{rr},
\beta_{\theta\theta},
\beta_{\phi\phi}
\right),
\end{align}
where
\begin{align}
\beta_{rr}
&=
\frac{2Gm}{r}
-
\frac{2GQ}{r^3}P_2
-
\frac{J^2}{m^2r^2}\sin^2\theta,
\\
\beta_{\theta\theta}
&=
\frac{J^2}{m^2}\cos^2\theta
-
\frac{2GQ}{r}P_2,
\\
\beta_{\phi\phi}
&=
\left(
\frac{J^2}{m^2}
-
\frac{2GQ}{r}P_2
\right)\sin^2\theta .
\end{align}
As shown above, these fields solve the NLO weak Galilean equations (see again \eqref{eq:weak-mixed-nlo-data}). At NNLO, the vector sector starts contributing through
\begin{equation}
  A_\phi
  =
  -\frac{2GJ}{r}\sin^2\theta\,.
\end{equation}
The direct combination of the Kerr-type and extended Hartle--Thorne-type NNLO data gives
\begin{equation}
P_{\rm sup}
=
-\frac{G^2m^2}{2r^2}
+
\frac{GJ^2}{mr^3}\cos^2\theta
+
\frac{2G^2mQ}{r^4}P_2
+
\frac{G^2Q^2}{2r^6}(P_2)^2,
\end{equation}
and
\begin{align}
\epsilon^{\rm sup}_{rr}
&=
\frac{
\left(
\frac{J^2}{m^2}-2Gmr
\right)
\left(
\frac{J^2}{m^2}\sin^2\theta-2Gmr
\right)
}{r^4}
-
\frac{10G^2mQ}{r^4}P_2
\nonumber\\
&\quad
+
\frac{G^2Q^2}{12r^6}
\left(
8(P_2)^2
-
16P_2
+
77
\right),
\\
\epsilon^{\rm sup}_{\theta\theta}
&=
-\frac{5G^2mQ}{r^2}P_2
+
\frac{G^2Q^2}{36r^4}
\left(
44(P_2)^2
+
8P_2
-
43
\right),
\\
\epsilon^{\rm sup}_{\phi\phi}
&=
\frac{2GJ^2}{mr}\sin^4\theta
-
\frac{5G^2mQ}{r^2}P_2\sin^2\theta
\nonumber\\
&\quad
+
\frac{G^2Q^2}{36r^4}
\left(
44(P_2)^2
+
8P_2
-
43
\right)\sin^2\theta .
\end{align}
However, this direct superposition does not solve the NNLO weak Galilean equations. The main reason for this failure is that the direct substitution of these data into the NNLO equations leaves residual mixed terms proportional to \(J^2Q\). To correct this superposition, we add the following ansatz
\bea
 P_{\rm mix} &=& P_{\rm sup} + \Delta P (r,\theta)\,, \nn \\
 \epsilon^{\rm mix}_{ij} &=& \epsilon^{\rm sup}_{ij} + \Delta\epsilon_{ij} (r,\theta)\,, 
\eea 
where $\Delta P(r,\theta)$ and $\Delta\epsilon_{ij} (r,\theta)$ are unknown correction terms to be fixed by the NNLO equations. Solving the NNLO field equations gives
\bea 
  \Delta P
  &=&
  -\frac{6GJ^2Q}{7m^2r^5}P_2\,, \nn \\
\Delta\epsilon_{ij}
&=&
\frac{GJ^2Q}{m^2r^3}\, 
\mathrm{diag}
\left(
0,
\mathcal F_\theta(\theta),
\sin^2\theta\,\mathcal F_\phi(\theta)
\right).
\label{eq:mixed-delta-epsilon}
\eea 
The angular functions are
\begin{align}
\mathcal F_\theta(\theta)
&=
-\frac{41}{175}
+\frac{22}{35}P_2
-\frac{54}{175}P_4,
\\
\mathcal F_\phi(\theta)
&=
-\frac{29}{175}
+\frac{2}{7}P_2
-\frac{6}{175}P_4.
\end{align}
Therefore the corrected mixed weak-branch ADM data are
\begin{equation}
  \left(
  M,\,
  A_\phi,\,
  \beta_{ij},\,
  P_{\rm mix},\,
  \epsilon^{\rm mix}_{ij}
  \right),
\end{equation}
with the components given above. These fields solve the NNLO weak Galilean
equations \footnote{In this work, ``solve'' means that the data are substituted into the corresponding stationary field equations obtained from the expanded ADM action, or
equivalently from the expanded ADM Hamilton equations.}. As in the previous weak-branch examples, these data should be understood as a representative in the ADM gauge used here, rather than as a gauge-invariant classification of mixed \(J^2Q\) configurations. The
important point is that the mixed spin--quadrupole sector is not obtained
by a naive superposition of the separately consistent Kerr-type and
extended Hartle--Thorne-type sectors. The NNLO equations generate
non-vanishing \(J^2Q\) sources, and these sources fix the additional terms
in \eqref{eq:mixed-delta-epsilon}. This statement is intrinsic to the truncated NNLO Galilean ADM equations
for the stationary ansatz and ADM gauge used here. It should not be
interpreted as a claim that the corrected mixed data determine a unique
full stationary axisymmetric solution of general relativity. Such a
relativistic completion, if it exists, may require additional higher-order
fields, higher-shift data, and further coordinate or gauge choices.

\subsubsection*{Metric reconstructed from the Galilean data.}
This corrected data set defines a consistent Galilean weak-branch NNLO vacuum solution. Using the inverse map of
Appendix~\ref{app:adm-conventions-read-off}, we reconstruct the spacetime
metric components fixed by the NNLO Galilean data. However, we emphasize that the NNLO Galilean equations do not fix
an independent higher-order shift coefficient \(Z_\phi\) (see the discussion below Table~\ref{tab:field_equations}). Such a term may
be supplied by matching to a particular relativistic uplift, for example
the weak Kerr expansion, but it is not part of the Galilean solution
itself.  Using the weak-branch inverse map (Appendix~\ref{app:adm-conventions-read-off}), one obtains
\begin{align}
  g_{tt}^{\rm mix}
  &=
  -c^2
  +\frac{2Gm}{r}
  -\frac{2GQ}{r^3}P_2
  \nonumber\\
  &\quad
  +\frac{1}{c^2}
  \left[
  -\frac{2GJ^2}{m r^3}\cos^2\theta
  -\frac{2G^2mQ}{r^4}P_2
  -\frac{2G^2Q^2}{r^6}P_2^2
  +\frac{12GJ^2Q}{7m^2r^5}P_2
  \right],
  \\
  g_{t\phi}^{\rm mix}
  &=
  -\frac{2GJ}{c^2 r}\sin^2\theta
  +\frac{1}{c^4}
  \left[
  -\frac{2GJ^3}{m^2r^3}
  +\frac{4G^2JQ}{r^4}P_2
  \right]\sin^2\theta,
  \\
  g_{rr}^{\rm mix}
  &=
  1
  +\frac{1}{c^2}
  \left[
  \frac{2Gm}{r}
  -\frac{2GQ}{r^3}P_2
  -\frac{J^2}{m^2r^2}\sin^2\theta
  \right]
  \nonumber\\
  &\quad
  +\frac{1}{c^4}
  \bigg[
  \frac{
  \left(
  \frac{J^2}{m^2}-2Gmr
  \right)
  \left(
  \frac{J^2}{m^2}\sin^2\theta-2Gmr
  \right)
  }{r^4}
  -\frac{10G^2mQ}{r^4}P_2\, \nn \\
  &+ \quad \frac{G^2Q^2}{12r^6}
  \left(
  8P_2^2-16P_2+77
  \right)
  \bigg],
  \\
  g_{\theta\theta}^{\rm mix}
  &=
  r^2
  +\frac{1}{c^2}
  \left[
  \frac{J^2}{m^2}\cos^2\theta
  -\frac{2GQ}{r}P_2
  \right]
  \nonumber\\
  &\quad
  +\frac{1}{c^4}
  \left[
  -\frac{5G^2mQ}{r^2}P_2
  +\frac{G^2Q^2}{36r^4}
  \left(
  44P_2^2+8P_2-43
  \right)
  +\frac{GJ^2Q}{m^2r^3}
  \mathcal F_\theta(\theta)
  \right],
  \\
  g_{\phi\phi}^{\rm mix}
  &=
  r^2\sin^2\theta
  +\frac{1}{c^2}
  \left[
  \frac{J^2}{m^2}
  -\frac{2GQ}{r}P_2
  \right]\sin^2\theta
  \nonumber\\
  &\quad
  +\frac{1}{c^4}
  \bigg[
  \frac{2GJ^2}{mr}\sin^4\theta
  -\frac{5G^2mQ}{r^2}P_2\sin^2\theta
  \nonumber\\
  &\qquad
  +\frac{G^2Q^2}{36r^4}
  \left(
  44P_2^2+8P_2-43
  \right)\sin^2\theta
  +\frac{GJ^2Q}{m^2r^3}
  \sin^2\theta\,\mathcal F_\phi(\theta)
  \bigg].
\end{align}
These are the metric components reconstructed from the NNLO Galilean ADM
data in the chosen ADM representative. They differ from both the
Hartle--Thorne metric and its post-linear quadrupole extension by
additional mixed $J^2Q$ terms. A natural open question is whether these
finite-order reconstructed data can be related to known stationary
axisymmetric multipolar geometries \cite{Ryan:1995wh}. Establishing such
a relation would require the missing higher-order completion data and an
explicit coordinate or gauge matching.

\paragraph{Frame dragging:} The finite-order reconstructed metric components can also be used to extract the frame-dragging
angular velocity,
\begin{equation}
  \Omega_{\rm drag}
  =
  -\frac{g_{t\phi}}{g_{\phi\phi}} .
\end{equation}
This quantity provides a simple comparison between the weak and strong
branches. In the corrected mixed weak-branch solution, the leading shift
vanishes and the NNLO Galilean data fix
\begin{equation}
  \Omega_{\rm drag}^{\rm mix,Gal}
  =
  \frac{2GJ}{c^2r^3}
  +\mathcal O(c^{-6}) .
\end{equation}
The absence of a \(c^{-4}\) correction follows from the inverse ADM map:
the \(\beta_{\phi\phi}A^\phi\) term in \(g_{t\phi}\) cancels the
corresponding correction coming from \(g_{\phi\phi}\).  Hence any
\(c^{-4}\) correction to the weak-branch frame dragging requires additional
higher-order shift data, such as a choice of \(Z_\phi\), and is therefore an
extra relativistic uplift choice rather than a consequence of the NNLO
Galilean equations alone.  The mixed \(J^2Q\) data enter the reconstructed
spatial metric at NNLO, but their direct contribution to
\(\Omega_{\rm drag}\) appears only at the next order, where such higher shift
data are also required.

\subsection{Strong branch}
In the strong branch the leading fields \((N,\gamma_{ij},N^i)\) are allowed to be nontrivial. Consequently, \(K_{ij}\) may also be nonzero already in the NLO field equations unlike the weak branch. Therefore, in the strong branch we take the following LO data 
\begin{equation}
  N=\sqrt f,
  \qquad
  \gamma_{ij}
  =
  \mathrm{diag}\left(f^{-1},r^2,r^2\sin^2\theta\right),
  \qquad
  f=1-\frac{2Gm}{r}. \label{LOstrondata}
\end{equation}
In contrast to the weak branch, the leading shift need not vanish. In the Kerr-type strong branch, the shift is
\begin{equation}
  N_\phi
  =
  -\frac{2GJ}{r}\sin^2\theta .
\end{equation}
Although \(N_\phi\) appears in the leading ADM data, it does not enter the LO Galilean variational equations. It contributes first at NLO through the extrinsic curvature
\begin{equation}
  K_{ij}
  =
  -N^{-1}\nabla_{(i}N_{j)} .
\end{equation}
Similarly, the next shift coefficient \(A_i\) contributes first at NNLO through \(L_{ij}\). Therefore, in the strong branch the dynamical hierarchy is naturally organized as
\begin{align}
\text{LO:}\qquad
&
N=\sqrt f,
\qquad
\gamma_{ij}
=
\mathrm{diag}\left(f^{-1},r^2,r^2\sin^2\theta\right),
\\
\text{NLO:}\qquad
&
N_\phi,
\qquad
M,
\qquad
\beta_{ij},
\\
\text{NNLO:}\qquad
&
A_\phi,
\qquad
P,
\qquad
\epsilon_{ij}.
\end{align}
The strong branch is adapted to expansions around Schwarzschild background.

\subsubsection{Strong Kerr-type solutions}

The strong Kerr-type sector is obtained by solving the Galilean equations around the nontrivial leading--order data \eqref{LOstrondata}. The NLO Galilean data are
\begin{align}
N_\phi
&=
-\frac{2GJ}{r}\sin^2\theta,
\\
M
&=
\frac{1}{\sqrt f}
\left(
\frac{GJ^2}{m r^3}\cos^2\theta
+
\frac{2G^2J^2}{r^4}\sin^2\theta
\right),
\\
\beta_{rr}
&=
\frac{J^2}{m^2r^2}
\left(
\frac{\cos^2\theta}{f}
-
\frac{1}{f^2}
\right),
\\
\beta_{\theta\theta}
&=
\frac{J^2}{m^2}\cos^2\theta,
\\
\beta_{\phi\phi}
&=
\frac{J^2}{m^2}\sin^2\theta
\left(
1+\frac{2Gm}{r}\sin^2\theta
\right).
\end{align}
These fields solve the NLO strong Galilean equations. Again we fix the integration constant by comparing the strong scaling of the Kerr metric (See details in Appendix~\ref{strongkerrapp}). The nonzero leading shift is not a removable uniform boost. Indeed, it is position dependent and represents the frame-dragging sector of the strong branch. Note that this is also consistent with the strong-field large-\(c\) expansion of Kerr discussed in Ref.~\cite{VandenBleeken:2019gqa}. The apparent absence of the higher-order spatial tensor in that description is a consequence of a particular gauge choice in KS decomposition. In the ADM parametrization used here the same sector is kept explicitly through \(\beta_{ij}\).

At NNLO, the shift correction \(A_\phi\), the lapse correction \(P\), and the spatial metric correction \(\epsilon_{ij}\) contribute. The NNLO data are
\begin{align}
A_\phi
&=
\frac{2GJ^3}{m^2r^3}\sin^2\theta
\left(
1+\cos^2\theta+\frac{2Gm}{r}\sin^2\theta
\right),
\\
\epsilon_{rr}
&=
\frac{J^4}{m^4r^4}
\left(
\frac{1}{f^3}
-
\frac{\cos^2\theta}{f^2}
\right),
\qquad
\epsilon_{\theta\theta}
=
0,
\\
\epsilon_{\phi\phi}
&=
-\frac{2GJ^4}{m^3r^3}
\sin^4\theta\cos^2\theta,
\\
P
&= 
-\frac{J^4}{2m^4\sqrt f}
\Bigg[
\frac{2Gm}{r^5}\cos^4\theta
\nonumber\\
&\qquad
+\frac{4G^2m^2}{r^6}\sin^2\theta
\left(
1+2\cos^2\theta+\frac{2Gm}{r}\sin^2\theta
\right)
\nonumber\\
&\qquad
+\frac{G^2m^2}{f r^6}
\left(
\cos^2\theta+\frac{2Gm}{r}\sin^2\theta
\right)^2
\Bigg].
\end{align}
Together with the NLO data above, these fields solve the NNLO strong Galilean equations. This solution can be considered as the strong gravity with slow rotational effects up to $\mathcal{O}(J^4)$\,.  This solution can also be obtained by the strong scaling of the Kerr metric and using the read-off data.

\subsubsection{Strong Hartle--Thorne-type NLO solution}

There is also a strong Hartle--Thorne-type NLO solution. It is obtained
around the same strong Schwarzschild leading--order data \eqref{LOstrondata}, but it contains an independent quadrupole parameter \(Q\) together with spin parameter $J$. The NLO data are
\begin{align}
N_\phi
&=
-\frac{2GJ}{r}\sin^2\theta,
\\
M
&=
\frac{1}{\sqrt f}
\left[
f H_1 P_2
+
\frac{G^2J^2}{r^4}
\right],
\\
\beta_{rr}
&=
-\frac{1}{f}
\left[
2H_2P_2
+
\frac{2G^2J^2}{fr^4}
\right],
\\
\beta_{\theta\theta}
&=
-2r^2H_3P_2,
\\
\beta_{\phi\phi}
&=
\beta_{\theta\theta}\sin^2\theta\,.
\end{align}
Here the functions \(H_1,H_2,H_3\) are given explicitly in Appendix~\ref{shtapp}. These fields solve the NLO strong Galilean equations in the exterior region \(r>2Gm\).

This solution also agrees with the data obtained from the strong scaling of
the Hartle--Thorne metric, as shown in Appendix~\ref{shtapp}. Moreover, the
Kerr-type solution is recovered from the special value
\begin{equation}
Q=\frac{J^2}{m},
\end{equation}
where the logarithmic Hartle--Thorne quadrupole terms drop out.

We did not include a strong Hartle--Thorne NNLO solution in this work.
The reason is that the ordinary relativistic Hartle--Thorne metric does
not provide independent strong \(c^{-4}\) spatial data under this scaling.
Thus a genuine NNLO extension would require solving the NNLO Galilean
equations directly, or starting from a higher-order relativistic
Hartle--Thorne-type metric. This is a nontrivial problem and may be
interesting to study separately.

\paragraph{Frame dragging:} In the strong Kerr branch the leading shift is nonzero, so the
frame-dragging angular velocity already starts at order \(c^0\). Using the strong scaling of the Kerr metric gives 
\begin{align}
  \Omega_{\rm drag}^{\rm Kerr,strong}
  &=
  \frac{2GJ}{r^3}
  -
  \frac{2GJ^3}{c^2m^2r^7}
  \mathcal A_{\rm K}(r,\theta)
  \nonumber\\
  &\quad
  +
  \frac{2GJ^5}{c^4m^4r^{11}}
  \left[
  \mathcal A_{\rm K}(r,\theta)^2
  -
  r^4\cos^2\theta
  \right]
  +\mathcal O(c^{-6}),
\end{align}
where
\begin{equation}
  \mathcal A_{\rm K}(r,\theta)
  =
  r^2(1+\cos^2\theta)
  +
  2Gmr\sin^2\theta .
\end{equation}
This agrees with the large-\(c\) expansion of the exact Kerr
frame-dragging angular velocity through NNLO.  Thus, unlike in the weak
branch, no additional uplift ambiguity enters the strong Kerr result at this
order. 

\section{Conclusions} \label{sec:conclusions}

In this work, we started from a unified \(\omega\)-expansion of the ADM variables. This unification provides a common framework for the two opposite non-Lorentzian limits of general relativity. We focused on the Galilean branch and analyzed its ADM action, field equations and stationary solutions up to NNLO since the Carrollian branch, together with its stationary solutions, has been studied in \cite{Bal:2026xup}.

The NLO Galilean theory is closely related to Galilean-invariant nonrelativistic gravity \cite{Hansen:2019pkl}. At NNLO, the expansion contains nonlinear corrections which are essential for going beyond the leading nonrelativistic regime. In
\cite{Elbistan:2022plu} similar nonrelativistic actions were obtained in a Kol--Smolkin-type parametrization and the connection with post-Newtonian physics was discussed up to NNLO. The aim of the present work was complementary: we used the ADM formulation to derive a practical
order-by-order set of equations and to construct finite-order stationary
solutions of the truncated Galilean theory.

The main results are the stationary weak- and strong-branch solutions.
In the weak branch, the leading data are flat and the solutions can be
built order by order. At NLO, the Kerr-type spin sector and the Hartle--Thorne-type quadrupolar sector solve the Galilean equations
separately, and they can also be combined at this order. At NNLO, the
weak Kerr-type sector remains a solution, and the Hartle--Thorne-type
quadrupolar sector becomes a solution after including the post-linear
\(Q^2\) completion \cite{Frutos-Alfaro:2015lua}. However, the naive
superposition of these two NNLO sectors is not a solution. The nonlinear
NNLO equations generate mixed \(J^2Q\) source terms, and these terms
require additional corrections to the NNLO lapse $P$ and NNLO spatial tensor $\epsilon_{ij}$. This
gives the post-linear mixed weak-branch solution constructed in this
paper as a finite-order solution of the truncated NNLO Galilean ADM
equations in the chosen ADM gauge. Its full relativistic completion is
not fixed by these truncated data alone.

The strong branch has a different structure, since the leading ADM data
are already nontrivial. In this branch, the Kerr-type data provide a
consistent solution through NNLO. In addition, the strong scaling of the
Hartle--Thorne metric gives an NLO solution with an independent
quadrupole parameter \(Q\). The Kerr relation \(Q=J^2/m\) is only a
special limit of this solution. We do not construct a generic strong Hartle--Thorne-type NNLO
solution here. The ordinary Hartle--Thorne metric does not supply
independent strong \(c^{-4}\) spatial data under this scaling, so such
an extension would require solving the NNLO equations directly or using
a higher-order relativistic Hartle--Thorne-type metric. Thus the weak
branch naturally supports a systematic multipolar construction around
flat leading data, whereas the strong branch is more directly tied to
controlled large-\(c\) limits of strong-field geometries.

The finite-order reconstructed metric components provide a useful bridge
back to relativistic geometry, but they should not be interpreted by
themselves as unique full relativistic completions. They allow us to compare the Galilean solutions with expanded
Kerr and Hartle--Thorne metrics and also provide a starting point for
simple physical diagnostics. For example, frame dragging can be computed
from
\(\Omega_{\rm drag}=-g_{t\phi}/g_{\phi\phi}\). In the corrected weak
mixed solution, the direct effect of the \(J^2Q\) correction appears only
at the \(\mathcal{O}(c^{-6})\) order, where higher shift data are also
required. Geodesic motion may provide a more sensitive probe. Although
the rotational shift does not contribute directly to the leading
Galilean geodesic equation, it enters the relativistic geodesic equation
of the reconstructed metric through \(g_{t\phi}\). Moreover, the
post-linear weak-branch solution introduces new \(J^2Q\) terms in
\(g_{tt}\) and \(g_{\phi\phi}\), which can affect circular-orbit
conditions, orbital frequencies and precession effects. This is also
related to the standard use of slowly rotating and multipolar compact
object metrics in relativistic diagnostics
\cite{Hartle:1967he,Hartle:1968si,Ryan:1995wh,Baubock:2013gna}. In this
sense, the Galilean ADM reconstruction gives a systematic
field-equation-level organization of the metric data that enter such
relativistic diagnostics.

There are several natural extensions of this work. First, it would be
important to develop the Hamiltonian formulation of the NLO and NNLO
Galilean ADM theories in more detail. This should clarify the constraint
structure, the role of the lapse and shift equations, the treatment of
boundary terms, and the possible definition of asymptotic charges such
as mass and angular momentum in the nonrelativistic expansion. Second,
it would be interesting to extend the present analysis to genuinely
time-dependent backgrounds. At NNLO, the field \(\epsilon_{ij}\) does
not acquire an independent evolution equation from the Galilean action;
its time derivative appears only at the next order. A Hamiltonian
treatment, including the associated primary constraints and higher-order
kinetic terms, should therefore play an important role in understanding
time-dependent Galilean backgrounds.

Finally, it would be useful to compare the post-linear mixed weak-branch
reconstruction with exact or approximate stationary axisymmetric
multipolar geometries. In particular, the Geroch--Hansen construction of multipole
moments for stationary space-times \cite{Geroch:1970cd,Hansen:1974zz}, the Ernst
formulation \cite{Ernst:1968,Ernst:1968II}, and the Manko--Novikov class
\cite{Manko:1992} provide natural reference points for understanding how
independent multipole moments are embedded in relativistic vacuum
solutions. It would also be interesting to understand whether the
reconstructed mixed weak-branch data can be related, possibly after
suitable coordinate or gauge transformations, to such known rotating
multipolar geometries. In the same direction, one may study
frame-dragging and geodesic effects in analogy with spacetime-mapping,
quasi-Kerr and bumpy-spacetime analyses
\cite{Ryan:1995wh,Collins:2004ex,Glampedakis:2006,Gair:2008}, and compare
the resulting expressions with post-Newtonian expansions. Such a
comparison, however, requires a careful separation between quantities
controlled by the NNLO Galilean ADM data and quantities that depend on
additional higher-order uplift data.

\section{Acknowledgements}
We would like to thank Mehmet Özkan, Oğuzhan Kaşıkçı, Ertuğrul Ekiz and Efe Hamamcı for their useful comments and feedback. U.Z. is supported by the Scientific and Technological Research Council of Türkiye (TÜBİTAK) under Grant Nos. 125F467 and 125F024\,. 
\appendix

\section{The expansion details} \label{expansionapp}

In this appendix we collect the technical details of $\omega$ and large-\(c\)
expansion of the ADM action and field equations.  We use the conventions of
the main text, but repeat some definitions for completeness so that the
appendix can be read independently.  We begin with the expansion of the ADM
variables and then present the action and field equations order by order.

\paragraph{}
We will start with the expansion of the ADM metric 
\begin{equation}
  \hat\gamma_{ij}
  =
  \gamma_{ij}
  +\omega^2\beta_{ij}
  +\omega^4\epsilon_{ij} +O(\omega^6).
  \label{admmetricexp}
\end{equation}
The expansion of the inverse spatial metric is determined by
\[
  \hat\gamma^{ik}\hat\gamma_{kj}=\delta^i{}_j.
\]
Using the ansatz for \(\hat\gamma_{ij}\), one obtains
\begin{equation}
  \hat\gamma^{ij}
  =
  \gamma^{ij}
  -\omega^2\beta^{ij}
  +\omega^4
  \left(
    \beta^i{}_k\beta^{kj}
    -\epsilon^{ij}
  \right)
  +O(\omega^6).
\end{equation}
In the action we have the factor $\hat N\sqrt{\hat\gamma}$, then we need first the expansion of the $\sqrt{\hat\gamma}$
\begin{equation}
  \sqrt{\hat\gamma}
  =
  \sqrt{\gamma}
  \left[
    1
    +\frac{\omega^2}{2}\beta
    +\omega^4
    \left(
      \frac{1}{2}\epsilon
      +\frac{1}{8}\beta^2
      -\frac{1}{4}\beta_{ij}\beta^{ij}
    \right)
  \right]
  +O(\omega^6)\,,
\end{equation}
and also the lapse ansatz
\[
  \hat N=N+\omega^2M+\omega^4P.
\]
Therefore
\begin{equation}
  \hat N\sqrt{\hat\gamma}
  =
  \sqrt{\gamma}
  \left(
    \mathcal{N}^{(0)}
    +\omega^2\mathcal{N}^{(2)}
    +\omega^4\mathcal{N}^{(4)}
  \right)
  +O(\omega^6),
  \label{eq:lapse-measure-expansion}
\end{equation}
where
\begin{equation}
  \begin{aligned}
    \mathcal{N}^{(0)}
    &=
    N, 
    \\
    \mathcal{N}^{(2)}
    &=
    M+\frac{1}{2}N\beta,
    \\
    \mathcal{N}^{(4)}
    &=
    P+\frac{1}{2}M\beta
    +N
    \left(
      \frac{1}{2}\epsilon
      +\frac{1}{8}\beta^2
      -\frac{1}{4}\beta_{ij}\beta^{ij}
    \right).
  \end{aligned}
\end{equation}

\paragraph{Expansion of the Connection:}

Let \(\Gamma^k{}_{ij}\) be the Levi-Civita connection of \(\gamma_{ij}\), and let
\(\hat\Gamma^k{}_{ij}\) be the Levi-Civita connection of \(\hat\gamma_{ij}\).  We write
\begin{equation}
  \hat\Gamma^k{}_{ij}
  =
  \Gamma^k{}_{ij}
  +\omega^2 C^k{}_{ij}
  +\omega^4 D^k{}_{ij}
  +O(\omega^6).
  \label{eq:connection-expansion}
\end{equation}
The first correction is
\begin{equation}
  C^k{}_{ij}
  =
  \frac{1}{2}
  \left(
    \nabla_i\beta^k{}_j
    +\nabla_j\beta^k{}_i
    -\nabla^k\beta_{ij}
  \right),
  \label{eq:c-connection}
\end{equation}
where $\nabla_i$ is the Levi-Civita connection for spatial metric $\gamma_{ij}$\,.  The second correction is
\begin{equation}
  D^k{}_{ij}
  = \frac{1}{2}
  \left(
    \nabla_i\epsilon^k{}_j
    +\nabla_j\epsilon^k{}_i
    -\nabla^k\epsilon_{ij}
  \right) -\frac{1}{2}\beta^{k\ell}
  \left(
    \nabla_i\beta_{j\ell}
    +\nabla_j\beta_{i\ell}
    -\nabla_\ell\beta_{ij}
  \right)\,. 
 \label{eq:d-connection}
\end{equation}

\paragraph{Expansion of the Ricci Tensor:}

The Ricci tensor expands as
\begin{equation}
  \hat R_{ij}
  =
  R_{ij}
  +\omega^2 R^{(2)}_{ij}
  +\omega^4 R^{(4)}_{ij}
  +O(\omega^6). \nn 
\end{equation}
The order-\(\omega^2\) term is
\begin{equation}
  R^{(2)}_{ij}
  =
  \nabla_k C^k{}_{ij}
  -\nabla_j C^k{}_{ik}. \nn 
\end{equation}
Equivalently, in terms of \(\beta_{ij}\),
\begin{equation}
  R^{(2)}_{ij}
  =
  \frac{1}{2}
  \left(
    \nabla_k\nabla_i\beta^k{}_j
    +\nabla_k\nabla_j\beta^k{}_i
    -\nabla^2\beta_{ij}
    -\nabla_i\nabla_j\beta
  \right).
  \label{eq:ricci-tensor-r2-expanded}
\end{equation}
The order-\(\omega^4\) term is
\begin{equation}
  R^{(4)}_{ij}
  =
  \nabla_k D^k{}_{ij}
  -\nabla_j D^k{}_{ik}
  +C^k{}_{ij}C^\ell{}_{k\ell}
  -C^\ell{}_{ik}C^k{}_{j\ell}.\nn
\end{equation}
Using the definition of \(D^k{}_{ij}\) and \(C^k{}_{ij}\), this can be written as
\begin{align}
  R^{(4)}_{ij}
  &=
  \frac12
  \left(
    \nabla_k\nabla_i\epsilon_j{}^k
    +\nabla_k\nabla_j\epsilon_i{}^k
    -\nabla^2\epsilon_{ij}
    -\nabla_i\nabla_j\epsilon
  \right)
  -\frac12\nabla_k
  \left[
    \beta^{k\ell}
    \left(
      \nabla_i\beta_{j\ell}
      +\nabla_j\beta_{i\ell}
      -\nabla_\ell\beta_{ij}
    \right)
  \right]
  \nonumber\\
  &\quad
  +\frac12\nabla_{(i}
  \left(
    \beta^{k\ell}\nabla_{j)}\beta_{k\ell}
  \right)
  +\frac14\nabla_\ell\beta
  \left(
    \nabla_i\beta_j{}^\ell
    +\nabla_j\beta_i{}^\ell
    -\nabla^\ell\beta_{ij}
  \right)
  \nonumber\\
  &\quad
  -\frac18\gamma^{km}\gamma^{\ell n}
  \left(
    \nabla_\ell\beta_{im}
    +\nabla_i\beta_{\ell m}
    -\nabla_m\beta_{\ell i}
  \right)
  \left(
    \nabla_j\beta_{kn}
    +\nabla_k\beta_{jn}
    -\nabla_n\beta_{jk}
  \right)
  \nonumber\\
  &\quad
  -\frac18\gamma^{km}\gamma^{\ell n}
  \left(
    \nabla_\ell\beta_{jm}
    +\nabla_j\beta_{\ell m}
    -\nabla_m\beta_{\ell j}
  \right)
  \left(
    \nabla_i\beta_{kn}
    +\nabla_k\beta_{in}
    -\nabla_n\beta_{ik}
  \right). \label{R4ij}
\end{align}

\paragraph{Expansion of the Ricci Scalar:}

The Ricci scalar is
\[
  \hat R=\hat\gamma^{ij}\hat R_{ij}.
\]
The expansion of $\hat R$ is then 
\begin{equation}
  \hat R
  =
  R
  +\omega^2 R^{(2)}
  +\omega^4 R^{(4)}
  +O(\omega^6).\nn
\end{equation}
The leading term is
\[
  R=\gamma^{ij}R_{ij}.
\]
The order-\(\omega^2\) term is
\begin{equation}
  R^{(2)}
  =
  \gamma^{ij}R^{(2)}_{ij}
  -\beta^{ij}R_{ij}.\nn
\end{equation}
Equivalently,
\begin{equation}
  R^{(2)}
  =
  \nabla_i\nabla_j\beta^{ij}
  -\nabla^2\beta
  -\beta^{ij}R_{ij}.
  \label{eq:ricci-scalar-r2-expanded}
\end{equation}
The order-\(\omega^4\) term is
\begin{equation}
  R^{(4)}
  =
  \gamma^{ij}R^{(4)}_{ij}
  -\beta^{ij}R^{(2)}_{ij}
  +\left(
    \beta^i{}_k\beta^{kj}
    -\epsilon^{ij}
  \right)R_{ij}.\nn
\end{equation}
Explicitly, 
\begin{align}
R^{(4)}
={}&\nabla_i\nabla_j \epsilon^{ij}-\nabla^2 \epsilon-\epsilon^{ij}R_{ij}
\nonumber\\
&+\nabla_k\!\left(
\beta^{ij}\nabla^k\beta_{ij}
-\beta^{ij}\nabla_i\beta_j{}^{k}
+\beta^{ki}\nabla_i\beta
-\beta^{k\ell}\nabla_i\beta^{i}{}_{\ell}
\right)
\nonumber\\
&+\frac12\,(\nabla_i\beta_{jk})(\nabla^j\beta^{ik})
-\frac14\,(\nabla_i\beta_{jk})(\nabla^i\beta^{jk})
-\frac14\,(\nabla_i\beta)(\nabla^i\beta)
+\beta^{i}{}_{k}\beta^{kj}R_{ij}\, , \label{eq:ricci-scalar-r4-expanded}
\end{align}
where we use the shorthand notation $\nabla^2 \equiv \nabla_i\nabla^i$. 

\paragraph{Expansion of \(\mathcal{M}\):}
Starting from the definition,
\bea
\cM &=& \hat N\,\sqrt{\hat \g}\, \hat R\,,
\eea 
we combine the measure expansion \eqref{eq:lapse-measure-expansion} with the Ricci scalar expansion given above, we obtain
\begin{equation}
  {\mathcal{M}}
  =
  \sqrt{\gamma}
  \left[
    \mathcal{M}^{(0)}
    +\omega^2\mathcal{M}^{(2)}
    +\omega^4\mathcal{M}^{(4)}
  \right]
  +O(\omega^6).
  \label{eq:magnetic-block-expansion}
\end{equation}
The three coefficients are
\begin{align}
  \mathcal{M}^{(0)}
  &=
  NR,
  \label{eq:magnetic-block-m0}
  \\
  \mathcal{M}^{(2)}
  &=
  \left(
    M+\frac{1}{2}N\beta
  \right)R
  +N R^{(2)},
  \label{eq:magnetic-block-m2}
  \\
  \mathcal{M}^{(4)}
  &=
  \left[
    P+\frac{1}{2}M\beta
    +N
    \left(
      \frac{1}{2}\epsilon
      +\frac{1}{8}\beta^2
      -\frac{1}{4}\beta_{ij}\beta^{ij}
    \right)
  \right]R
  \nonumber\\
  &\hspace{2em}
  +\left(
    M+\frac{1}{2}N\beta
  \right)R^{(2)}
  +N R^{(4)}.
  \label{eq:magnetic-block-m4}
\end{align}
Here \(R^{(2)}\) is given explicitly by
\eqref{eq:ricci-scalar-r2-expanded}, and \(R^{(4)}\) is given by
\eqref{eq:ricci-scalar-r4-expanded}. 

\paragraph{\texorpdfstring{Expansion of the Electric Block \(\mathcal{E}\):}{Expansion of the Electric Block E}}

The electric ADM block is
\begin{equation}
  {\mathcal{E}}
  :=
  \hat N\sqrt{\hat\gamma}
  \left(
    \hat K_{ij}\hat K^{ij}
    -\hat K^2
  \right),
  \qquad
  \hat K:=\hat\gamma^{ij}\hat K_{ij}.
  \label{eq:electric-block-hat}
\end{equation}
We now expand this through order \(\omega^4\), using the ansatz
\begin{equation}
  \hat K_{ij}
  =
  K_{ij}
  +\omega^2L_{ij}
  +\omega^4F_{ij}.
  \label{eq:electric-k-ansatz-repeat}
\end{equation}
We also define the traces
\begin{equation}
  K:=\gamma^{ij}K_{ij},
  \qquad
  L:=\gamma^{ij}L_{ij},
  \qquad
  F:=\gamma^{ij}F_{ij}.
  \label{eq:k-l-f-traces}
\end{equation}
The coefficients are
\begin{align}
  K_{ij}
  &=
  N^{-1}
  \left(
    \frac{1}{2}\partial_t\gamma_{ij}
    -\nabla_{(i}N_{j)}
  \right),
  \label{eq:omega-ansatz-kapp}
  \\
  L_{ij}
  &=
  N^{-1}
  \left(
    \frac{1}{2}\partial_t\beta_{ij}
    -\nabla_{(i}A_{j)}
    -\beta_{k(i}\nabla_{j)}N^k
    -\frac{1}{2}N^k\nabla_k\beta_{ij}
    -M K_{ij}
  \right),
  \label{eq:omega-ansatz-lapp}
  \\
  F_{ij}
  &=
  N^{-1}
  \left[
    \frac{1}{2}\partial_t\epsilon_{ij}
    -\nabla_{(i}Z_{j)}
    -\beta_{k(j}\nabla_{i)}A^k
    -\epsilon_{k(j}\nabla_{i)}N^k
  \right.
  \nonumber\\
  &\hspace{3.6em}\left.
    -\frac{1}{2}N^k\nabla_k\epsilon_{ij}
    -\frac{1}{2}A^k\nabla_k\beta_{ij}
    -\left(L_{ij}M+K_{ij}P\right)
  \right].
  \label{eq:omega-ansatz-fapp}
\end{align}

\paragraph{\texorpdfstring{Expansion of the Trace \(\hat K\):}{Expansion of the Trace K}}

Using
\[
  \hat\gamma^{ij}
  =
  \gamma^{ij}
  -\omega^2\beta^{ij}
  +\omega^4
  \left(
    \beta^i{}_k\beta^{kj}
    -\epsilon^{ij}
  \right)
  +O(\omega^6),
\]
we obtain
\begin{equation}
  \hat K
  =
  K
  +\omega^2K^{(2)}
  +\omega^4K^{(4)}
  +O(\omega^6),
  \label{eq:k-trace-expansion}
\end{equation}
where
\begin{align}
  K^{(2)}
  &=
  L-\beta^{ij}K_{ij},
  \label{eq:k-trace-order2}
  \\
  K^{(4)}
  &=
  F-\beta^{ij}L_{ij}
  +\left(
    \beta^i{}_k\beta^{kj}
    -\epsilon^{ij}
  \right)K_{ij}.
  \label{eq:k-trace-order4}
\end{align}
Therefore
\begin{equation}
  \hat K^2
  =
  K^2
  +\omega^2\left(2KK^{(2)}\right)
  +\omega^4
  \left[
    \left(K^{(2)}\right)^2
    +2KK^{(4)}
  \right]
  +O(\omega^6).
  \label{eq:k-trace-square-expansion}
\end{equation}

\paragraph{\texorpdfstring{Expansion of \(\hat K_{ij}\hat K^{ij}\):}{Expansion of Kij Kij}}

Next expand
\[
  \hat K_{ij}\hat K^{ij}
  =
  \hat\gamma^{ik}\hat\gamma^{j\ell}
  \hat K_{ij}\hat K_{k\ell}.
\]
We write
\begin{equation}
  \hat K_{ij}\hat K^{ij}
  =
  S^{(0)}
  +\omega^2S^{(2)}
  +\omega^4S^{(4)}
  +O(\omega^6).
  \label{eq:k-norm-expansion}
\end{equation}
The leading term is
\begin{equation}
  S^{(0)}
  =
  K_{ij}K^{ij}.
  \label{eq:k-norm-s0}
\end{equation}
The order-\(\omega^2\) term 
\begin{equation}
  S^{(2)}
  =
  2K^{ij}L_{ij}
  -2\beta^{ik}K_i{}^jK_{kj}.
  \label{eq:k-norm-s2}
\end{equation}
Finally, the order-\(\omega^4\) term 
\begin{align}
  S^{(4)}
  &=
  2K^{ij}F_{ij}
  +L_{ij}L^{ij}
  -4\beta^{ik}K_i{}^jL_{kj}
  \nonumber\\
  &\quad
  +2
  \left(
    \beta^i{}_m\beta^{mk}
    -\epsilon^{ik}
  \right)
  K_i{}^jK_{kj}
  +\beta^{ik}\beta^{j\ell}K_{ij}K_{k\ell}.
  \label{eq:k-norm-s4}
\end{align}

\paragraph{\texorpdfstring{Expansion of the Kinetic Combination:}{Expansion of the Kinetic Combination}}

Define 
\begin{equation}
  \hat{\mathcal{Q}}
  :=
  \hat K_{ij}\hat K^{ij}-\hat K^2.
  \label{eq:qhat-definition}
\end{equation}
Then
\begin{equation}
  \hat{\mathcal{Q}}
  =
  \mathcal{Q}^{(0)}
  +\omega^2\mathcal{Q}^{(2)}
  +\omega^4\mathcal{Q}^{(4)}
  +O(\omega^6)\,,
  \label{eq:qhat-expansion}
\end{equation}
where the explicit form of the terms are
\begin{align}
 \mathcal{Q}^{(0)}
  &=
  K_{ij}K^{ij}
  -
  K^2,
  \nonumber\\
  \mathcal{Q}^{(2)}
  &=
  2K^{ij}L_{ij}
  -
  2KL
  -
  2K_i{}^kK^{ij}\beta_{jk}
  +
  2KK^{ij}\beta_{ij},
  \nonumber\\
  \mathcal{Q}^{(4)}
  &=
  2
  \left(
    F^{ij}K_{ij}
    -
    FK
  \right)
  +
  \left(
    L_{ij}L^{ij}
    -
    L^2
  \right)
  \nonumber\\
  &\quad
  -
  2
  \left(
    K_i{}^kK^{ij}
    -
    KK^{jk}
  \right)
  \epsilon_{jk}
  -
  2
  \left(
    2K^{ij}L_i{}^k
    -
    LK^{jk}
    -
    KL^{jk}
  \right)
  \beta_{jk}
  \nonumber\\
  &\quad
  -
  \left(
    K^{ij}K^{kl}\beta_{ij}\beta_{kl}
    -
    K^{ij}K^{kl}\beta_{ik}\beta_{jl}
  \right)
  +
  2
  \left(
    K_i{}^kK^{ij}
    -
    KK^{jk}
  \right)
  \beta_j{}^\ell\beta_{k\ell}.
  \label{eq:q-coefficients}
\end{align}
Using the measure expansion derived in \eqref{eq:lapse-measure-expansion} and multiplying this by the expansion of \(\hat{\mathcal{Q}}\), we find
\begin{equation}
  \hat{\mathcal{E}}
  =
  \sqrt{\gamma}
  \left[
    \mathcal{E}^{(0)}
    +\omega^2\mathcal{E}^{(2)}
    +\omega^4\mathcal{E}^{(4)}
  \right]
  +O(\omega^6).
  \label{eq:electric-block-expansion}
\end{equation}
The coefficients are
\begin{equation}
  \hat{\mathcal{E}}
  =
  \sqrt{\gamma}
  \left[
    \mathcal{E}^{(0)}
    +\omega^2\mathcal{E}^{(2)}
    +\omega^4\mathcal{E}^{(4)}
  \right]
  +O(\omega^6).
\end{equation}
The coefficients are
\begin{align}
  \mathcal{E}^{(0)}
  &=
  N\mathcal{Q}^{(0)},
  \nonumber\\
  \mathcal{E}^{(2)}
  &=
  \left(
    M+\frac{1}{2}N\beta
  \right)\mathcal{Q}^{(0)}
  +N\mathcal{Q}^{(2)},
  \nonumber\\
  \mathcal{E}^{(4)}
  &=
  \left[
    P+\frac{1}{2}M\beta
    +N
    \left(
      \frac{1}{2}\epsilon
      +\frac{1}{8}\beta^2
      -\frac{1}{4}\beta_{ij}\beta^{ij}
    \right)
  \right]\mathcal{Q}^{(0)}
  \nonumber\\
  &\quad
  +\left(
    M+\frac{1}{2}N\beta
  \right)\mathcal{Q}^{(2)}
  +N\mathcal{Q}^{(4)}.
  \label{eq:electric-coefficients}
\end{align}
This completes the electric block expansion through order \(\omega^4\).  The full ADM
Lagrangian density can now be organized as
\[
  \hat N\sqrt{\hat\gamma}
  \left[
    \hat R
    +\omega^\alpha
    \left(
      \hat K_{ij}\hat K^{ij}
      -\hat K^2
    \right)
  \right]
  =
  \hat{\mathcal{M}}+\omega^\alpha\hat{\mathcal{E}},
\]
with \(\hat{\mathcal{M}}\) and \(\hat{\mathcal{E}}\) both expanded to the same order.  The extra \(\omega^\alpha\) multiplying \(\hat{\mathcal{E}}\) is the explicit scaling that measures the distinction between large- and small-$c$ expansion. 
\section{Field equations} \label{eomexpapp}
In this appendix, we collect the field equations and their expansion in powers of small $\omega$. We begin with the relativistic field equations. Varying the action \eqref{ADMAction} with respect to the ADM fields yields
\bea
\delta\, S_{ADM}=  \int\, \sqrt{\hat \gamma}\, d^{d+1}\,x\,\left(  \delta \hat N\, E_{0} + \delta\, \hat N^i\, E_i + \delta\, \hat \gamma^{ij}\, E_{ij}\right)\,, \nn
\eea
where the field equations are explicitly given by
\bea
E_0&:& \left( \hat R - \hat K_{ij}\, \hat K^{ij} +  \, \hat K^2 \,\, \right)\,\,,\nn \\ 
E_i\, &:&\,\hat \nabla^{j}\left( \hat K_{ij} - \hat \gamma_{ij}\, \hat K\right)\,,\nn \\
E_{ij} &:& \hat N\,\hat G_{ij}\,  -\hat \nabla_{i}\,\hat \nabla_{j}\hat N +\hat \g_{ij}\, \hat \nabla^2\, \hat N -2\, \hat K_{k(i}\hat \nabla_{j)}\hat N^k\, - \hat N^k \hat \nabla_k\, \hat K_{ij}\, \nn \\
&&+  \hat N\left(-2\, \hat K_{i}{}^{k}\, \hat K_{jk}\, +\, \hat K\,\hat K_{ij}\, +\, \frac12\left(3\, \hat K_{kl}\, \hat K^{kl}-\hat K^2\right)\, \hat \gamma_{ij}\right) \, \nn \\
&&+2\, \hat \gamma_{ij}\, \hat K_{kl} \hat \nabla^l \hat N^k
+c^{-1}\,\partial_t \hat K_{ij}
- c^{-1}\,\hat \gamma_{ij} \hat \gamma^{kl}\partial_t \hat K_{kl}
+\,\hat \gamma_{ij}\hat N^k \hat \nabla_k \hat K\,. 
\label{relADMEOM}
\eea
There are several equivalent ways to present the field equations of the ADM action, especially in relation to the constraint and evolution equations in the Hamiltonian formulation. In this work,  we adopt the constraint and Hamilton's field equations. These equations are related to each other and the equality is shown in the following way. Starting from the constraints, the lapse and shift variations give
\begin{align}
  \boxed{
  \hat R
  -
  \hat K_{ij}\hat K^{ij}
  +
  \hat K^2
  =
  0,
  }
  \label{eq:hamiltonian-constraint}
  \\
  \boxed{
  \hat\nabla^j
  \left(
    \hat K_{ij}
    -
    \hat\gamma_{ij}\hat K
  \right)
  =
  0.
  }
  \label{eq:momentum-constraint}
\end{align}
Now, take the trace-reversed spatial equation 
\begin{equation}
  E_{ij}
  -
  \frac12
  \hat\gamma_{ij}\hat\gamma^{kl}E_{kl}
  =
  0.
  \label{eq:trace-reversed-Eij}
\end{equation}
We set \(c=1\). And in three spatial dimensions, direct substitution of $E_{ij} $ in \eqref{relADMEOM} gives
\begin{align}
  \partial_t\hat K_{ij}
  &=
  \hat\nabla_i\hat\nabla_j\hat N
  -
  \hat N\hat R_{ij}
  -
  \hat N\hat K\hat K_{ij}
  +
  2\hat N\hat K_i{}^k\hat K_{jk}
  \nonumber\\
  &\quad
  +
  \hat N^k\hat\nabla_k\hat K_{ij}
  +
  2\hat K_{k(i}\hat\nabla_{j)}\hat N^k
  +
  \frac14
  \hat N\hat\gamma_{ij}
  \left(
    \hat R
    -
    \hat K_{kl}\hat K^{kl}
    +
    \hat K^2
  \right).
  \label{eq:Kij-before-E0}
\end{align}
Using the Hamiltonian constraint \(E_0=0\), the last line in parentheses vanishes.  Therefore
\begin{equation}
  \boxed{
  \partial_t\hat K_{ij}
  =
  \hat\nabla_i\hat\nabla_j\hat N
  -
  \hat N
  \left(
    \hat R_{ij}
    +
    \hat K\hat K_{ij}
    -
    2\hat K_i{}^k\hat K_{jk}
  \right)
  +
  \hat N^k\hat\nabla_k\hat K_{ij}
  +
  2\hat K_{k(i}\hat\nabla_{j)}\hat N^k.
  }
  \label{eq:Kij-evolution-expanded}
\end{equation}
Equivalently,
\begin{equation}
  \boxed{
  \left(
    \partial_t
    -
    \mathcal L_{\hat N}
  \right)
  \hat K_{ij}
  =
  \hat\nabla_i\hat\nabla_j\hat N
  -
  \hat N
  \left(
    \hat R_{ij}
    +
    \hat K\hat K_{ij}
    -
    2\hat K_i{}^k\hat K_{jk}
  \right).
  }
  \label{eq:Kij-evolution-Lie}
\end{equation}
The other Hamiltonian evolution equation is the defining ADM relation
\begin{equation}
  \boxed{
  \partial_t\hat\gamma_{ij}
  =
  2\hat N\hat K_{ij}
  +
  2\hat\nabla_{(i}\hat N_{j)}.
  }
  \label{eq:gamma-evolution}
\end{equation}
Finally the set of equations are given by (\(c=1\))
\begin{align}
  &\hat R
  -
  \hat K_{ij}\hat K^{ij}
  +
  \hat K^2
  =
  0,
  \\
  &\hat\nabla^j
  \left(
    \hat K_{ij}
    -
    \hat\gamma_{ij}\hat K
  \right)
  =
  0,
  \\
  &\partial_t\hat\gamma_{ij} -  2\hat N\hat K_{ij}
  -
  2\hat\nabla_{(i}\hat N_{j)}
  = 0\,,
  \\
  &\partial_t\hat K_{ij} -  \hat\nabla_i\hat\nabla_j\hat N
  + 
  \hat N
  \left(
    \hat R_{ij}
    +
    \hat K\hat K_{ij}
    -
    2\hat K_i{}^k\hat K_{jk}
  \right) -
  \hat N^k\hat\nabla_k\hat K_{ij}
  -
  2\hat K_{k(i}\hat\nabla_{j)}\hat N^k
  =0\,.
\end{align}
The Hamiltonian constraint is split as
\[
  \hat{\mathcal{H}}
  =
  \hat{\mathcal{H}}^{\rm M}
  +\omega^\alpha \hat{\mathcal{H}}^{\rm E}
  =
  0,
\]
where
\[
  \hat{\mathcal{H}}^{\rm M}
  :=
  \hat R,
  \qquad
  \hat{\mathcal{H}}^{\rm E}
  :=
  \left(\hat K^2 - 
    \hat K_{ij}\hat K^{ij}
  \right)
  = - \hat{\mathcal{Q}}.
\]
The magnetic and electric sectors are expanded independently:
\begin{align}
  \hat{\mathcal{H}}^{\rm M}
  &=
  \overset{(0)}{\mathcal{H}}{}^{\rm M}
  +\omega^2\overset{(2)}{\mathcal{H}}{}^{\rm M}
  +\omega^4\overset{(4)}{\mathcal{H}}{}^{\rm M}
  +O(\omega^6),
  \\
  \hat{\mathcal{H}}^{\rm E}
  &=
  \overset{(0)}{\mathcal{H}}{}^{\rm E}
  +\omega^2\overset{(2)}{\mathcal{H}}{}^{\rm E}
  +\omega^4\overset{(4)}{\mathcal{H}}{}^{\rm E}
  +O(\omega^6).
\end{align}
Using the curvature and extrinsic-curvature expansions already derived,
\begin{align}
  \overset{(0)}{\mathcal{H}}{}^{\rm M}
  &:=
  R,
  &
  \overset{(2)}{\mathcal{H}}{}^{\rm M}
  &:=
  R^{(2)},
  &
  \overset{(4)}{\mathcal{H}}{}^{\rm M}
  &:=
  R^{(4)},
  \\
  \overset{(0)}{\mathcal{H}}{}^{\rm E}
  &:=
  -\mathcal{Q}^{(0)},
  &
  \overset{(2)}{\mathcal{H}}{}^{\rm E}
  &:=
  -\mathcal{Q}^{(2)},
  &
  \overset{(4)}{\mathcal{H}}{}^{\rm E}
  &:=
  -\mathcal{Q}^{(4)}.
\end{align}
The momentum constraint is
\[
\mathcal{H}_i
  :=
  \hat\nabla_j
  \left(
    \hat K^j{}_{i}
    -\delta^j_i\hat K
  \right)
  =
  0.
\]
Define
\[
  \hat\Pi^j{}_i
  :=
  \hat K^j{}_i-\delta^j_i\hat K.
\]
We expand
\[
  \hat\Pi^j{}_i
  =
  \Pi^{j(0)}{}_i
  +\omega^2\Pi^{j(2)}{}_i
  +\omega^4\Pi^{j(4)}{}_i
  +O(\omega^6),
\]
where
\begin{align}
  \Pi^{j(0)}{}_i
  &:=
  K^j{}_i-\delta^j_iK,
  \\
  \Pi^{j(2)}{}_i
  &:=
  L^j{}_i-\beta^{jk}K_{ki}
  -\delta^j_i
  \left(
    L-\beta^{kl}K_{kl}
  \right),
  \\
  \Pi^{j(4)}{}_i
  &:=
  F^j{}_i-\beta^{jk}L_{ki}
  +\left(
    \beta^j{}_m\beta^{mk}
    -\epsilon^{jk}
  \right)K_{ki}
  \nonumber\\
  &\quad
  -\delta^j_i
  \left[
    F-\beta^{kl}L_{kl}
    +\left(
      \beta^k{}_m\beta^{ml}
      -\epsilon^{kl}
    \right)K_{kl}
  \right].
\end{align}
Therefore
\begin{equation}
  \mathcal{H}_i
  =
  \mathcal{H}^{(0)}_i
  +\omega^2\mathcal{H}^{(2)}_i
  +\omega^4\mathcal{H}^{(4)}_i
  +O(\omega^6),
\end{equation}
with
\begin{align}
  \mathcal{H}^{(0)}_i
  &:=
  \nabla_j\Pi^{j(0)}{}_i,
  \\
  \mathcal{H}^{(2)}_i
  &:=
  \nabla_j\Pi^{j(2)}{}_i
  +C^j{}_{jm}\Pi^{m(0)}{}_i
  -C^m{}_{ji}\Pi^{j(0)}{}_m,
  \\
  \mathcal{H}^{(4)}_i
  &:=
  \nabla_j\Pi^{j(4)}{}_i
  +C^j{}_{jm}\Pi^{m(2)}{}_i
  -C^m{}_{ji}\Pi^{j(2)}{}_m
  \nonumber\\
  &\quad
  +D^j{}_{jm}\Pi^{m(0)}{}_i
  -D^m{}_{ji}\Pi^{j(0)}{}_m .
\end{align}
Using the connection coefficients
\eqref{eq:c-connection}--\eqref{eq:d-connection}, the explicit form of the
momentum constraints is
\begin{align}
 \mathcal{H}^{(0)}_i
  &=
  \nabla_j K_i{}^j-\nabla_i K,
  \nonumber\\[4pt]
 \mathcal{H}^{(2)}_i
  &=
  \nabla_j L_i{}^j-\nabla_i L
  +\beta^{jk}\left(\nabla_i K_{jk}-\nabla_k K_{ij}\right)
  +\frac12 K^{jk}\nabla_i\beta_{jk}
  \nonumber\\
  &\quad
  +\frac12 K_i{}^j\nabla_j\beta
  -K_i{}^j\nabla_k\beta_j{}^k,
  \nonumber\\[4pt]
  \mathcal{H}^{(4)}_i
  &=
  \nabla_j F_i{}^j-\nabla_i F
  +\epsilon^{jk}\left(\nabla_i K_{jk}-\nabla_k K_{ij}\right)
  -\beta_j{}^l\beta^{jk}\nabla_i K_{kl}
  +\beta_j{}^l\beta^{jk}\nabla_l K_{ik}
  \nonumber\\
  &\quad
  +\frac12 K^{jk}\nabla_i\epsilon_{jk}
  +\frac12 L^{jk}\nabla_i\beta_{jk}
  +\beta^{jk}\left(\nabla_i L_{jk}-\nabla_k L_{ij}\right)
  +\frac12 K_i{}^j\nabla_j\epsilon
  +\frac12 L_i{}^j\nabla_j\beta
  \nonumber\\
  &\quad
  -K_i{}^j\nabla_k\epsilon_j{}^k
  -L_i{}^j\nabla_k\beta_j{}^k
  -K^{jk}\beta_j{}^l\nabla_i\beta_{kl}
  -\frac12 K_i{}^j\beta^{kl}\nabla_j\beta_{kl}
  -\frac12 K_i{}^j\beta_j{}^k\nabla_k\beta
  \nonumber\\
  &\quad
  +K_i{}^j\beta^{kl}\nabla_l\beta_{jk}
  +K_i{}^j\beta_j{}^k\nabla_l\beta_k{}^l .
\end{align}
Therefore, the momentum equations through order \(\omega^4\) are 
\[
  \boxed{
  \mathcal{H}^{(0)}_i=0,
  \qquad
  \mathcal{H}^{(2)}_i=0,
  \qquad
  \mathcal{H}^{(4)}_i=0.
  }
\]
The defining ADM relation is
\[
  \partial_t\hat\gamma_{ij}
  =
  2\hat N\hat K_{ij}
  +\hat\nabla_i\hat N_j
  +\hat\nabla_j\hat N_i\,. 
\]
Equivalently, we can define 
\[
  \hat{\mathcal{B}}_{ij}
  :=
  \partial_t\hat\gamma_{ij}
  -2\hat N\hat K_{ij}
  -2\hat\nabla_{(i}\hat N_{j)}
  =
  0.
\]
From the expansion, we get
\[
  \hat{\mathcal{B}}_{ij}
  =
  \mathcal{B}^{(0)}_{ij}
  +\omega^2\mathcal{B}^{(2)}_{ij}
  +\omega^4\mathcal{B}^{(4)}_{ij}
  +O(\omega^6).
\]
The first three equations are
\begin{align}
  \mathcal{B}^{(0)}_{ij}
  &:=
  \partial_t\gamma_{ij}
  -2NK_{ij}
  -2\nabla_{(i}N_{j)}
  =
  0,
  \\
  \mathcal{B}^{(2)}_{ij}
  &:=
  \partial_t\beta_{ij}
  -2NL_{ij}
  -2MK_{ij}
  -2\nabla_{(i}A_{j)}
  \nonumber\\
  &\quad
  -2\beta_{k(i}\nabla_{j)}N^k
  -N^k\nabla_k\beta_{ij}
  =
  0,
  \\
  \mathcal{B}^{(4)}_{ij}
  &:=
  \partial_t\epsilon_{ij}
  -2NF_{ij}
  -2ML_{ij}
  -2PK_{ij}
  -2\nabla_{(i}Z_{j)}
  \nonumber\\
  &\quad
  -2\beta_{k(j}\nabla_{i)}A^k
  -2\epsilon_{k(j}\nabla_{i)}N^k
  -N^k\nabla_k\epsilon_{ij}
  -A^k\nabla_k\beta_{ij}
  =
  0.
\end{align}
These equations are equivalent to the definitions of \(K_{ij}\), \(L_{ij}\), and
\(F_{ij}\) given earlier. Therefore, Galilean and Carrollian equations are 
\bea
\overset{(0)}{\mathcal{\mathcal{B}}}{}^{G}_{ij} =\overset{(0)}{\mathcal{\mathcal{B}}}{}^{C}_{ij}  &=&  \mathcal{B}^{(0)}_{ij} \,, \nn \\
\overset{(2)}{\mathcal{\mathcal{B}}}{}^{G}_{ij} = \overset{(2)}{\mathcal{\mathcal{B}}}{}^{C}_{ij}   &=& \mathcal{B}^{(2)}_{ij}\,, \nn \\
\overset{(4)}{\mathcal{\mathcal{B}}}{}^{G}_{ij} = \overset{(4)}{\mathcal{\mathcal{B}}}{}^{C}_{ij} &=&  \mathcal{B}^{(4)}_{ij} \,.
\eea 
For the \(K_{ij}\) evolution equation we use the split
\[
  \hat{\mathcal{A}}_{ij}
  =
  \hat{\mathcal{A}}^{\rm M}_{ij}
  +\omega^{\alpha}\hat{\mathcal{A}}^{\rm E}_{ij}
  =
  0,
\]
with
\begin{align}
  \hat{\mathcal{A}}^{\rm M}_{ij}
  &:=\hat N\hat R_{ij} - \hat\nabla_i\hat\nabla_j\hat N,\\
  \hat{\mathcal{A}}^{\rm E}_{ij}
  &:=\mathcal{D}_t\hat K_{ij} + \hat N
  \left(
    \hat K\hat K_{ij}
    -2\hat K_i{}^k\hat K_{jk}
  \right)\,,
\end{align}
where we define 
\bea
\mathcal{D}_t\hat K_{ij} = \left( \partial_t \hat K_{ij}  
-  \mathcal{L}_{\vec{N}}\,\hat K_{ij} \right)\,. 
\eea 
The magnetic block has the expansion
\begin{align}
  \hat{\mathcal{A}}^{\rm M}_{ij}
  &=
  \overset{(0)}{\mathcal{A}}{}^{\rm M}_{ij}
  +\omega^2\overset{(2)}{\mathcal{A}}{}^{\rm M}_{ij}
  +\omega^4\overset{(4)}{\mathcal{A}}{}^{\rm M}_{ij}
  +O(\omega^6),
\end{align}
where
\begin{align}
  \overset{(0)}{\mathcal{A}}{}^{\rm M}_{ij}
  &=NR_{ij} - \nabla_i\nabla_jN\, \nn 
  \\
  \overset{(2)}{\mathcal{A}}{}^{\rm M}_{ij}
  &= M R_{ij} - \nabla_i\nabla_jM
  \, \nn \\
  & + \frac12
  \left(
    \nabla_i\beta_j{}^k
    +\nabla_j\beta_i{}^k
    -\nabla^k\beta_{ij}
  \right)\nabla_kN
  \nonumber\\
  &\quad
   +\frac{N}{2}
  \left(
    \nabla_k\nabla_i\beta_j{}^k
    +\nabla_k\nabla_j\beta_i{}^k
    -\nabla^2\beta_{ij}
    -\nabla_i\nabla_j\beta
  \right). \nn 
  \\
  \overset{(4)}{\mathcal{A}}{}^{\rm M}_{ij}
  &= P R_{ij} - \nabla_i\nabla_jP\,  + N R^{(4)}_{ij}\,
  \nonumber\\
  &\quad
  + \frac12
  \left(
    \nabla_i\beta_j{}^k
    +\nabla_j\beta_i{}^k
    -\nabla^k\beta_{ij}
  \right)\nabla_kM
  \nonumber\\
  &\quad
  + \frac12
  \left(
    \nabla_i\epsilon_j{}^k
    +\nabla_j\epsilon_i{}^k
    -\nabla^k\epsilon_{ij}
  \right)\nabla_kN
  \nonumber\\
  &\quad
  - \frac12\beta^{k\ell}
  \left(
    \nabla_i\beta_{j\ell}
    +\nabla_j\beta_{i\ell}
    -\nabla_\ell\beta_{ij}
  \right)\nabla_kN
  \nonumber\\
  &\quad
  + \frac{M}{2}
  \left(
    \nabla_k\nabla_i\beta_j{}^k
    +\nabla_k\nabla_j\beta_i{}^k
    -\nabla^2\beta_{ij}
    -\nabla_i\nabla_j\beta
  \right)\, \label{Kmagevolexpeom}
\end{align}
where $R^{(4)}_{ij}$ is given in Eqn. \eqref{R4ij}\,. 

In order to expand the electric term, we need to define some shorthand notation. Hence, for any covariant two-tensor \(X_{ij}\), define
\[
  \mathcal D_NX_{ij}
  :=
  \partial_tX_{ij}-\mathcal L_NX_{ij},
\]
where
\[
  \mathcal L_NX_{ij}
  =
  N^k\nabla_kX_{ij}
  +X_{ik}\nabla_jN^k
  +X_{kj}\nabla_iN^k.
\]
Also define the order-by-order shift corrections 
\[
  \Delta_AX_{ij}
  :=
  A^k\nabla_kX_{ij}
  +X_{ik}\nabla_jA^k
  +X_{kj}\nabla_iA^k,
\]
\[
  \Delta_ZX_{ij}
  :=
  Z^k\nabla_kX_{ij}
  +X_{ik}\nabla_jZ^k
  +X_{kj}\nabla_iZ^k.
\]
Then
\[
  D_t\hat K_{ij}
  =
  \mathcal D_NK_{ij}
  +\omega^2\left(\mathcal D_NL_{ij}-\Delta_AK_{ij}\right)
  +\omega^4\left(\mathcal D_NF_{ij}
  -\Delta_AL_{ij}-\Delta_ZK_{ij}\right)
  +O(\omega^6).
\]
Therefore, the expansion of the electric part is
\begin{align}
  \hat{\mathcal{A}}^{\rm E}_{ij}
  &=
  \overset{(0)}{\mathcal{A}}{}^{\rm E}_{ij}
  +\omega^2\overset{(2)}{\mathcal{A}}{}^{\rm E}_{ij}
  +\omega^4\overset{(4)}{\mathcal{A}}{}^{\rm E}_{ij}
  +O(\omega^6),
\end{align}
where
\begin{align}
  \overset{(0)}{\mathcal{A}}{}^{\rm E}_{ij}
  &= \mathcal D_NK_{ij} + N\left(KK_{ij}-2K_i{}^kK_{jk}\right). \nn \\
  \overset{(2)}{\mathcal{A}}{}^{\rm E}_{ij}
  &=
\mathcal D_NL_{ij}-\Delta_AK_{ij} + M\left(KK_{ij}-2K_i{}^kK_{jk}\right)
\nonumber\\
& + N\Big[
(L-\beta^{k\ell}K_{k\ell})K_{ij}
+KL_{ij}
-2L_i{}^kK_{jk}
-2K_i{}^kL_{jk}
+2\beta^{k\ell}K_{i\ell}K_{jk}
\Big]. \nn \\
  \overset{(4)}{\mathcal{A}}{}^{\rm E}_{ij}
  &= \mathcal D_NF_{ij}
-\Delta_AL_{ij}
-\Delta_ZK_{ij}
 + P\left(KK_{ij}-2K_i{}^kK_{jk}\right)
\nonumber\\
& + M\Big[
(L-\beta^{k\ell}K_{k\ell})K_{ij}
+KL_{ij}
-2L_i{}^kK_{jk}
-2K_i{}^kL_{jk}
+2\beta^{k\ell}K_{i\ell}K_{jk}
\Big]
\nonumber\\
& + N\Big[
( F-\beta^{k\ell}L_{k\ell}
  +B^{k\ell}K_{k\ell})K_{ij}
+(L-\beta^{k\ell}K_{k\ell})L_{ij}
+KF_{ij} 
-2F_i{}^kK_{jk} \nonumber\\
&
-2L_i{}^kL_{jk}
-2K_i{}^kF_{jk}
+2\beta^{k\ell}
\left(L_{i\ell}K_{jk}+K_{i\ell}L_{jk}\right)
-2 (\beta^k{}_m\beta^{ml}-\epsilon^{kl})K_{i\ell}K_{jk}
\Big]. \label{Kelevalexeom}
\end{align}
Having completed the expansion of the action and Hamilton's field equations, we now collect the Galilean field equations up to NNLO.

\subsection{Galilean field equations}

We list the Galilean field equations in this subsection for convenience. These equations are obtained by choosing \(\alpha=2\) in the unified \(\omega\)-expansion and collecting the coefficients order by order. They are the equations used in the construction of the stationary vacuum solutions in the main text. The fully expanded expressions for the individual magnetic and electric blocks were given above; here we only summarize their organization into LO, NLO and NNLO Galilean theories.

\subsubsection{Galilean LO gravity equations}
\bea
\overset{(0)}{\mathcal{H}}{}^{\rm Gal}
&:=&
\overset{(0)}{\mathcal{H}}{}^{\rm M}
=
0\,,
\nn\\
\overset{(0)}{\mathcal{H}}{}^{\rm Gal}_{i}
&:=&
0\,,
\qquad
\text{(no vector)}
\nn\\
\overset{(0)}{\mathcal{B}}{}^{\rm Gal}_{ij}
&:=&
\mathcal{B}^{(0)}_{ij}
=
0\,,
\nn\\
\overset{(0)}{\mathcal{A}}{}^{\rm Gal}_{ij}
&:=&
\overset{(0)}{\mathcal{A}}{}^{\rm M}_{ij}
=
0\,.
\label{LOeom}
\eea

\subsubsection{Galilean NLO gravity equations}
\bea
\overset{(2)}{\mathcal{H}}{}^{\rm Gal}
&:=&
\overset{(2)}{\mathcal{H}}{}^{\rm M}
+
\overset{(0)}{\mathcal{H}}{}^{\rm E}
=
0\,,
\nn\\
\overset{(2)}{\mathcal{H}}{}^{\rm Gal}_{i}
&:=&
\mathcal{H}^{(0)}_{i}
=
0\,,
\nn\\
\overset{(2)}{\mathcal{B}}{}^{\rm Gal}_{ij}
&:=&
\mathcal{B}^{(2)}_{ij}
=
0\,,
\nn\\
\overset{(2)}{\mathcal{A}}{}^{\rm Gal}_{ij}
&:=&
\overset{(2)}{\mathcal{A}}{}^{\rm M}_{ij}
+
\overset{(0)}{\mathcal{A}}{}^{\rm E}_{ij}
=
0\,.
\label{NLOeom}
\eea

\subsubsection{Galilean NNLO gravity equations}
\bea
\overset{(4)}{\mathcal{H}}{}^{\rm Gal}
&:=&
\overset{(4)}{\mathcal{H}}{}^{\rm M}
+
\overset{(2)}{\mathcal{H}}{}^{\rm E}
=
0\,,
\nn\\
\overset{(4)}{\mathcal{H}}{}^{\rm Gal}_{i}
&:=&
\mathcal{H}^{(2)}_{i}
=
0\,,
\nn\\
\overset{(4)}{\mathcal{B}}{}^{\rm Gal}_{ij}
&:=&
\mathcal{B}^{(4)}_{ij}
=
0\,,
\nn\\
\overset{(4)}{\mathcal{A}}{}^{\rm Gal}_{ij}
&:=&
\overset{(4)}{\mathcal{A}}{}^{\rm M}_{ij}
+
\overset{(2)}{\mathcal{A}}{}^{\rm E}_{ij}
=
0\,.
\label{NNLOeom}
\eea

\subsection{Stationary Galilean field equations}
\label{stationaryredeom}

In this subsection we collect the stationary form of the Galilean field equations used in the solution analysis. These equations are obtained from the general blocks above by setting all explicit time derivatives to zero. The ADM relations \(\mathcal B^{(n)}_{ij}=0\) are used only to define \(K_{ij}\), \(L_{ij}\), and \(F_{ij}\). All expressions in this subsection should be understood in the stationary sense, namely after setting all explicit time-derivative terms to zero.

\paragraph{LO:}
\begin{align}
\overset{(0)}{\mathcal H}{}^{\rm Gal}
&=
R
=
0,
\\
\overset{(0)}{\mathcal A}{}^{\rm Gal}_{ij}
&=
N R_{ij}
-
\nabla_i\nabla_j N
=
0. \label{LOstationaryeom}
\end{align}

\paragraph{NLO:}
\begin{align}
\overset{(2)}{\mathcal H}{}^{\rm Gal}
&=
\nabla_i\nabla_j\beta^{ij}
  -\nabla^2\beta
  -\beta^{ij}R_{ij}  + K^2
- K_{ij}K^{ij}
=
0,
\\
\overset{(2)}{\mathcal H}{}^{\rm Gal}_{i}
&=
\left(\nabla_j K_i{}^j-\nabla_i K\right)
=
0,
\label{eq:stat-nlo-mom}
\\
\overset{(2)}{\mathcal A}{}^{\rm Gal}_{ij}
&=
M R_{ij} - \nabla_i\nabla_jM
  \,  + \frac12
  \left(
    \nabla_i\beta_j{}^k
    +\nabla_j\beta_i{}^k
    -\nabla^k\beta_{ij}
  \right)\nabla_kN
  \nonumber\\
  &\quad
   +\frac{N}{2}
  \left(
    \nabla_k\nabla_i\beta_j{}^k
    +\nabla_k\nabla_j\beta_i{}^k
    -\nabla^2\beta_{ij}
    -\nabla_i\nabla_j\beta
  \right)\nn \\
  &\quad -\mathcal L_{\vec N}K_{ij}
+
N
\left(
K K_{ij}
-
2K_i{}^kK_{jk}
\right)
=
0. \label{NNLOstationaryeom}
\end{align}

\paragraph{NNLO:}
\begin{align}
\overset{(4)}{\mathcal H}{}^{\rm Gal}
&=
\nabla_i\nabla_j \epsilon^{ij}-\nabla^2 \epsilon-\epsilon^{ij}R_{ij}
\nonumber\\
&\quad +\nabla_k\!\left(
\beta^{ij}\nabla^k\beta_{ij}
-\beta^{ij}\nabla_i\beta_j{}^{k}
+\beta^{ki}\nabla_i\beta
-\beta^{k\ell}\nabla_i\beta^{i}{}_{\ell}
\right)
\nonumber\\
&\quad +\frac12\,(\nabla_i\beta_{jk})(\nabla^j\beta^{ik})
-\frac14\,(\nabla_i\beta_{jk})(\nabla^i\beta^{jk})
-\frac14\,(\nabla_i\beta)(\nabla^i\beta)
+\beta^{i}{}_{k}\beta^{kj}R_{ij} \nn \\
&\quad  -(2K^{ij}L_{ij}
  -
  2KL
  -
  2K_i{}^kK^{ij}\beta_{jk}
  +
  2KK^{ij}\beta_{ij})
=
0, \nn 
\\
\overset{(4)}{\mathcal H}{}^{\rm Gal}_{i}
&=
 2\beta^{jk}\left(\nabla_i K_{jk}-\nabla_k K_{ij}\right)
  +K^{jk}\nabla_i\beta_{jk}
  +2\left(\nabla_j L_i{}^j-\nabla_i L\right)\, \nn \\
  &\quad +K_i{}^j\nabla_j\beta
  -2K_i{}^j\nabla_k\beta_j{}^k
=
0, \nn 
\\
\overset{(4)}{\mathcal A}{}^{\rm Gal}_{ij}
&=
P R_{ij} - \nabla_i\nabla_jP\,  + N R^{(4)}_{ij}\, + \frac12
  \left(
    \nabla_i\beta_j{}^k
    +\nabla_j\beta_i{}^k
    -\nabla^k\beta_{ij}
  \right)\nabla_kM
  \nonumber\\
  &\quad
  + \frac12
  \left(
    \nabla_i\epsilon_j{}^k
    +\nabla_j\epsilon_i{}^k
    -\nabla^k\epsilon_{ij}
  \right)\nabla_kN
  - \frac12\beta^{k\ell}
  \left(
    \nabla_i\beta_{j\ell}
    +\nabla_j\beta_{i\ell}
    -\nabla_\ell\beta_{ij}
  \right)\nabla_kN
  \nonumber\\
  &\quad
  + \frac{M}{2}
  \left(
    \nabla_k\nabla_i\beta_j{}^k
    +\nabla_k\nabla_j\beta_i{}^k
    -\nabla^2\beta_{ij}
    -\nabla_i\nabla_j\beta
  \right)\,\nn \\
  &+\mathcal D_NL_{ij}-\Delta_AK_{ij} + M\left(KK_{ij}-2K_i{}^kK_{jk}\right)
\nonumber\\
& + N\Big[
(L-\beta^{k\ell}K_{k\ell})K_{ij}
+KL_{ij}
-2L_i{}^kK_{jk}
-2K_i{}^kL_{jk}
+2\beta^{k\ell}K_{i\ell}K_{jk}
\Big]=
0.
\end{align}

\subsection{Weak-branch reduction} \label{weakredeom}
We can simplify the field equations even further by setting the following weak-branch ansatz 
\begin{equation}
N=1,
\qquad
N^i=0,
\qquad
\gamma_{ij}=\mathrm{diag}\left(1,r^2,r^2\sin^2\theta\right).
\end{equation}
Therefore, in field equations, we can set 
\begin{equation}
K_{ij}=0,
\qquad
R_{ij}=0,
\qquad
R=0,
\qquad
L_{ij}=-\nabla_{(i}A_{j)}.
\end{equation}

\paragraph{LO:}
The LO equations are identically satisfied by the weak leading data,
\begin{align}
\overset{(0)}{\mathcal H}{}^{\rm Gal}
&=
R
=
0,
\\
\overset{(0)}{\mathcal A}{}^{\rm Gal}_{ij}
&=
R_{ij}
=
0.
\end{align}

\paragraph{NLO:}
The NLO equations reduce to
\begin{align}
\overset{(2)}{\mathcal H}{}^{\rm Gal}
&=
\nabla_i\nabla_j\beta^{ij}
-
\nabla^2\beta
=
0\,, \nn 
\\
\overset{(2)}{\mathcal H}{}^{\rm Gal}_{i}
&=
0\,, \nn \\
\overset{(2)}{\mathcal A}{}^{\rm Gal}_{ij}
&=
-\nabla_i\nabla_jM
+\frac12
\left(
\nabla_k\nabla_i\beta_j{}^k
+\nabla_k\nabla_j\beta_i{}^k
-\nabla^2\beta_{ij}
-\nabla_i\nabla_j\beta
\right)
=
0.
\label{eq:weak-nlo-eom}
\end{align}
Thus the NLO momentum constraint is identically satisfied in the weak branch since $N^i = 0$. The NLO shift \(A_i\) first enters the Galilean equations at NNLO through the momentum constraint. 

\paragraph{NNLO:}
The NNLO equations reduce to
\begin{align}
\overset{(4)}{\mathcal H}{}^{\rm Gal}
&=
\nabla_i\nabla_j \epsilon^{ij}
-
\nabla^2 \epsilon
\nonumber\\
&\quad
+\nabla_k\!\left(
\beta^{ij}\nabla^k\beta_{ij}
-\beta^{ij}\nabla_i\beta_j{}^{k}
+\beta^{ki}\nabla_i\beta
-\beta^{k\ell}\nabla_i\beta^{i}{}_{\ell}
\right)
\nonumber\\
&\quad
+\frac12\,(\nabla_i\beta_{jk})(\nabla^j\beta^{ik})
-\frac14\,(\nabla_i\beta_{jk})(\nabla^i\beta^{jk})
-\frac14\,(\nabla_i\beta)(\nabla^i\beta)
=
0\,, \nn \\
\overset{(4)}{\mathcal H}{}^{\rm Gal}_{i}
&=
2\left(
\nabla_jL_i{}^j
-
\nabla_iL
\right)
=
0\,, \nn \\
\overset{(4)}{\mathcal A}{}^{\rm Gal}_{ij}
&=
-\nabla_i\nabla_jP
+
R^{(4)}_{ij}
+\frac12
\left(
\nabla_i\beta_j{}^k
+\nabla_j\beta_i{}^k
-\nabla^k\beta_{ij}
\right)\nabla_kM
\nonumber\\
&\quad
+\frac{M}{2}
\left(
\nabla_k\nabla_i\beta_j{}^k
+\nabla_k\nabla_j\beta_i{}^k
-\nabla^2\beta_{ij}
-\nabla_i\nabla_j\beta
\right)
=
0.
\label{eq:weak-nnlo-eom}
\end{align}

\subsection{Carrollian field equations}

For completeness, we also list the Carrollian expansion of the field equations. These equations will not be used in the solution analysis, but they make the symmetry between the Galilean and Carrollian hierarchies manifest.

\subsubsection{Carrollian LO gravity equations}
\bea
\overset{(0)}{\mathcal{H}}{}^{\rm Car}
&:=&
\overset{(0)}{\mathcal{H}}{}^{\rm E}
=
0\,,
\nn\\
\overset{(0)}{\mathcal{H}}{}^{\rm Car}_{i}
&:=&
\mathcal{H}^{(0)}_{i}
=
0\,,
\nn\\
\overset{(0)}{\mathcal{B}}{}^{\rm Car}_{ij}
&:=&
\mathcal{B}^{(0)}_{ij}
=
0\,,
\nn\\
\overset{(0)}{\mathcal{A}}{}^{\rm Car}_{ij}
&:=&
\overset{(0)}{\mathcal{A}}{}^{\rm E}_{ij}
=
0\,.
\label{CLOeom}
\eea

\subsubsection{Carrollian NLO gravity equations}
\bea
\overset{(2)}{\mathcal{H}}{}^{\rm Car}
&:=&
\overset{(2)}{\mathcal{H}}{}^{\rm E}
+
\overset{(0)}{\mathcal{H}}{}^{\rm M}
=
0\,,
\nn\\
\overset{(2)}{\mathcal{H}}{}^{\rm Car}_{i}
&:=&
\mathcal{H}^{(2)}_{i}
=
0\,,
\nn\\
\overset{(2)}{\mathcal{B}}{}^{\rm Car}_{ij}
&:=&
\mathcal{B}^{(2)}_{ij}
=
0\,,
\nn\\
\overset{(2)}{\mathcal{A}}{}^{\rm Car}_{ij}
&:=&
\overset{(0)}{\mathcal{A}}{}^{\rm M}_{ij}
+
\overset{(2)}{\mathcal{A}}{}^{\rm E}_{ij}
=
0\,.
\label{CNLOeom}
\eea

\subsubsection{Carrollian NNLO gravity equations}
\bea
\overset{(4)}{\mathcal{H}}{}^{\rm Car}
&:=&
\overset{(2)}{\mathcal{H}}{}^{\rm M}
+
\overset{(4)}{\mathcal{H}}{}^{\rm E}
=
0\,,
\nn\\
\overset{(4)}{\mathcal{H}}{}^{\rm Car}_{i}
&:=&
\mathcal{H}^{(4)}_{i}
=
0\,,
\nn\\
\overset{(4)}{\mathcal{B}}{}^{\rm Car}_{ij}
&:=&
\mathcal{B}^{(4)}_{ij}
=
0\,,
\nn\\
\overset{(4)}{\mathcal{A}}{}^{\rm Car}_{ij}
&:=&
\overset{(2)}{\mathcal{A}}{}^{\rm M}_{ij}
+
\overset{(4)}{\mathcal{A}}{}^{\rm E}_{ij}
=
0\,.
\label{CNNLOeom}
\eea

\section{ADM form of the NLO non-relativistic gravity action}
\label{newtonianaction}

We now compare the NLO ADM action with the non-relativistic gravity action of
Hansen--Hartong--Obers, given in Eq.~(3.29) of \cite{Hansen:2020pqs}. To make the comparison explicit, we express their non-relativistic fields in an adapted ADM parametrization suited to the large-\(c\) expansion used here. In this gauge, the map is
\begin{align}
  v^\mu
  &=
  N^{-1}
  \left(
    \delta^\mu_0-\delta^\mu_iN^i
  \right),
  &
  h^{ij}
  &=
  \gamma^{ij},
  \\
  M^\mu
  &=
  N^{-1}
  \left[
    -\frac{M}{N}\delta^\mu_0
    +\delta^\mu_i
    \left(
      \frac{M}{N}N^i-A^i
    \right)
  \right],
  &
  \Phi^{ij}
  &=
  -\beta^{ij}. \label{mapADMNRG1}
\end{align}
Equivalently, the covariant component relations are
\begin{align}
  h_{ij}
  &=
  \gamma_{ij},
  &
  h_{0i}
  &=
  \gamma_{ij}N^j,
  &
  h_{00}
  &=
  \gamma_{ij}N^iN^j,
  \\
  \tau_\mu
  &=
  -N\delta^0_\mu,
  &
  m_\mu
  &=
  -M\delta^0_\mu,
  \\
  \Phi_{ij}
  &=
  \beta_{ij},
  &
  \Phi_{i0}
  &=
  \beta_{ij}N^j+\gamma_{ij}A^j,
  \\
  \Phi_{00}
  &=
  \beta_{ij}N^iN^j
  +\gamma_{ij}A^iN^j
  +\gamma_{ij}N^iA^j. \label{mapADMNRG2}
\end{align}
\paragraph{NLO ADM action:} The Galilean NLO ADM Lagrangian is 
\bea
\mathcal L^{\rm Gal}_{\rm NLO} = \sqrt{\gamma}\left[ \mathcal{M}^{(2)} + \mathcal{E}^{(0)} \right]\,.
\eea 
Explicitly, 
\begin{align}
  \mathcal{L}^{\rm ADM}_{\rm NLO}
  &=
  \sqrt{\gamma}
  \left[
    \left(
      M+\frac12N\beta
    \right)R
    +NR^{(2)}
    +N
    \left(
      K_{ij}K^{ij}-K^2
    \right)
  \right].
  \label{eq:ADM-NLO-action-before-parts}
\end{align}
Here
\begin{equation}
  \beta:=\gamma^{ij}\beta_{ij},
  \qquad
  R^{(2)}
  =
  \nabla_i\nabla_j\beta^{ij}
  -\nabla^2\beta
  -\beta^{ij}R_{ij}.
\end{equation}
\paragraph{NRG action:} In the notation of  \cite{Hansen:2020pqs}, the covariant NRG Lagrangian is
\begin{align}
  \mathcal{L}_{\rm NRG}
  &=
  \frac{e}{16\pi G_N}
  \Bigg[
    h^{\mu\rho}h^{\nu\sigma}K_{\mu\nu}K_{\rho\sigma}
    -\left(
      h^{\mu\nu}K_{\mu\nu}
    \right)^2
    -2m_\nu
    \left(
      h^{\mu\rho}h^{\nu\sigma}
      -h^{\mu\nu}h^{\rho\sigma}
    \right)
    \check\nabla_\mu K_{\rho\sigma}\nn \\
    &+\Phi h^{\mu\nu}\check R_{\mu\nu}
    +\frac14h^{\mu\rho}h^{\nu\sigma}F_{\mu\nu}F_{\rho\sigma}
    +\frac12\zeta_{\rho\sigma}
    h^{\mu\rho}h^{\nu\sigma}
    \left(
      \partial_\mu\tau_\nu-\partial_\nu\tau_\mu
    \right)
    \nonumber\\
  &\quad
    -\Phi_{\rho\sigma}h^{\mu\rho}h^{\nu\sigma}
    \Big(
      \check R_{\mu\nu}
      -\check\nabla_\mu a_\nu
      -a_\mu a_\nu
      -\frac12h_{\mu\nu}h^{\kappa\lambda}\check R_{\kappa\lambda}
      +h_{\mu\nu}e^{-1}
      \partial_\kappa
      \left(
        eh^{\kappa\lambda}a_\lambda
      \right)
    \Big)
  \Bigg].
  \label{eq:NRG-action-329-copy}
\end{align}
Using the maps \eqref{mapADMNRG1} and \eqref{mapADMNRG2}, we show that these two actions are equivalent up to a boundary term. The Newtonian potential is
\begin{equation}
  \Phi:=-v^\mu m_\mu=\frac{M}{N},
  \qquad
  e=N\sqrt{\gamma}.
\end{equation}
Furthermore, in the adapted ADM gauge,
\begin{equation}
  \tau=-Ndt,
\end{equation}
so that
\begin{equation}
  \tau\wedge d\tau=0,
  \qquad
  h^{\mu\rho}h^{\nu\sigma}
  \left(
    \partial_\mu\tau_\nu-\partial_\nu\tau_\mu
  \right)
  =
  0.
\end{equation}
Thus the Lagrange-multiplier term in \eqref{eq:NRG-action-329-copy} vanishes in this gauge. The possible overall sign difference between the paper's convention
\(K_{\mu\nu}:=-\frac12\mathcal{L}_v h_{\mu\nu}\) and our ADM convention drops out because this term is quadratic. The remaining terms also vanish in the adapted ADM gauge,
\begin{equation}
  h^{\mu\rho}h^{\nu\sigma}F_{\mu\nu}F_{\rho\sigma}=0,
  \qquad
  h^{\mu\rho}h^{\nu\sigma}
  \left(
    \partial_\mu\tau_\nu-\partial_\nu\tau_\mu
  \right)=0,
\end{equation}
because \(m_i=0\) and \(\tau=-Ndt\). Also,
\begin{equation}
  m_\nu
  \left(
    h^{\mu\rho}h^{\nu\sigma}
    -h^{\mu\nu}h^{\rho\sigma}
  \right)
  \check\nabla_\mu K_{\rho\sigma}=0,
\end{equation}
because \(m_\nu\) has only a time component while \(h^{0\mu}=0\) in this parametrization. Putting these reductions together gives
\begin{equation}
  \mathcal{L}_{\rm NRG}
  \doteq
  \frac{1}{16\pi G_N}
  \mathcal{L}^{\rm ADM}_{\rm NLO},
\end{equation}
up to the common normalization and boundary terms. Hence the NLO action \eqref{eq:ADM-NLO-action-before-parts} obtained from the ADM expansion is the adapted ADM-gauge form of \eqref{eq:NRG-action-329-copy}.

\section{ADM conventions and unified read-off map}
\label{app:adm-conventions-read-off}

We use the ADM decomposition
\begin{equation}
  \dd s^2
  =
  -c^2\hat N^2\dd t^2
  +\hat\gamma_{ij}
  \left(\dd x^i+\hat N^i\dd t\right)
  \left(\dd x^j+\hat N^j\dd t\right).
  \label{eq:adm-decomposition}
\end{equation}
Therefore the spacetime metric components are
\begin{equation}
  g_{tt}
  =
  -c^2\hat N^2+\hat\gamma_{ij}\hat N^i\hat N^j,
  \qquad
  g_{ti}
  =
  \hat\gamma_{ij}\hat N^j,
  \qquad
  g_{ij}
  =
  \hat\gamma_{ij}.
  \label{eq:adm-components}
\end{equation}
We expand the ADM variables as
\begin{align}
  \hat\gamma_{ij}
  &=
  \gamma_{ij}
  +\frac{1}{c^2}\beta_{ij}
  +\frac{1}{c^4}\epsilon_{ij}
  +\cO(c^{-6}),
  \label{eq:adm-spatial-expansion}
  \\
  \hat N
  &=
  N+\frac{1}{c^2}M+\frac{1}{c^4}P
  +\cO(c^{-6}),
  \label{eq:adm-lapse-expansion}
  \\
  \hat N^i
  &=
  N^i+\frac{1}{c^2}A^i+\frac{1}{c^4}Z^i
  +\cO(c^{-6}).
  \label{eq:adm-shift-expansion}
\end{align}
Indices on the expansion fields are raised and lowered with the leading spatial
metric \(\gamma_{ij}\). In particular,
\begin{equation}
  N_i:=\gamma_{ij}N^j,
  \qquad
  A_i:=\gamma_{ij}A^j,
  \qquad
  Z_i:=\gamma_{ij}Z^j .
  \label{eq:lowered-shifts}
\end{equation}
The inverse spatial metric is
\begin{equation}
  \hat\gamma^{ij}
  =
  \gamma^{ij}
  -\frac{1}{c^2}\beta^{ij}
  +\frac{1}{c^4}
  \left(
    \beta^i{}_{k}\beta^{kj}
    -\epsilon^{ij}
  \right)
  +\cO(c^{-6}).
  \label{eq:inverse-spatial-expansion}
\end{equation}

\subsection*{Read-off map from the spacetime metric} 

Suppose the spacetime metric components admit the expansion
\begin{align}
  g_{tt}
  &=
  c^2 t^{(2)}
  +t^{(0)}
  +\frac{1}{c^2}t^{(-2)}
  +\cO(c^{-4}),
  \label{eq:gtt-expansion}
  \\
  g_{ti}
  &=
  s_i^{(0)}
  +\frac{1}{c^2}s_i^{(2)}
  +\frac{1}{c^4}s_i^{(4)}
  +\cO(c^{-6}),
  \label{eq:gti-expansion}
  \\
  g_{ij}
  &=
  g_{ij}^{(0)}
  +\frac{1}{c^2}g_{ij}^{(2)}
  +\frac{1}{c^4}g_{ij}^{(4)}
  +\cO(c^{-6}).
  \label{eq:gij-expansion}
\end{align}
The spatial fields are read off directly:
\begin{equation}
  \boxed{
  \gamma_{ij}=g_{ij}^{(0)},
  \qquad
  \beta_{ij}=g_{ij}^{(2)},
  \qquad
  \epsilon_{ij}=g_{ij}^{(4)}.
  }
  \label{eq:spatial-read-off}
\end{equation}
For the shift, using \(g_{ti}=\hat\gamma_{ij}\hat N^j\), one obtains
\begin{align}
  g_{ti}
  &=
  N_i
  +\frac{1}{c^2}
  \left(
    A_i+\beta_{ij}N^j
  \right)
  \nonumber\\
  &\quad
  +\frac{1}{c^4}
  \left(
    Z_i+\beta_{ij}A^j+\epsilon_{ij}N^j
  \right)
  +\cO(c^{-6}).
  \label{eq:shift-forward-map}
\end{align}
Hence the lowered shift fields are
\begin{align}
  \boxed{
  N_i=s_i^{(0)}
  },
  \label{eq:N-lowered-read-off}
  \\
  \boxed{
  A_i=s_i^{(2)}-\beta_{ij}N^j
  },
  \label{eq:A-lowered-read-off}
  \\
  \boxed{
  Z_i=s_i^{(4)}-\beta_{ij}A^j-\epsilon_{ij}N^j
  }.
  \label{eq:Z-lowered-read-off}
\end{align}
The lapse is obtained from
\begin{equation}
  c^2\hat N^2
  =
  -g_{tt}
  +\hat\gamma^{ij}g_{ti}g_{tj},
\end{equation}
or equivalently
\begin{equation}
  \hat N^2
  =
  c^{-2}
  \left(
    -g_{tt}
    +\hat\gamma^{ij}g_{ti}g_{tj}
  \right).
  \label{eq:lapse-master-map}
\end{equation}
Using \eqref{eq:inverse-spatial-expansion}, this gives
\begin{align}
  \hat N^2
  &=
  -t^{(2)}
  +\frac{1}{c^2}
  \left(
    \gamma^{ij}s_i^{(0)}s_j^{(0)}-t^{(0)}
  \right)
  \nonumber\\
  &\quad
  +\frac{1}{c^4}
  \left(
    2\gamma^{ij}s_i^{(0)}s_j^{(2)}
    -\beta^{ij}s_i^{(0)}s_j^{(0)}
    -t^{(-2)}
  \right)
  +\cO(c^{-6}).
  \label{eq:lapse-square-expansion}
\end{align}
Comparing this with
\begin{equation}
  \hat N^2
  =
  N^2
  +\frac{2NM}{c^2}
  +\frac{M^2+2NP}{c^4}
  +\cO(c^{-6}),
\end{equation}
one obtains the unified lapse read-off map
\begin{align}
  \boxed{
  N
  =
  \sqrt{-t^{(2)}}
  },
  \nonumber\\
  \boxed{
  M
  =
  \frac{
  \gamma^{ij}s_i^{(0)}s_j^{(0)}-t^{(0)}
  }{2N}
  },
  \nonumber\\
  \boxed{
  P
  =
  \frac{
  2\gamma^{ij}s_i^{(0)}s_j^{(2)}
  -\beta^{ij}s_i^{(0)}s_j^{(0)}
  -t^{(-2)}
  -M^2
  }{2N}
  }.
  \label{eq:lapsmap}
\end{align}

\subsection*{Inverse map: reconstructing the spacetime metric}

Conversely, starting from the expanded ADM variables
\eqref{eq:adm-spatial-expansion}--\eqref{eq:adm-shift-expansion}, the
spacetime metric coefficients are reconstructed as
\begin{align}
  g_{ij}^{(0)}
  &=
  \gamma_{ij},
  &
  g_{ij}^{(2)}
  &=
  \beta_{ij},
  &
  g_{ij}^{(4)}
  &=
  \epsilon_{ij},
  \label{eq:inverse-spatial-map}
\end{align}
and
\begin{align}
  s_i^{(0)}
  &=
  N_i,
  \label{eq:inverse-shift-map-0}
  \\
  s_i^{(2)}
  &=
  A_i+\beta_{ij}N^j,
  \label{eq:inverse-shift-map-2}
  \\
  s_i^{(4)}
  &=
  Z_i+\beta_{ij}A^j+\epsilon_{ij}N^j .
  \label{eq:inverse-shift-map-4}
\end{align}
For the time-time component one finds
\begin{align}
  t^{(2)}
  &=
  -N^2,
  \label{eq:inverse-lapse-map-2}
  \\
  t^{(0)}
  &=
  -2NM+\gamma_{ij}N^iN^j,
  \label{eq:inverse-lapse-map-0}
  \\
  t^{(-2)}
  &=
  -\left(M^2+2NP\right)
  +2\gamma_{ij}N^iA^j
  +\beta_{ij}N^iN^j .
  \label{eq:inverse-lapse-map-minus2}
\end{align}
Equivalently, the reconstructed components are
\begin{align}
  g_{tt}
  &=
  -c^2N^2
  -2NM
  +\gamma_{ij}N^iN^j
  \nonumber\\
  &\quad
  +\frac{1}{c^2}
  \left[
    -\left(M^2+2NP\right)
    +2\gamma_{ij}N^iA^j
    +\beta_{ij}N^iN^j
  \right]
  +\cO(c^{-4}),
  \label{eq:inverse-gtt-map}
  \\
  g_{ti}
  &=
  N_i
  +\frac{1}{c^2}
  \left(
    A_i+\beta_{ij}N^j
  \right)
  \nonumber\\
  &\quad
  +\frac{1}{c^4}
  \left(
    Z_i+\beta_{ij}A^j+\epsilon_{ij}N^j
  \right)
  +\cO(c^{-6}),
  \label{eq:inverse-gti-map}
  \\
  g_{ij}
  &=
  \gamma_{ij}
  +\frac{1}{c^2}\beta_{ij}
  +\frac{1}{c^4}\epsilon_{ij}
  +\cO(c^{-6}) .
  \label{eq:inverse-gij-map}
\end{align}
The expression \eqref{eq:inverse-gtt-map} reconstructs \(g_{tt}\) up to the
order fixed by the lapse expansion \eqref{eq:adm-lapse-expansion}. The full
\(\cO(c^{-4})\) term in \(g_{tt}\) would require the next coefficient in
\(\hat N=N+c^{-2}M+c^{-4}P+c^{-6}T+\cdots\).

\subsection*{Weak branch}

The weak branch is obtained by taking
\begin{equation}
  t^{(2)}=-1,
  \qquad
  s_i^{(0)}=0,
  \qquad
  \gamma_{ij}
  =
  \operatorname{diag}
  \left(
    1,r^2,r^2\sin^2\theta
  \right).
  \label{eq:weak-branch-conditions}
\end{equation}
The read-off map then reduces to
\begin{equation}
  N=1,
  \qquad
  N_i=0,
  \qquad
  A_i=s_i^{(2)},
  \qquad
  Z_i=s_i^{(4)}-\beta_{ij}A^j,
  \label{eq:weak-shift-read-off}
\end{equation}
and
\begin{equation}
  M=-\frac12 t^{(0)},
  \qquad
  P=-\frac12\left(t^{(-2)}+M^2\right).
  \label{eq:weak-lapse-read-off}
\end{equation}
Conversely, if one starts from the weak-branch ADM data
\begin{equation}
  N=1,
  \qquad
  N^i=0,
\end{equation}
the reconstructed metric components become
\begin{align}
  g_{tt}
  &=
  -c^2
  -2M
  -\frac{1}{c^2}
  \left(
    M^2+2P
  \right)
  +\cO(c^{-4}),
  \label{eq:weak-inverse-gtt-map}
  \\
  g_{ti}
  &=
  \frac{1}{c^2}A_i
  +\frac{1}{c^4}
  \left(
    Z_i+\beta_{ij}A^j
  \right)
  +\cO(c^{-6}),
  \label{eq:weak-inverse-gti-map}
  \\
  g_{ij}
  &=
  \gamma_{ij}
  +\frac{1}{c^2}\beta_{ij}
  +\frac{1}{c^4}\epsilon_{ij}
  +\cO(c^{-6}) .
  \label{eq:weak-inverse-gij-map}
\end{align}
For an axisymmetric configuration, the relevant mixed component is therefore
\begin{equation}
  g_{t\phi}
  =
  \frac{1}{c^2}A_\phi
  +\frac{1}{c^4}
  \left(
    Z_\phi+\beta_{\phi\phi}A^\phi
  \right)
  +\cO(c^{-6}),
  \qquad
  A^\phi=\gamma^{\phi\phi}A_\phi .
  \label{eq:weak-inverse-gtphi-map}
\end{equation}
The NNLO Galilean equations do not fix an independent higher-order shift
coefficient \(Z_i\). Therefore, in the weak-branch reconstruction, the
metric components fixed directly by the NNLO Galilean data are obtained
by setting \(Z_i=0\), unless a separate relativistic uplift is specified.

\section{Kerr metric} \label{kerrapp}
In this appendix, we expand the Kerr metric order by order.  Let us begin with the
Kerr metric in Boyer–Lindquist coordinates \cite{Kerr:1963ud}
\begin{equation}
\begin{aligned}
ds^{2}
= {}&
-\left(1-\frac{2Gm r}{\Sigma c^{2}}\right)c^{2}dt^{2}
-\frac{4Gm a r \sin^{2}\theta}{\Sigma c^{2}}\,c\,dt\,d\phi\, \nn \\
&
+\frac{\Sigma}{\Delta}\,dr^{2}
+\Sigma\,d\theta^{2}
+\left(r^{2}+a^{2}
+\frac{2Gm a^{2} r \sin^{2}\theta}{\Sigma c^{2}}\right)
\sin^{2}\theta\,d\phi^{2},
\end{aligned}
\end{equation}
where
\bea
\Sigma = r^{2}+a^{2}\cos^{2}\theta,
\qquad
\Delta = r^{2}-\frac{2Gm r}{c^{2}}+a^{2}, \qquad a = \frac{J}{m c}.
\eea 
\subsection{Weak Kerr branch}

The weak branch keeps \(G,m,J\) fixed and expands around flat leading ADM data.  The
Kerr metric components through NNLO are
\begin{align}
  g_{tt}
  &=
  -c^{2}
  +\frac{2Gm}{r}
  -c^{-2}\frac{2GJ^2}{m r^3}\cos^2\theta
  +c^{-4}\frac{2GJ^4}{m^3r^5}\cos^4\theta
  +O(c^{-6}), \label{eq:weak-kerr-metric-components}
  \\
  g_{t\phi}
  &=
  -c^{-2}\frac{2GJ}{r}\sin^2\theta
  +c^{-4}\frac{2GJ^3}{m^2r^3}\sin^2\theta\cos^2\theta
  +O(c^{-6}),
  \\
  g_{rr}
  &=
  1+c^{-2}
  \left(
    \frac{2Gm}{r}
    -\frac{J^2}{m^2r^2}\sin^2\theta
  \right)
  \nonumber\\
  &\quad
  +c^{-4}
  \frac{
  \left(\frac{J^2}{m^2}-2Gmr\right)
  \left(\frac{J^2}{m^2}\sin^2\theta-2Gmr\right)}
  {r^4}
  +O(c^{-6}),
  \\
  g_{\theta\theta}
  &=
  r^2+c^{-2}\frac{J^2}{m^2}\cos^2\theta+O(c^{-6}),
  \\
  g_{\phi\phi}
  &=
  r^2\sin^2\theta
  +c^{-2}\frac{J^2}{m^2}\sin^2\theta
  +c^{-4}\frac{2GJ^2}{m r}\sin^4\theta
  +O(c^{-6}).
\end{align}
\paragraph{Comparison with Ref.~\cite{Hartong:2023yxo}.}
A large-$c$ expansion of the Kerr metric was also studied in
Ref.~\cite{Hartong:2023yxo}. The expressions above are obtained by
expanding the Kerr metric directly in Boyer--Lindquist variables and then
reading off the ADM fields. These variables correspond to the oblate
spheroidal variables $(R,\Theta,\phi)$ used in Ref.~\cite{Hartong:2023yxo}.
To avoid confusion with the notation used above, let us denote the
ordinary spherical variables of Ref.~\cite{Hartong:2023yxo} by
$(r',\theta',\phi)$. At the first relevant order the relation is
\begin{align}
R^2
&=
r'^2-\frac{J^2}{m^2c^2}\sin^2\theta'
+\mathcal O(c^{-4}),
\\
\Theta
&=
\theta'
-\frac{J^2}{2m^2r'^2c^2}\sin\theta'\cos\theta'
+\mathcal O(c^{-4}) .
\end{align}
Since $a=J/(mc)$, this coordinate transformation starts at order $c^{-2}$
and mixes the leading spatial metric with the subleading fields.

For example, rewriting the $g_{tt}$ component displayed above in terms of
the Boyer--Lindquist variables $(R,\Theta)$ gives
\begin{equation}
g_{tt}
=
-c^2
+\frac{2Gm}{R}
-\frac{2GJ^2}{mR^3c^2}\cos^2\Theta
+\mathcal O(c^{-4}) .
\end{equation}
Using
\begin{equation}
R
=
r'
-\frac{J^2}{2m^2r'c^2}\sin^2\theta'
+\mathcal O(c^{-4}),
\end{equation}
the monopole term becomes
\begin{equation}
\frac{2Gm}{R}
=
\frac{2Gm}{r'}
+
\frac{GJ^2}{mr'^3c^2}\sin^2\theta'
+\mathcal O(c^{-4}) .
\end{equation}
Therefore the complete spin-squared coefficient in $g_{tt}$ is
\begin{equation}
\frac{GJ^2}{mr'^3}
\left(
\sin^2\theta'-2\cos^2\theta'
\right)
=
-\frac{2GJ^2}{mr'^3}P_2(\cos\theta') .
\end{equation}
Thus, after the $c$-dependent coordinate transformation, the $g_{tt}$
component takes the quadrupolar form appearing in Eq.~(4.10) of
Ref.~\cite{Hartong:2023yxo}.

The leading frame-dragging component is unchanged at this order. Indeed,
the direct expansion gives
\begin{equation}
g_{t\phi}
=
-\frac{2GJ}{Rc^2}\sin^2\Theta
+\mathcal O(c^{-4}) .
\end{equation}
Since the coordinate corrections in $R$ and $\Theta$ start at order
$c^{-2}$, they only affect this term at order $c^{-4}$. Therefore, at the
order displayed here,
\begin{equation}
g_{t\phi}
=
-\frac{2GJ}{r'c^2}\sin^2\theta'
+\mathcal O(c^{-4}) .
\end{equation}
The apparent difference between the two expansions is therefore a
coordinate/read-off effect, not a physical disagreement.

One can use the read-off map in the previous section to obtain the ADM data as the following. 
\paragraph{Leading ADM read-off data:}
\begin{equation}
  N=1,
  \qquad
  \gamma_{ij}
  =
  \operatorname{diag}\left(1,r^2,r^2\sin^2\theta\right).
\end{equation}

\paragraph{Order-\(c^{-2}\) ADM read-off data:}
\begin{align}
  M
  &=
  -\frac{Gm}{r},
  &
  A_\phi
  &=
  -\frac{2GJ}{r}\sin^2\theta,
  \\
  \beta_{rr}
  &=
  \frac{2Gm}{r}
  -\frac{J^2}{m^2r^2}\sin^2\theta,
  &
  \beta_{\theta\theta}
  &=
  \frac{J^2}{m^2}\cos^2\theta,
  &
  \beta_{\phi\phi}
  &=
  \frac{J^2}{m^2}\sin^2\theta.
\end{align}
\paragraph{Order-\(c^{-4}\) ADM read-off data:}
\begin{align}
  P
  &=
  \frac{GJ^2}{m r^3}\cos^2\theta
  -\frac{G^2m^2}{2r^2}\,, \qquad 
  Z_\phi
  =
  \frac{2GJ^3}{m^2r^3}\sin^2\theta
  \left(1+\cos^2\theta\right),
  \\ 
  \epsilon_{rr}
  &=
  \frac{
  \left(\frac{J^2}{m^2}-2Gmr\right)
  \left(\frac{J^2}{m^2}\sin^2\theta-2Gmr\right)}
  {r^4}\,, \qquad \epsilon_{\theta\theta}
  =
  0\,, \qquad 
  \epsilon_{\phi\phi}
  =
  \frac{2GJ^2}{m r}\sin^4\theta.
\end{align}

\subsection{Strong Kerr branch}\label{strongkerrapp}

The strong branch keeps the Schwarzschild factor
\[
  f(r)=1-\frac{2Gm}{r}
\]
at leading order with the following scaling of the parameters
\bea
m \to c\, m\,, \qquad G\to c\, G\,, \qquad J \to c\, J\,.  
\eea 
The metric components through NNLO are
\begin{align}
  g_{tt}
  &=
  -c^{2}f
  -\frac{2GJ^2}{m r^3}\cos^2\theta
  +c^{-2}\frac{2GJ^4}{m^3r^5}\cos^4\theta
  -c^{-4}\frac{2GJ^6}{m^5r^7}\cos^6\theta
  +O(c^{-6}),
  \\
  g_{t\phi}
  &=
  -\frac{2GJ}{r}\sin^2\theta
  +c^{-2}\frac{2GJ^3}{m^2r^3}\sin^2\theta\cos^2\theta -c^{-4}\frac{2GJ^5}{m^4r^5}\sin^2\theta\cos^4\theta
  +O(c^{-6}),
  \\
  g_{rr}
  &=
  \frac{1}{f}
  +c^{-2}\frac{J^2}{m^2r^2}
  \left(
    \frac{\cos^2\theta}{f}
    -\frac{1}{f^2}
  \right) +c^{-4}\frac{J^4}{m^4r^4}
  \left(
    \frac{1}{f^3}
    -\frac{\cos^2\theta}{f^2}
  \right)
  +O(c^{-6}),
  \\
  g_{\theta\theta}
  &=
  r^2+c^{-2}\frac{J^2}{m^2}\cos^2\theta+O(c^{-6}),
  \\
  g_{\phi\phi}
  &=
  r^2\sin^2\theta
  +c^{-2}\frac{J^2}{m^2}\sin^2\theta
  \left(
    1+\frac{2Gm}{r}\sin^2\theta
  \right)
  \nonumber\\
  &\quad
  -c^{-4}\frac{2GJ^4}{m^3r^3}
  \sin^4\theta\cos^2\theta
  +O(c^{-6}).
\end{align}
Using the read-off map from the previous section, we can infer the following ADM variables. 
\paragraph{Leading ADM read-off data:}
\[
  N=\sqrt f,
  \qquad
  \gamma_{ij}
  =
  \operatorname{diag}\left(f^{-1},r^2,r^2\sin^2\theta\right),
  \qquad
  N_\phi=-\frac{2GJ}{r}\sin^2\theta.
\]
\paragraph{Order-\(c^{-2}\) ADM read-off data:}
\begin{align}
  M
  &=
  \frac{1}{\sqrt f}
  \left(
    \frac{GJ^2}{m r^3}\cos^2\theta
    +\frac{2G^2J^2}{r^4}\sin^2\theta
  \right),
  \\
  A_\phi
  &=
  \frac{2GJ^3}{m^2r^3}\sin^2\theta
  \left(
    1+\cos^2\theta+\frac{2Gm}{r}\sin^2\theta
  \right),
  \\
  \beta_{rr}
  &=
  \frac{J^2}{m^2r^2}
  \left(
    \frac{\cos^2\theta}{f}
    -\frac{1}{f^2}
  \right),\qquad
  \beta_{\theta\theta}
  =
  \frac{J^2}{m^2}\cos^2\theta,
  \\
  \beta_{\phi\phi}
  &=
  \frac{J^2}{m^2}\sin^2\theta
  \left(
    1+\frac{2Gm}{r}\sin^2\theta
  \right).
\end{align}
\paragraph{Order-\(c^{-4}\) ADM read-off data:}
\begin{align}
  \epsilon_{rr}
  &=
  \frac{J^4}{m^4r^4}
  \left(
    \frac{1}{f^3}
    -\frac{\cos^2\theta}{f^2}
  \right),\qquad 
  \epsilon_{\theta\theta}
  =
  0,
  \\
  \epsilon_{\phi\phi}
  &=
  -\frac{2GJ^4}{m^3r^3}
  \sin^4\theta\cos^2\theta,
  \\
  P
  &=
  -\frac{J^4}{2m^4\sqrt f}
  \Bigg[
  \frac{2Gm}{r^5}\cos^4\theta
  \nonumber\\
  &\qquad
  +\frac{4G^2m^2}{r^6}\sin^2\theta
  \left(
    1+2\cos^2\theta+\frac{2Gm}{r}\sin^2\theta
  \right)
  \nonumber\\
  &\qquad
  +\frac{G^2m^2}{f r^6}
  \left(
    \cos^2\theta+\frac{2Gm}{r}\sin^2\theta
  \right)^2
  \Bigg],
  \\
  Z_\phi
  &=
  -\frac{2GJ^5}{m^4r^5}\sin^2\theta
  \Bigg[
  1+\cos^2\theta+\cos^4\theta
  \nonumber\\
  &\qquad
  +\frac{4Gm}{r}\sin^2\theta\left(1+\cos^2\theta\right)
  +\frac{4G^2m^2}{r^2}\sin^4\theta
  \Bigg].
\end{align}

\section{Hartle-Thorne metric} \label{htapp}

The Hartle–Thorne metric is an approximate solution of the vacuum Einstein equations describing the exterior field of a slowly rotating, weakly deformed compact object \cite{Hartle:1967he,Hartle:1968si}. The  metric is 
\begin{align}
ds^{2} &=
-\left(1-\frac{2mG}{c^2 r}\right)\!\Bigg[1+2k_{1}P_{2}(\cos\theta)
+2\left(1-\frac{2mG}{c^2r}\right)^{-1}\frac{G^2J^{2}}{c^6r^{4}}\,(2\cos^{2}\theta-1)\Bigg]\,c^2 dt^{2} \notag\\
&\quad
+\left(1-\frac{2m G}{c^2 r}\right)^{-1}\!\Bigg[1-2k_{2}P_{2}(\cos\theta)
-2\left(1-\frac{2mG}{c^2r}\right)^{-1}\frac{G^2J^{2}}{c^6r^{4}}\Bigg]\,dr^{2} \notag\\
&\quad
+r^{2}\,[1-2k_{3}P_{2}(\cos\theta)]\,(d\theta^{2}+\sin^{2}\theta\,d\phi^{2})
-\frac{4GJ}{c^2r}\sin^{2}\theta\, dt\,d\phi  \label{HT}
\end{align}
where
\begin{align}
k_{1} &= \frac{G J^{2}}{c^4 m r^{3}}\!\left(1+\frac{Gm}{c^2r}\right)
+\frac{5c^2}{8}\,\frac{GQc^2-\tfrac{GJ^{2}}{m}}{G^3m^{3}}\,
Q^{2}_{2}\!\left(\frac{c^2r}{Gm}-1\right)\,, \nn \\
k_{2} &= k_{1}-\frac{6G^2J^{2}}{c^6r^{4}}\,,  \nn \\
k_{3} &= k_{1}+\frac{G^2J^{2}}{c^6r^{4}}
+\frac{5c^2}{4}\,\frac{GQ-\tfrac{GJ^{2}}{c^2m}}{G^2m^{2}\sqrt{r^{2}-\tfrac{2Gmr}{c^2}}}\,
Q^{1}_{2}\!\left(\frac{c^2r}{Gm}-1\right)\,, \nn \\
P_{2}(\cos\theta) &= \tfrac12\,(3\cos^{2}\theta-1)\,, \nn \\
Q^{1}_{2}(x) &= (x^{2}-1)^{1/2}\!\left[\frac{3x}{2}\ln\!\frac{x+1}{x-1}
-\frac{3x^{2}-2}{x^{2}-1}\right]\,, \nn \\
Q^{2}_{2}(x) &= (x^{2}-1)\!\left[\frac{3}{2}\ln\!\frac{x+1}{x-1}
-\frac{3x^{3}-5x}{(x^{2}-1)^{2}}\right].
\end{align}
\subsection{Weak Hartle--Thorne branch}
Keeping \(G,m,J,Q\) fixed and expanding \eqref{HT} for \(c\to\infty\) gives
\begin{align}
  g_{tt}
  &=
  -c^2
  +\left(\frac{2Gm}{r}-\frac{2GQ\Ptwo}{r^3}\right)
  -\frac{1}{c^2}\frac{2G^2mQ\Ptwo}{r^4}
  +\cO(c^{-4}),
  \label{eq:weak-ht-gtt}\\
  g_{t\phi}
  &=
  -\frac{1}{c^2}\frac{2GJ}{r}\sin^2\theta\,,
  \label{eq:weak-ht-gtphi}\\
  g_{rr}
  &=
  1+\frac{1}{c^2}
  \left(
  \frac{2Gm}{r}-\frac{2GQ\Ptwo}{r^3}
  \right)
  +\frac{1}{c^4}
  \left(
  \frac{4G^2m^2}{r^2}
  -\frac{10G^2mQ\Ptwo}{r^4}
  \right)
  +\cO(c^{-6}),
  \label{eq:weak-ht-grr}\\
  g_{\theta\theta}
  &=
  r^2
  -\frac{1}{c^2}\frac{2GQ\Ptwo}{r}
  -\frac{1}{c^4}\frac{5G^2mQ\Ptwo}{r^2}
  +\cO(c^{-6}),
  \label{eq:weak-ht-gthetatheta}\\
  g_{\phi\phi}
  &=
  \left(
  r^2
  -\frac{1}{c^2}\frac{2GQ\Ptwo}{r}
  -\frac{1}{c^4}\frac{5G^2mQ\Ptwo}{r^2}
  \right)\sin^2\theta
  +\cO(c^{-6}).
  \label{eq:weak-ht-gphiphi}
\end{align}
Again we can use the read-off map, and infer the following ADM expansions. 
\paragraph{Order-\(c^{-2}\) ADM read-off data:}
\begin{align}
  M
  &=
  -\frac{Gm}{r}+\frac{GQ\Ptwo}{r^3},
  &
  A_\phi
  &=
  -\frac{2GJ}{r}\sin^2\theta,
  \label{eq:weak-ht-nlo-scalars}\\
  \beta_{rr}
  &=
  \frac{2Gm}{r}-\frac{2GQ\Ptwo}{r^3},
  &
  \beta_{\theta\theta}
  &=
  -\frac{2GQ\Ptwo}{r},
  &
  \beta_{\phi\phi}
  &=
  -\frac{2GQ\Ptwo}{r}\sin^2\theta.
  \label{eq:weak-ht-beta}
\end{align}
The NLO Hartle-Thorne weak expansion is the exact solution of the NLO Galilean gravity theory.  
\paragraph{Order-\(c^{-4}\) ADM read-off data:}
The NNLO fields read from the same seed as
\begin{align}
  P
  &=-\frac{1}{2}
\left(
-\frac{Gm}{r}
+
\frac{GQ}{r^3}P_2(\cos\theta)
\right)^2
+
\frac{G^2mQ}{r^4}P_2(\cos\theta), \nn \\
  \epsilon_{rr}
  &=
  \frac{4G^2m^2}{r^2}
  -\frac{10G^2mQ\Ptwo}{r^4}, \nn 
  \\
  \epsilon_{\theta\theta}
  &=
  -\frac{5G^2mQ\Ptwo}{r^2}, \nn 
  \\
  \epsilon_{\phi\phi}
  &=
  -\frac{5G^2mQ\Ptwo}{r^2}\sin^2\theta.
  \label{eq:weak-ht-epsilon}
\end{align}
With the standard Geroch--Hansen convention: 
\(M_2^{\rm Kerr}=-J^2/m\) and \(Q=-M_2\).

\subsection{Strong scaling of the Hartle--Thorne metric}
\label{shtapp}

In this appendix we present the strong-scaling read-off of the ordinary
Hartle--Thorne metric.  For the strong large-\(c\) expansion we use
\begin{equation}
  m\to c\,m,
  \qquad
  G\to c\,G,
  \qquad
  J\to c\,J,
  \qquad
  Q\to c^{-1}\,Q .
  \label{eq:strong-scaling-HT}
\end{equation}
With this scaling,
\begin{equation}
  f(r)=1-\frac{2Gm}{r},
  \qquad
  x=\frac{r}{Gm}-1 .
\end{equation}
Therefore the logarithmic functions appearing in the Hartle--Thorne metric
are not expanded.  For example,
\begin{equation}
  \log\frac{x+1}{x-1}
  =
  \log\frac{r}{r-2Gm}.
\end{equation}
The Hartle--Thorne functions scale as
\begin{equation}
  k_a=c^{-2}H_a,
  \qquad
  a=1,2,3,
\end{equation}
where
\begin{align}
  H_1
  &=
  \frac{GJ^2}{mr^3}
  \left(
  1+\frac{Gm}{r}
  \right)
  +
  \frac{5}{8}
  \frac{GQ-\frac{GJ^2}{m}}{G^3m^3}
  Q_2^2(x),
  \\
  H_2
  &=
  H_1-\frac{6G^2J^2}{r^4},
  \\
  H_3
  &=
  H_1
  +
  \frac{G^2J^2}{r^4}
  +
  \frac{5}{4}
  \frac{
  GQ-\frac{GJ^2}{m}
  }{
  G^2m^2\sqrt{r^2-2Gmr}
  }
  Q_2^1(x).
\end{align}
The strong-scaled Hartle--Thorne metric then takes the form
\begin{align}
  g_{tt}
  &=
  -c^2 f
  -2fH_1P_2(\cos\theta)
  -\frac{2G^2J^2}{r^4}
  \left(
  2\cos^2\theta-1
  \right),
  \\
  g_{t\phi}
  &=
  -\frac{2GJ}{r}\sin^2\theta,
  \\
  g_{rr}
  &=
  f^{-1}
  -
  \frac{1}{c^2}f^{-1}
  \left[
  2H_2P_2(\cos\theta)
  +
  \frac{2G^2J^2}{fr^4}
  \right],
  \\
  g_{\theta\theta}
  &=
  r^2
  -
  \frac{2}{c^2}r^2H_3P_2(\cos\theta),
  \\
  g_{\phi\phi}
  &=
  \left[
  r^2
  -
  \frac{2}{c^2}r^2H_3P_2(\cos\theta)
  \right]\sin^2\theta .
\end{align}
Note that under this scaling the terms at the order of $\mathcal{O} (c^{-4})$ do not show up. As a result we do not have the NNLO for the strong Hartle--Thorne metric. We work in the exterior region where 
\begin{equation}
r>2Gm .
\end{equation}
\paragraph{Leading ADM read-off data:}
\[
  N=\sqrt f,
  \qquad
  \gamma_{ij}
  =
  \operatorname{diag}\left(f^{-1},r^2,r^2\sin^2\theta\right),
  \qquad
  N_\phi=-\frac{2GJ}{r}\sin^2\theta.
\]

\paragraph{Order-\(c^{-2}\) ADM read-off data:}
\begin{align}
  M
  &=
  \frac{1}{\sqrt{f(r)}}
\left[
f(r)H_1P_2(\cos\theta)
+
\frac{G^2J^2}{r^4}
\right],
  \\
  A_\phi
  &=
  -\frac{4GJ}{r}
H_3P_2(\cos\theta)\sin^2\theta,
  \\
  \beta_{rr}
  &=
  -\frac{1}{f(r)}
\left[
2H_2P_2(\cos\theta)
+
\frac{2G^2J^2}{f(r)r^4}
\right],\qquad
  \beta_{\theta\theta}
  =
 -2r^2H_3P_2(\cos\theta),
  \\
  \beta_{\phi\phi}
  &= \beta_{\theta\theta}
  \sin^2\theta.
\end{align}

\section{Extended HT: Post-linear quadrupole-quadrupole metric}
To control the quadrupole-squared sector, we use the post-linear
quadrupole--quadrupole Hartle--Thorne metric of
\cite{Frutos-Alfaro:2015lua}. Restoring \(G\) and \(c\), and using our
quadrupole convention, its components may be written as

\begin{align}
  g_{tt}^{\rm PL}
  &=
  -c^2
  \left[
  1-\frac{2Gm}{c^2r}
  +\frac{2GQ}{c^2r^3}\Ptwo
  \right.
  \nonumber\\
  &\qquad\left.
  +\frac{2G^2mQ}{c^4r^4}\Ptwo
  +\frac{2G^2Q^2}{c^4r^6}\Ptwo^2
  -\frac{2G^2J^2}{3c^6r^4}\left(2\Ptwo+1\right)
  \right],
  \label{eq:pl-gtt}\\
  g_{t\phi}^{\rm PL}
  &=
  -\frac{2GJ}{c^2r}\sin^2\theta,
  \label{eq:pl-gtphi}\\
  g_{rr}^{\rm PL}
  &=
  1+\frac{2Gm}{c^2r}
  +\frac{4G^2m^2}{c^4r^2}
  -\frac{2GQ}{c^2r^3}\Ptwo
  -\frac{10G^2mQ}{c^4r^4}\Ptwo
  \nonumber\\
  &\quad
  +\frac{G^2Q^2}{12c^4r^6}
  \left(8\Ptwo^2-16\Ptwo+77\right)
  \nonumber\\
  &\quad
  +\frac{2G^2J^2}{c^6r^4}\left(8\Ptwo-1\right),
  \label{eq:pl-grr}\\
  g_{\theta\theta}^{\rm PL}
  &=
  r^2
  \left[
  1-\frac{2GQ}{c^2r^3}\Ptwo
  -\frac{5G^2mQ}{c^4r^4}\Ptwo
  \right.
  \nonumber\\
  &\qquad\left.
  +\frac{G^2Q^2}{36c^4r^6}
  \left(44\Ptwo^2+8\Ptwo-43\right)
  +\frac{G^2J^2}{c^6r^4}\Ptwo
  \right],
  \label{eq:pl-gthetatheta}\\
  g_{\phi\phi}^{\rm PL}
  &=
  g_{\theta\theta}^{\rm PL}\sin^2\theta.
  \label{eq:pl-gphiphi}
\end{align}
The \(Q^2\)-terms are those of \cite{Frutos-Alfaro:2015lua}, while the
\(J^2\)-terms are the standard Hartle--Thorne slow-rotation
contributions included for comparison. For the NNLO lapse coefficient \(P\), only the \(c^{-2}\) coefficient of \(g_{tt}\) is
needed.
Equation \eqref{eq:pl-gtt} gives
\begin{equation}
  t^{(-2)}
  =
  -\frac{2G^2mQ}{r^4}\Ptwo
  -\frac{2G^2Q^2}{r^6}\Ptwo^2.
  \label{eq:pl-t2}
\end{equation}
Using the map given in Section~\ref{app:adm-conventions-read-off}, while keeping the same NLO field
\[
  M=-\frac{Gm}{r}+\frac{GQ\Ptwo}{r^3},
\]
one obtains the post-linear NNLO lapse coefficient
\begin{align}
  P &=
  -\frac{G^2m^2}{2r^2}
  +\frac{2G^2mQ}{r^4}\Ptwo
  +\frac{G^2Q^2}{2r^6}\Ptwo^2.
  \label{eq:pl-p-final}
\end{align}
This is precisely the term required by the NNLO weak Galilean equations. The post-linear metric also corrects the spatial NNLO field
\eqref{eq:weak-ht-epsilon} by
\begin{align}
  \Delta_{Q^2}\epsilon_{rr}
  &=
  \frac{G^2Q^2}{12r^6}
  \left(8\Ptwo^2-16\Ptwo+77\right),
  \\
  \Delta_{Q^2}\epsilon_{\theta\theta}
  &=
  \frac{G^2Q^2}{36r^4}
  \left(44\Ptwo^2+8\Ptwo-43\right),
  \\
  \Delta_{Q^2}\epsilon_{\phi\phi}
  &=
  \frac{G^2Q^2}{36r^4}
  \left(44\Ptwo^2+8\Ptwo-43\right)\sin^2\theta.
\end{align}
The NLO fields \(M,A_i,\beta_{ij}\) are unchanged and given in \eqref{eq:weak-ht-nlo-scalars}- \eqref{eq:weak-ht-beta}.

\newpage

\bibliographystyle{JHEP}
\bibliography{bsld}
\end{document}